\pdftrailerid{}
\PassOptionsToPackage{unicode}{hyperref}
\PassOptionsToPackage{hyphens}{url}
\PassOptionsToPackage{dvipsnames,svgnames,x11names}{xcolor}
\documentclass[12pt]{article}

\newif\ifanon

\anonfalse

\newif\ifauthorchecks
\authorchecksfalse

\newif\ifdlmarks
\dlmarksfalse
\newcommand{\dlfill}[1]{\ifdlmarks{\color{dlblue}#1}\else#1\fi}

\newcommand{\dlnum}[1]{\ifdlmarks{\color{dlblue}#1}\else#1\fi}

\usepackage{amsmath,amssymb,amsthm,mathtools,bm}
\usepackage[utf8]{inputenc}
\usepackage{textcomp}
\IfFileExists{lmodern.sty}{%
  \usepackage[T1]{fontenc}
  \usepackage{lmodern}
  \IfFileExists{microtype.sty}{\usepackage{microtype}
    \UseMicrotypeSet[protrusion]{basicmath}}{}
}{
  \IfFileExists{microtype.sty}{\usepackage[expansion=false]{microtype}}{}
}
\IfFileExists{upquote.sty}{\usepackage{upquote}}{}
\usepackage{xcolor}
\definecolor{dlblue}{RGB}{0,82,155}
\definecolor{dlorange}{RGB}{186,85,0}
\usepackage{booktabs,array,multirow}
\usepackage{graphicx}
\usepackage{caption}
\usepackage{float}
\usepackage{algorithm}
\usepackage{algpseudocode}
\usepackage{enumitem}
\setlist[enumerate,1]{label=(\alph*)}
\setlist[enumerate,2]{label=(\roman*)}

\usepackage[round]{natbib}
\usepackage{xr-hyper}
\IfFileExists{SM.aux}{}
  {\IfFileExists{SM-labels.tex}{\makeatletter
\newlabel{appendix:A1}{{S1}{1}{Nuisance estimation, cross-fitting, and variance estimation}{section.1}{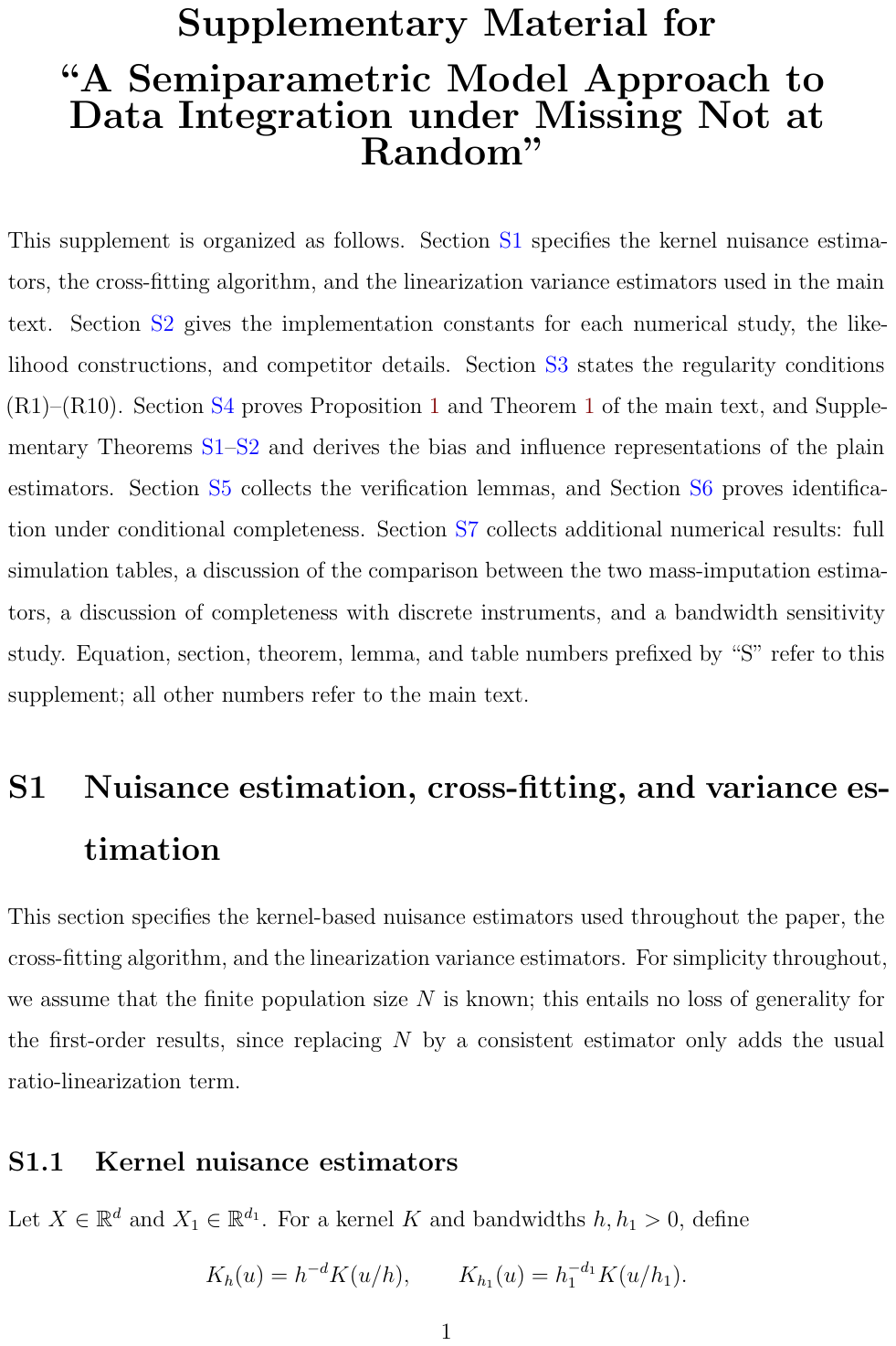}}
\newlabel{app:nuisance-kernel}{{S1.1}{1}{Kernel nuisance estimators}{subsection.1.1}{paper_SM.pdf}}
\newlabel{eq:mhat}{{S1}{2}{Kernel nuisance estimators}{equation.1.1}{paper_SM.pdf}}
\newlabel{eq:ghat-app}{{S2}{2}{Kernel nuisance estimators}{equation.1.2}{paper_SM.pdf}}
\newlabel{eq:ag-hat}{{S3}{3}{Kernel nuisance estimators}{equation.1.3}{paper_SM.pdf}}
\newlabel{app:cross-fit}{{S1.2}{3}{Cross-fitting algorithm}{subsection.1.2}{paper_SM.pdf}}
\newlabel{appendix:variance}{{S1.3}{4}{Linearization variance estimators}{subsection.1.3}{paper_SM.pdf}}
\newlabel{eq:Jhat}{{S4}{5}{Linearization variance estimators}{equation.1.4}{paper_SM.pdf}}
\newlabel{rmk:mar-var}{{S1}{8}{The $\gamma $-correction under the MAR submodel}{remark.1}{paper_SM.pdf}}
\newlabel{supp:impl}{{S2}{9}{Implementation details}{section.2}{paper_SM.pdf}}
\newlabel{supp:bw-rates}{{S2.1}{9}{Bandwidth rates}{subsection.2.1}{paper_SM.pdf}}
\newlabel{supp:constants}{{S2.2}{9}{Constants used in the numerical studies}{subsection.2.2}{paper_SM.pdf}}
\newlabel{supp:likelihood}{{S2.3}{12}{Likelihood constructions}{subsection.2.3}{paper_SM.pdf}}
\newlabel{sec:param-ps}{{S2.3}{12}{Likelihood constructions}{subsection.2.3}{paper_SM.pdf}}
\newlabel{eq:param-ps}{{S5}{12}{Likelihood constructions}{equation.2.5}{paper_SM.pdf}}
\newlabel{eq:pseudo-loglik}{{S6}{13}{Likelihood constructions}{equation.2.6}{paper_SM.pdf}}
\newlabel{eq:profile-loglik}{{S7}{13}{Likelihood constructions}{equation.2.7}{paper_SM.pdf}}
\newlabel{eq:gamma-pl}{{S8}{13}{Likelihood constructions}{equation.2.8}{paper_SM.pdf}}
\newlabel{supp:competitors}{{S2.4}{13}{Additional competing estimators}{subsection.2.4}{paper_SM.pdf}}
\newlabel{eq:smz-score}{{S9}{13}{Additional competing estimators}{equation.2.9}{paper_SM.pdf}}
\newlabel{appendix:regularity}{{S3}{14}{Regularity conditions}{section.3}{paper_SM.pdf}}
\newlabel{appendix:proofs}{{S4}{21}{Proofs of the main results}{section.4}{paper_SM.pdf}}
\newlabel{eq:dPhi0}{{S10}{22}{Proof of Proposition~\ref {prop:orth}: tilt score}{equation.4.10}{paper_SM.pdf}}
\newlabel{eq:dPhig}{{S11}{22}{Proof of Proposition~\ref {prop:orth}: tilt score}{equation.4.11}{paper_SM.pdf}}
\newlabel{eq:dPhim}{{S12}{22}{Proof of Proposition~\ref {prop:orth}: tilt score}{equation.4.12}{paper_SM.pdf}}
\newlabel{eq:mi-deriv}{{S13}{22}{Proof of Proposition~\ref {prop:orth}: mean score}{equation.4.13}{paper_SM.pdf}}
\newlabel{eq:mi-orth-if-app}{{S14}{25}{Proof of Theorem~\ref {thm:joint}: population mean}{equation.4.14}{paper_SM.pdf}}
\newlabel{eq:mi-orth-if-def-app}{{S15}{25}{Proof of Theorem~\ref {thm:joint}: population mean}{equation.4.15}{paper_SM.pdf}}
\newlabel{app:plain-representations}{{S4.1}{25}{Bias and influence representations}{subsection.4.1}{paper_SM.pdf}}
\newlabel{thm:ipw}{{S1}{25}{Plain IPW with first-order smoothing bias}{theorem.1}{paper_SM.pdf}}
\newlabel{eq:ipw-bias-limit}{{S16}{25}{Plain IPW with first-order smoothing bias}{equation.4.16}{paper_SM.pdf}}
\newlabel{thm:mi}{{S2}{26}{Plain MI with first-order smoothing bias}{theorem.2}{paper_SM.pdf}}
\newlabel{eq:mi-bias-limit}{{S17}{26}{Plain MI with first-order smoothing bias}{equation.4.17}{paper_SM.pdf}}
\newlabel{eq:ipw-bias-detail}{{S18}{26}{Bias and influence representations}{equation.4.18}{paper_SM.pdf}}
\newlabel{eq:ipw-if-detail}{{S19}{26}{Bias and influence representations}{equation.4.19}{paper_SM.pdf}}
\newlabel{eq:ipw-expansion}{{S20}{26}{Bias and influence representations}{equation.4.20}{paper_SM.pdf}}
\newlabel{eq:outcome-law-detail}{{S21}{27}{Bias and influence representations}{equation.4.21}{paper_SM.pdf}}
\newlabel{eq:mi-bias-detail}{{S22}{27}{Bias and influence representations}{equation.4.22}{paper_SM.pdf}}
\newlabel{eq:mi-if-detail}{{S23}{27}{Bias and influence representations}{equation.4.23}{paper_SM.pdf}}
\newlabel{eq:mi-expansion}{{S24}{27}{Bias and influence representations}{equation.4.24}{paper_SM.pdf}}
\newlabel{eq:plain-variance-detail}{{S25}{27}{Bias and influence representations}{equation.4.25}{paper_SM.pdf}}
\newlabel{appendix:verification}{{S5}{30}{Verification lemmas}{section.5}{paper_SM.pdf}}
\newlabel{supp:criterion}{{S5.1}{30}{Profile discrepancy criterion}{subsection.5.1}{paper_SM.pdf}}
\newlabel{lem:remainder}{{S1}{31}{Population and evaluation remainders}{lemma.1}{paper_SM.pdf}}
\newlabel{lem:fold-control}{{S2}{32}{Fold-conditional bounds for two reference designs}{lemma.2}{paper_SM.pdf}}
\newlabel{lem:kernel-route}{{S3}{33}{A sufficient kernel route for the plain linearizations}{lemma.3}{paper_SM.pdf}}
\newlabel{lem:mi-common-if}{{S4}{36}{Common first-order fluctuation of the two MI estimators}{lemma.4}{paper_SM.pdf}}
\newlabel{appendix:identification}{{S6}{37}{Identification under conditional completeness}{section.6}{paper_SM.pdf}}
\newlabel{lem:completeness-identification}{{S5}{38}{Identification under conditional completeness}{lemma.5}{paper_SM.pdf}}
\newlabel{supp:additional}{{S7}{39}{Additional numerical results}{section.7}{paper_SM.pdf}}
\newlabel{supp:full-tables}{{S7.1}{39}{Full simulation summaries}{subsection.7.1}{paper_SM.pdf}}
\newlabel{tab:sim-designs}{{S1}{39}{The three designs of Section~\ref {sec:sim-designs}. Population values under $X_1\sim \mathrm {Unif}(-1.5,1.5)$ and $X_2\sim N(0,1)$. The residual is that from projecting $g(X_1)$ on $X_1$; $\mathrm {corr}(\text {resid}, m_1)$ is its correlation with the respondent conditional mean $m_1(X)$}{table.caption.7}{paper_SM.pdf}}
\newlabel{supp:mi-comparison}{{S7.2}{39}{Comparison of the two mass-imputation estimators}{subsection.7.2}{paper_SM.pdf}}
\newlabel{supp:completeness}{{S7.3}{40}{Completeness and discrete instruments}{subsection.7.3}{paper_SM.pdf}}
\newlabel{supp:floor}{{S7.4}{40}{Sensitivity to the propensity floor}{subsection.7.4}{paper_SM.pdf}}
\newlabel{tab:floor}{{S2}{41}{\ifdlmarks \color {dlblue}\fi Sensitivity to the propensity floor in the study of Section~\ref {sec:app}, design $\mathrm {M}_1$, over the same 500 realized samples at each floor. RB is the relative bias of the average estimated standard error against the Monte Carlo standard deviation}{table.caption.8}{paper_SM.pdf}}
\newlabel{supp:bw}{{S7.5}{41}{Bandwidth sensitivity}{subsection.7.5}{paper_SM.pdf}}
\newlabel{tab:bandwidth}{{S3}{42}{Bandwidth sensitivity in the study of Section~\ref {sec:app}, design $\mathrm {M}_1$. Both $h_1$ and $h_2$ are multiplied by the factor shown. Based on 500 Monte Carlo replicates with the same realized samples throughout. ``$\sum \delta /\hat {\pi }$'' is $N^{-1}\sum _i \delta _i/\hat {\pi }^B_i$, which estimates one. True values: $\theta = 4.687$, $\gamma = 1.047$. \dlfill {Bias, Monte Carlo SD and the median absolute deviation are computed over all 500 replicates; the quantities involving the estimated standard error, including coverage, are computed over the replicates for which it is defined. The count of non-positive denominators is over all replicates and all evaluation points, of which there are about $1.8\times 10^{6}$.}}{table.caption.9}{paper_SM.pdf}}
\makeatother}{}}
\usepackage{bookmark}
\usepackage{pdfpages}
\IfFileExists{xurl.sty}{\usepackage{xurl}}{}
\hypersetup{
  pdftitle={A Semiparametric Model Approach to Data Integration under Missing Not at Random},
  pdfauthor={},
  pdfkeywords={cross-fitting; instrumental variable; mass imputation; Neyman orthogonality; non-probability sample; propensity score},
  colorlinks=true,
  linkcolor={blue},
  filecolor={Maroon},
  citecolor={Blue},
  urlcolor={Blue},
  pdfcreator={LaTeX}}
\usepackage{etoolbox}
\AtBeginEnvironment{thebibliography}{%
  \hypersetup{citecolor=black,linkcolor=black,urlcolor=black}%
  \color{black}%
}

\theoremstyle{plain}
\newtheorem{theorem}{Theorem}

\newtheorem{proposition}{Proposition}
\theoremstyle{definition}

\theoremstyle{remark}

\newcommand{\estlab}{OP}
\newcommand{\bX}{\boldsymbol{X}}
\newcommand{\bx}{\bm x}

\def\spacingset#1{\renewcommand{\baselinestretch}{#1}\small\normalsize}
\AtBeginEnvironment{table}{\spacingset{1}}
\AtBeginEnvironment{algorithm}{\spacingset{1}}

\begin{document}
\spacingset{1}

\ifanon
\else
  \title{\bf A Semiparametric Model Approach to\\
  Data Integration\\
  under Missing Not at Random}
  \author{Danhyang Lee\\
    Department of Statistics and Data Science,\\
    Southern Methodist University\\
    and\\
    Kosuke Morikawa\\
    Department of Statistics, Iowa State University\\
    and\\
    Jae Kwang Kim\thanks{The research of Kim was partially supported by a grant from the US National Science Foundation (2242820), a grant from the Iowa Agriculture and Home Economics Experiment Station (Project No.\ IOW05806), and grants from the U.S. Department of Agriculture's National Resources Inventory, Cooperative Agreement NR203A750023C006, and Great Rivers CESU 68-3A75-18-504.}\\
    Department of Statistics, Iowa State University}
  \date{}
  \maketitle
\fi

\begin{abstract}
Integrating probability and non-probability samples requires accounting for
an unknown selection mechanism that may depend on the outcome. We develop a
semiparametric approach in which an unspecified function captures covariate
dependence and a scalar tilting parameter governs outcome dependence.
Identification follows from an instrumental-variable exclusion restriction
and conditional completeness; a zero tilt gives a restricted
missing-at-random submodel. We profile out the covariate component and
construct cross-fitted, Neyman-orthogonal estimating equations for both the
tilting parameter and the finite population mean. The resulting
orthogonalized mass-imputation estimator supports centered root-$N$
inference under nuisance product-rate conditions. Its linearization variance
accounts for the probability sampling design and estimation of the tilting
parameter. Simulations illustrate the cost of flexibility under correct
parametric specification, protection against misspecification of the
covariate component, and the consequences of invalid instruments or a
misspecified tilt. A further simulation using 50,000 national
health-screening records finds smaller bias but higher root mean squared
error than a parametric competitor under the baseline instrument set. The
proposed estimator varies less across alternative instrument sets, although
weak instruments and smoothing choices affect precision and interval
coverage.
\end{abstract}

\vfill
\newpage
\spacingset{1.8} 

\section{Introduction}

Probability sampling provides a scientific foundation for inference about a
finite population \citep{cochran1977sampling,fuller2011sampling}, but declining
response rates and rising collection costs have increased interest in
non-probability sources, including administrative records, volunteer surveys,
and transaction data. These sources can complement probability samples,
but their unknown selection mechanisms can introduce substantial bias
\citep{baker2013summary}. A common strategy combines a non-probability sample
that observes the outcome with a reference probability sample that observes
shared auxiliary variables.

Most methods for this setting assume that selection is ignorable given the
auxiliary variables, corresponding to missing at random (MAR)
\citep{rubin1976}. Under this assumption, available approaches include mass
imputation \citep{kim2021}, calibration weighting
\citep{dever2016general,elliott2017inference}, doubly robust pseudo-likelihood
and pseudo empirical-likelihood estimation
\citep{chen2020doubly,chen2022pseudo}, and multiply robust procedures
\citep{chen2023general}. Reviews are given by
\citet{rao2021making} and \citet{yang2020statistical}.

Ignorability can be difficult to justify when participation depends on an
outcome that the reference sample does not observe \citep{wu2022statistical}.
Methods allowing non-ignorable selection include \citet{kim2021data}, the
parametric pseudo-likelihood approach of \citet{liu2024statistical}, and the
empirical-likelihood approach of \citet{kim2023empirical}. Their reliance on
a parametric propensity score motivates a more flexible treatment of the
covariate component. Semiparametric selection models of the form used here
have been developed for missing-data problems
\citep{kim2011semiparametric,shao2016,uehara2023statistical,shetty2025robust};
in particular, \citet{shetty2025robust} construct estimating equations
whose consistency does not require estimating the unknown covariate
function. Those methods take a single sample in which the covariates are
observed for every unit. Data integration differs in three respects that
drive the present paper: the population covariate distribution is available
only through a design-weighted reference sample, so every nuisance
regression and the profiling identity itself must combine two samples with
different inclusion mechanisms; the target is a finite-population mean,
whose variance has separate reference-design and selection components; and
an orthogonal estimating equation for the tilting parameter does not by
itself deliver an orthogonal estimator of the mean.

We consider a semiparametric logistic selection model with an unspecified
covariate function $g(x_1)$ and a scalar outcome-tilting parameter $\gamma$.
An instrumental-variable exclusion restriction and conditional completeness
of the respondent outcome law provide sufficient identification conditions
\citep{miao2024identification}. The model contains a restricted MAR submodel
at $\gamma=0$, so the analyst does not have to decide in advance whether
selection is ignorable: $\gamma$ is estimated from the data and the
resulting estimator is consistent in either case, provided the exclusion
restriction holds. Within this framework, our contributions are as follows.
\begin{enumerate}
\item \textbf{Profiled estimation with an orthogonal score.} We profile
out $g$ through a population identity that combines the two samples and
estimate $\gamma$ from a Neyman-orthogonal estimating equation that uses
only the auxiliary variables observed in both samples, building on the
pseudo-likelihood construction of \citet{liu2024statistical}. Cross-fitting
removes the first-order effect of nonparametric nuisance estimation under
the stated rate and sampling conditions, and the resulting estimator is
$\sqrt N$-consistent and asymptotically normal.
\item \textbf{An orthogonalized mass-imputation estimator of the mean.}
Orthogonality of the score for $\gamma$ does not transfer to a plug-in
estimator of the mean, which can still inherit smoothing bias. We
therefore orthogonalize the mean estimating equation itself, yielding a
mass-imputation estimator with centered root-$N$ inference under
product-rate conditions. Its linearization variance accounts for both
sampling sources and for the estimation of $\gamma$. Plain
inverse-probability-weighting and mass-imputation estimators serve as
benchmarks; their first-order smoothing biases are retained explicitly in
the supplementary theory.
\item \textbf{Numerical characterization of when the flexibility is
useful.} Synthetic designs separate nonlinear covariate selection from
violations of the exclusion or tilt assumptions, and a simulation built on
national health-screening records examines performance under an empirical
joint distribution, including sensitivity to the choice of instruments.
They show a bias--variance trade-off, not robustness to invalid
instruments.
\end{enumerate}

Section~\ref{sec:prelim} introduces the model and identification conditions.
Sections~\ref{sec:method}--\ref{sec:theory} develop estimation and inference;
Sections~\ref{sec:sim}--\ref{sec:app} report the numerical studies.
Proofs, detailed implementation, and additional results are in the
Supplementary Material.

\section{Setup and identification}\label{sec:prelim}
\subsection{Observed data and selection model}\label{sec:setup}

Let $U=\{1,\ldots,N\}$ index a finite population, with outcome $y_i$ and
$p$-dimensional auxiliary vector $x_i$ for unit $i$. The target is
$\theta=N^{-1}\sum_{i=1}^N y_i$. Sample $A$ is a probability sample with
indicator $I_i^A=I(i\in A)$, known inclusion probability $\pi_i^A$, and
design weight $d_i^A=1/\pi_i^A$. Sample $B$ is a non-probability sample with
indicator $\delta_i=I(i\in B)$. The samples are selected independently;
$x_i$ is observed in both, and $y_i$ only in $B$. Their sizes are $n_A$ and
$n_B$. The population identities used below do not require identifying
overlap between samples; cross-fitted estimation additionally requires the
overlap and fold conditions specified in (R6).

Partition $x=(x_1^\top,x_2^\top)^\top$ and assume
\begin{equation}\label{eq:semi-ps}
\pi^B(x_1,y;g,\gamma)=P(\delta=1\mid x_1,x_2,y)
=\operatorname{expit}\{g(x_1)-\gamma y\},
\end{equation}
where $g$ is unspecified and $\gamma$ governs outcome dependence. The
instrument $X_2$ is associated with $Y$ given $X_1$ but is excluded from
the selection model. This restriction must be justified substantively.

\subsection{Identification and interpretation}\label{sec:identification}

Relevance and exclusion alone do not identify $(g,\gamma)$. Following
\citet{miao2024identification}, assume conditional completeness of the
respondent outcome law: for almost every $x_1$, every measurable $h$ with
$E\{h(Y)^2\mid X_1=x_1,\delta=1\}<\infty$ satisfies
\[
E\{h(Y)\mid X_1=x_1,X_2,\delta=1\}=0
\quad\Longrightarrow\quad h(Y)=0\quad\text{almost surely}.
\]
Assume also that $Y\mid X_1=x_1,\delta=1$ is nondegenerate,
$E\{\exp(2\gamma Y)\mid X_1=x_1,\delta=1\}<\infty$ for each $\gamma$ in
a compact candidate set $\Gamma$, and selection probabilities are positive.
Together these conditions identify $(g,\gamma)$ and the population law of
$(X,Y)$; Supplementary Lemma~\ref{lem:completeness-identification} gives the
proof. Completeness distinguishes selection-odds functions through their
conditional averages over instrument values. It is a sufficient condition,
stronger than relevance. The separated-root and derivative conditions for
the estimating equation are imposed separately in (R2) and (R8).

The argument compares two candidate pairs that generate the same observed
laws: equality of their conditional selection odds among respondents,
together with completeness, forces
$(\gamma_1-\gamma_2)Y=g_1(X_1)-g_2(X_1)$ almost surely, and conditional
nondegeneracy of $Y$ then distinguishes the tilts. This identifies the
population model, not each unobserved finite-population outcome, and
identification is separate from regular estimation by a particular scalar
score.

Write $f_1(y\mid x)=f(y\mid x,\delta=1)$. Model~\eqref{eq:semi-ps} implies
the exponential-tilting identity
\begin{equation}\label{eq:tilt}
f(y\mid x,\delta=0)
=\frac{f_1(y\mid x)\exp(\gamma y)}
{\int f_1(y\mid x)\exp(\gamma y)\,d\mu(y)}.
\end{equation}
Thus, once $\gamma$ is identified, the nonrespondent outcome law is
recovered from the respondent law. At $\gamma=0$ the two conditional laws
coincide, giving an MAR mechanism. This submodel is restricted:
$P(\delta=1\mid X,Y)=\operatorname{expit}\{g(X_1)\}$ cannot depend on
$X_2$. A general MAR mechanism depending on both $X_1$ and $X_2$ lies
outside the model.

The marginal propensity follows from the inverse-propensity identity
\citep{pfeffermann1999parametric,morikawa2018},
\begin{equation}\label{eq:ps-identity}
\pi_m^B(x;g,\gamma)
=\left[\int\{\pi^B(x_1,y;g,\gamma)\}^{-1}f_1(y\mid x)\,d\mu(y)\right]^{-1}.
\end{equation}
The parametric pseudo-likelihood benchmark of \citet{liu2024statistical}
uses a linear covariate predictor and an estimated respondent distribution
in this identity. We leave the covariate predictor unspecified and estimate
the required respondent moments nonparametrically. Supplementary
Section~\ref{supp:likelihood} describes the parametric benchmark and the
corresponding semiparametric pseudo-profile likelihood.

\section{Proposed method}\label{sec:method}
\subsection{Profiling the selection model}\label{sec:profile}

For each candidate $\gamma$, the identity
$E\{\delta/\pi^B(X_1,Y;g,\gamma)-1\mid X_1\}=0$ gives
\begin{equation}\label{eq:g-identity}
\exp\{-g_\gamma(x_1)\}
=\frac{E(1-\delta\mid x_1)}{E\{\delta\exp(\gamma Y)\mid x_1\}}.
\end{equation}
The true covariate function is $g_{\gamma_0}$. Replacing the population
covariate total by its design-weighted estimate from $A$ yields
\begin{equation}\label{eq:ghat}
\exp\{-\hat g_\gamma(x_1)\}
=\frac{\sum_{i\in A}d_i^A K_{h_1}(x_1-x_{1i})-
\sum_{i\in B}K_{h_1}(x_1-x_{1i})}
{\sum_{i\in B}\exp(\gamma y_i)K_{h_1}(x_1-x_{1i})},
\end{equation}
where $K_{h_1}$ is a kernel with bandwidth $h_1$. The numerator estimates
the number of non-selected units near $x_1$ without resolving sample
overlap. The fitted conditional propensity is
\begin{equation}\label{eq:pi-cond}
\hat\pi^B(x_1,y;\gamma)
=\{1+\exp[-\hat g_\gamma(x_1)+\gamma y]\}^{-1}.
\end{equation}
Define $m(x;\gamma)=\log E\{\exp(\gamma Y)\mid X=x,\delta=1\}$. By
\eqref{eq:ps-identity}, the fitted marginal propensity is
\begin{equation}\label{eq:pi-marg}
\hat\pi_m^B(x;\gamma)
=\{1+\exp[-\hat g_\gamma(x_1)+\hat m(x;\gamma)]\}^{-1},
\end{equation}
where $\hat m$ is estimated from $B$. These quantities are recomputed for
each candidate $\gamma$. Supplementary Section~\ref{app:nuisance-kernel}
gives the nuisance estimators and profile derivatives.

\subsection{Orthogonal estimation of the tilting parameter}\label{sec:orth}

To motivate the estimating equation, consider the profile criterion
\[
\mathcal Q_N(\gamma)=\frac1N\left[
\sum_{i\in B}\{\hat\pi_m^B(x_i;\gamma)\}^{-1}
+\sum_{i\in A}d_i^A\log\hat\pi_m^B(x_i;\gamma)\right].
\]
Writing $\pi^B_{m,0}(X)=P_0(\delta=1\mid X)$, its population counterpart
has the contrast
\[
\mathcal Q(\gamma)-\mathcal Q(\gamma_0)
=E\left[\frac{\pi^B_{m,0}(X)}{\pi_m^B(X;\gamma)}
-\log\frac{\pi^B_{m,0}(X)}{\pi_m^B(X;\gamma)}-1\right]\ge0,
\]
whenever the expectations are finite, with equality precisely when the two
marginal propensities agree almost surely. Because the profile identity
\eqref{eq:g-identity} gives $E\{\pi^B_{m,0}(X)/\pi^B_m(X;\gamma)\mid X_1\}=1$,
the contrast is the Kullback--Leibler divergence from the population
covariate distribution to the respondent covariate distribution reweighted
by the inverse candidate propensity. Under completeness it identifies
$\gamma_0$ as the unique population minimizer; Supplementary
Section~\ref{supp:criterion} gives the derivation. The criterion motivates
the estimating equation without identifying it with the Bernoulli
pseudo log-likelihood of Supplementary Section~\ref{supp:likelihood}.
Differentiating the sample criterion leads to the following
unorthogonalized score.

The profile-based estimating function is
\begin{equation}\label{eq:score-naive}
\hat s(\gamma)=\frac1N\sum_{i=1}^N
\{\delta_i-I_i^Ad_i^A\hat\pi_m^B(x_i;\gamma)\}\hat c_i(\gamma),
\end{equation}
where $\hat c_i$ estimates
\[
c(x;\gamma)=\partial_\gamma\{\pi_m^B(x;\gamma)\}^{-1}
=\{\nabla_\gamma m(x;\gamma)-\nabla_\gamma g(x_1;\gamma)\}
\exp\{m(x;\gamma)-g(x_1;\gamma)\}.
\]
Here and below $g(x_1;\gamma)=g_\gamma(x_1)$.
The identity $\partial_\gamma\log\hat\pi_m^B=-\hat\pi_m^B\hat c$
gives the reference-sample term in \eqref{eq:score-naive}. Unique population
minimization does not exclude other stationary roots, so the root
separation in (R2) is imposed separately.

Plugging nonparametric fits into \eqref{eq:score-naive} can introduce
first-order nuisance bias. We remove this sensitivity through Neyman
orthogonalization and use cross-fitting to evaluate the score on records
held out from nuisance training \citep{chernozhukov2018double}. Write
$\pi^B_{m,0}(x)=E(\delta\mid X=x)$ and
$\varpi^B_{g,0}(x_1)=E(\delta\mid X_1=x_1)$. The orthogonal equation is
\begin{equation}\label{eq:score-orth}
\hat{s}_\perp(\gamma)
:= \frac{1}{N} \sum_{i=1}^{N}
\Bigl[\{\delta_i - I_i^A d_i^A \hat{\pi}^B_m(\bx_i; \gamma)\}\hat{c}_i(\gamma)
+ \hat{\phi}_i(\gamma) + \hat{\psi}_i(\gamma)\Bigr] = 0,
\end{equation}
We define $\hat\gamma$ as an interior approximate root of
\eqref{eq:score-orth}, selected by the grid-and-bracket rule in
Section~\ref{sec:implementation}, with score residual $o_p(N^{-1/2})$.
The rule uses only the sample estimating function; consistency for
$\gamma_0$ follows from Condition~(R2).
The augmentation terms are
\begin{align}
\hat{\phi}_i(\gamma)
&= \frac{\hat{\xi}(\bx_{1i};\gamma)}{1-\hat{\varpi}^B_{g,0}(\bx_{1i})}
\Bigl\{(I_i^A d_i^A - \delta_i)
- \delta_i \exp\{\gamma y_i - \hat{g}(\bx_{1i};\gamma)\}\Bigr\},
\label{eq:phi}\\[4pt]
\hat{\psi}_i(\gamma)
&= \delta_i \,
\frac{\hat{c}_i(\gamma)\{\hat{\pi}^B_m(\bx_i; \gamma)\}^2}
{\exp\{\hat{g}(\bx_{1i};\gamma)\}\,\hat{\pi}^B_{m,0}(\bx_i)}
\Bigl\{\exp(\gamma y_i) - \exp\{\hat{m}(\bx_i;\gamma)\}\Bigr\}.
\label{eq:psi}
\end{align}
The function $\xi(\bx_1; \gamma)$ is
\[
\xi(\bx_1; \gamma)
= E\!\left[c(X; \gamma)\,\pi^B_m(X; \gamma)\{1 - \pi^B_m(X; \gamma)\}
\,\middle|\, \mathbf{X}_1 = \bx_1\right],
\]
and is estimated by kernel smoothing.

The score components can be calculated through conditional moments. For
$k=0,1$, define
\[
r_m^{(k)}(x;\gamma)=E\{\delta Y^k e^{\gamma Y}\mid X=x\},
\qquad
r_g^{(k)}(x_1;\gamma)=E\{\delta Y^k e^{\gamma Y}\mid X_1=x_1\}.
\]
The profile identities and their derivatives are
\[
e^{m(x;\gamma)}=\frac{r_m^{(0)}(x;\gamma)}{\pi^B_{m,0}(x)},
\qquad e^{g(x_1;\gamma)}=
\frac{r_g^{(0)}(x_1;\gamma)}{1-\varpi^B_{g,0}(x_1)},
\]
\[
\nabla_\gamma m(x;\gamma)=\frac{r_m^{(1)}(x;\gamma)}{r_m^{(0)}(x;\gamma)},
\qquad
\nabla_\gamma g(x_1;\gamma)=\frac{r_g^{(1)}(x_1;\gamma)}{r_g^{(0)}(x_1;\gamma)}.
\]
Thus nuisance estimation requires regressions of observed respondent
quantities and design-weighted covariate information; it does not require
specifying the full respondent density. Positivity of the fitted
denominators and suitable weighted convergence rates are nevertheless
needed, especially when fitted propensities are small.

The residual in \eqref{eq:phi} has conditional mean zero given $X_1$ at
the truth; its coefficient cancels the first-order effect of estimating
$g_\gamma$. Likewise, the respondent residual in \eqref{eq:psi} has
conditional mean zero given $(X,\delta=1)$ and cancels the effect of
estimating $m$. The coefficient $\xi$ is estimated from $A$ with design
weights, as specified in Supplementary Section~\ref{app:nuisance-kernel}.

Let $W=(Y,\delta,I^A,d^A,X)$ and let
$\Phi_\perp(W;\gamma,\eta)=\Phi_0+\Phi_g+\Phi_m$ denote the population
counterparts of the three summands in \eqref{eq:score-orth}. At the truth
the three terms collapse to a single weighted contrast. Writing
$p(X)=\pi^B_{m,0}(X)$, $t(X_1)=\xi(X_1;\gamma_0)/\{1-\varpi^B_{g,0}(X_1)\}$
and $H(X)=p(X)c(X;\gamma_0)-t(X_1)$, and using
$1/\pi^B(X_1,Y)=1+\exp\{\gamma_0Y-g(X_1)\}$ and
$\exp\{m(X)-g(X_1)\}=\{1-p(X)\}/p(X)$, a direct calculation gives
\begin{equation}\label{eq:score-collapsed}
\Phi_\perp(W;\gamma_0,\eta_0)
=H(X)\left\{\frac{\delta}{\pi^B(X_1,Y;\gamma_0)}-I^Ad^A\right\}.
\end{equation}
The orthogonal score therefore compares the inverse-propensity-weighted
respondent measure with the design-weighted reference measure, both of
which estimate the population covariate measure, and weights the
discrepancy by $H(X)$. The two augmentation terms adjust the weight from
$p(X)c(X;\gamma_0)$ to $H(X)$ and move the residual from the marginal
propensity to the conditional one; this is what removes the first-order
sensitivity to $\hat g_\gamma$ and $\hat m$. The representation is also
the source of the conditional centering used for variance estimation in
Section~\ref{sec:theory}.

The nuisance collection $\eta$ includes the displayed conditional moments
and their profile derivatives. We estimate it on each outer training set,
average the held-out score contributions, and solve one pooled equation
(DML2). In the numerical tables, \estlab{} (orthogonal profile) denotes this
estimator; it is not the pseudo-likelihood maximizer in
Supplementary Section~\ref{supp:likelihood}.

\subsection{Orthogonal estimation of the population mean}
\label{sec:mean}\label{sec:orth-mi}

Given $\hat\gamma$, the plain inverse-probability-weighting (IPW) and
mass-imputation (MI) estimators are
\begin{align}
\hat\theta_{\mathrm{IPW}}&=\frac1N\sum_{i\in B}
\frac{y_i}{\hat\pi^B(x_{1i},y_i;\hat\gamma)},\label{eq:ipw}\\
\hat\theta_{\mathrm{MI}}&=\frac1N\sum_{i\in A}d_i^A\hat\mu(x_i;\hat\gamma).
\label{eq:mi}
\end{align}
If $N$ is unknown, a H\'ajek IPW estimator replaces it by
$\sum_{i\in B}\{\hat\pi^B(x_{1i},y_i;\hat\gamma)\}^{-1}$.
At the truth, $\mu(x;\gamma_0)=E(Y\mid X=x)$. The tilt identity gives
\begin{equation}\label{eq:mu-decomp}
\hat\mu(x;\gamma)=\hat\pi_m^B(x;\gamma)\hat\mu_1(x)
+\{1-\hat\pi_m^B(x;\gamma)\}\hat\mu_0(x;\gamma),
\end{equation}
where $\mu_1(x)=E(Y\mid X=x,\delta=1)$ and
\[
\mu_0(x;\gamma)=\frac{E\{Y\exp(\gamma Y)\mid X=x,\delta=1\}}
{E\{\exp(\gamma Y)\mid X=x,\delta=1\}}.
\]
Both are estimated from $B$; at $\gamma_0$, $\mu_0$ is the conditional
mean among non-selected units.

Orthogonality of the score for $\gamma$ does not make these plug-in means
orthogonal. For example, the derivative of the IPW functional with respect
to $g$ in direction $h$ is
$-E\{\delta O^B(X_1,Y;\gamma_0)Yh(X_1)\}$, which need not vanish, where
$O^B=(1-\pi^B)/\pi^B$. The usual augmented inverse-probability score
\citep{robins1994estimation},
$\Psi_0=\theta-I^Ad^A\mu(X;\gamma)
-\{\delta/\pi^B(X_1,Y;\gamma)\}\{Y-\mu(X;\gamma)\}$,
removes sensitivity to the outcome regression because
$E\{\delta/\pi^B\mid X\}=E(I^Ad^A\mid X)=1$ at the truth, but its
derivative with respect to $g$ involves the conditional tilted residual
$E\{\delta O^B(Y-\mu)\mid X_1\}$. Under MNAR, a correct outcome
regression alone does not protect against a misspecified selection model;
the correction below centers precisely this residual.

To orthogonalize the mean score, define
\begin{equation}\label{eq:ag}
a_g(\bx_1)
= \frac{E\{\delta\, O^B(\bX_1, Y; \gamma)\,\{Y - \mu(\bX; \gamma)\}\mid X_1 =
\bx_1\}}
{E\{\delta\, O^B(\bX_1, Y; \gamma)\mid \bX_1 =
\bx_1\}}.
\end{equation}
The correction $a_g(X_1)\{I^Ad^A-\delta/\pi^B(X_1,Y;\gamma)\}$ has mean
zero at the truth and removes the first-order effect of estimating $g$.
The resulting score is
\begin{equation}\label{eq:mi-orth-score}
\Psi_{\mathrm{MI},\perp}(W; \theta, \eta)
= \theta - \!\left[\frac{\delta}{\pi^B(\bX_1, Y; \gamma)} Y
+ \left(I^A d^A - \frac{\delta}{\pi^B(\bX_1, Y; \gamma)}\right)
\{\mu(\bX; \gamma) + a_g(\bX_1)\}\right].
\end{equation}
\begin{proposition}[Orthogonality of the estimating equations]
\label{thm:orth-gamma}\label{thm:orth-mi}\label{prop:orth}
Under the regularity conditions in Supplementary
Section~\ref{appendix:regularity}, let $\eta_0$ collect the true nuisance
functions and $\theta_0=E(Y)$. Both
$E\{\Phi_\perp(W;\gamma_0,\eta_0)\}$ and
$E\{\Psi_{\mathrm{MI},\perp}(W;\theta_0,\eta_0)\}$ are zero. For every
regular path $t\mapsto\eta_t$ through $\eta_0$, with $\gamma=\gamma_0$
held fixed,
\[
\left.\frac d{dt}E\{\Phi_\perp(W;\gamma_0,\eta_t)\}\right|_{t=0}=0,
\qquad
\left.\frac d{dt}E\{\Psi_{\mathrm{MI},\perp}(W;\theta_0,\eta_t)\}\right|_{t=0}=0.
\]
\end{proposition}

Solving the mean equation yields our recommended estimator,
\begin{equation}\label{eq:mi-orth}
\hat\theta_{\mathrm{MI},\perp}=\frac1N\left[
\sum_{i\in A}d_i^A(\hat\mu_i+\hat a_{g,i})+
\sum_{i\in B}\frac{Y_i-\hat\mu_i-\hat a_{g,i}}{\hat\pi_i^B}
\right],
\end{equation}
where $\hat\pi_i^B=\hat\pi^B(X_{1i},Y_i;\hat\gamma)$ and
$\hat\mu_i=\hat\mu(X_i;\hat\gamma)$. This is the plain MI estimator plus
a weighted outcome-residual correction and the correction for the profiled
selection component. All nuisance values are evaluated out of fold. Every
intermediate fit used to construct $\hat a_{g,i}$ must use the same outer
training set; Supplementary equation~\eqref{eq:ag-hat} gives the precise
definition. The finite-dimensional uncertainty in $\hat\gamma$ is retained
in the joint expansion below.

The last term of \eqref{eq:mi-orth} uses the same discrepancy between
the design-weighted reference measure and the inverse-propensity-weighted
respondent measure that appears in \eqref{eq:score-collapsed}; its
conditional expectation is zero at the true model, while its coefficient
cancels the first-order effect of estimating $g$. This is a local
orthogonality property; it does not supply the usual MAR double-robustness
claim under an arbitrary misspecified selection model.

Setting $\mu\equiv0$ in \eqref{eq:ag} gives
$a_g(X_1)=E(Y\mid X_1,\delta=0)$ and an orthogonalized IPW score.
We retain the plain estimators \eqref{eq:ipw}--\eqref{eq:mi} as benchmarks
for assessing the effect of mean-level orthogonalization.

\subsection{Implementation}\label{sec:implementation}

For each candidate $\gamma$ and each of $K$ outer folds, we fit all nuisance
functions on the training records and evaluate the score on the held-out
records, solve the pooled equation \eqref{eq:score-orth}, and then assemble
\eqref{eq:mi-orth} and its variance from out-of-fold predictions.
Algorithm~\ref{alg:ppml} summarizes the procedure; the nuisance fits are
recomputed at each candidate $\gamma$, so the dominant cost is
$O\{KG(n_A+n_B)^2\}$ for $G$ grid candidates. Linked records from
overlapping samples must share a fold.

\begin{algorithm}[tb]
\caption{Cross-fitted orthogonal profile estimation}\label{alg:ppml}
\begin{algorithmic}[1]
\State Assign $K$ outer folds as in (R6), keeping linked records from the
       same population unit together.
\For{each candidate $\gamma$ on the grid $\Gamma$ and each fold $k$}
  \State On the training records $D_{-k}$, compute the kernel nuisance
         fits of Supplementary Section~\ref{app:nuisance-kernel}.
  \State Evaluate $\hat\pi_m^B(x_i;\gamma)$ and
         $\hat\Phi_{\perp,i}(\gamma)$ on the held-out fold $U^{(k)}$.
\EndFor
\State Pool the score contributions across folds, \dlfill{locate the
       sign-change brackets} of $N^{-1}\sum_i\hat\Phi_{\perp,i}(\gamma)=0$,
       and select $\hat\gamma$ by the root-selection rule below.
\State Evaluate $\hat\mu_i$ and $\hat a_{g,i}(\hat\gamma)$ out of fold;
       every intermediate regression entering $\hat a_{g,i}$ uses the
       same outer training set.
\State Assemble $\hat\theta_{\mathrm{IPW}}$, $\hat\theta_{\mathrm{MI}}$,
       $\hat\theta_{\mathrm{MI},\perp}$ and their linearization variances
       from Supplementary Section~\ref{appendix:variance}.
\end{algorithmic}
\end{algorithm}

The numerical studies
use $K=5$ membership-stratified folds, whose theoretical validity requires
the conditional evaluation bounds in (R6). \dlfill{Folds are assigned to
records, stratified on $\delta_i$, with the two records of a unit appearing
in both samples forced into a common fold; Supplementary
Section~\ref{supp:constants} reports the number of such units.}

The score need not be monotone. We locate sign changes on a grid in
$\Gamma$ \dlfill{and refine by Brent's method. When more than one interior root
is found, one is selected by a rule fixed in advance.} Endpoint outcomes and searches without an interior root are recorded
separately. The selection rule is a finite-sample device and carries no
asymptotic weight: Condition~(R2) requires $\gamma_0$ to be the only zero
of the population moment $M(\gamma)$ on $\Gamma$, with
$\inf_{|\gamma-\gamma_0|\ge\epsilon}|M(\gamma)|>0$, together with uniform
convergence of the sample moment. Any sequence of sample roots that stayed
$\epsilon$ away from $\gamma_0$ would then violate uniform convergence, so
every interior root converges to $\gamma_0$ and the rule only decides
among roots that are asymptotically indistinguishable. A related
finite-sample issue is that the numerator of
\eqref{eq:ghat}, a design-weighted count of non-selected units near $x_1$,
can be negative at a point where the reference sample is sparse. The
fitted profile is undefined at such a point, and its treatment is a
separate implementation choice from the propensity floor described in
Section~\ref{sec:app}. \dlfill{The implementation truncates the numerator
at machine epsilon, keeps the evaluation point, and counts the occurrences;
Supplementary Section~\ref{supp:bw} reports them, and they are rare at the
bandwidth used.}

The numerical nuisance fits use Gaussian local-constant smoothers with
bandwidths of order $\hat\sigma n^{-1/(d+4)}$ for $d$ continuous coordinates
and exact matching for discrete coordinates; the smoothness--dimension
condition this implies is discussed in
Section~\ref{sec:theory}.
Supplementary Sections~\ref{supp:impl} and \ref{appendix:verification}
give the constants and conditions, and Supplementary
Section~\ref{supp:bw} examines bandwidth sensitivity.

\section{Asymptotic properties}\label{sec:theory}

We distinguish the superpopulation mean $\theta_0=E(Y)$ from the target
$\theta_N=N^{-1}\sum_{i=1}^N Y_i$. Under (R1), the reference and
non-probability sample sizes are of order $N$ and their selection mechanisms
are independent conditional on the finite population. Conditions
(R3)--(R6) control weighted nuisance errors and cross-fitted evaluation;
(R2) and (R8) ensure a separated, regular score root; and (R9)--(R10)
give the joint sampling CLT and valid design-specific variance estimation.
All conditions are stated in Supplementary
Section~\ref{appendix:regularity}.

A sufficient version of the rate condition is that all nuisance
coordinates used by the two scores, including derived ratios and
derivative coefficients, have weighted errors $O_p(r_N)$ with
$r_N=o(N^{-1/4})$. Orthogonality makes their population contribution
quadratic, while the conditional evaluation bound controls the held-out
empirical term. The resulting remainder is
$O_p(r_N^2+r_N/\sqrt N)=o_p(N^{-1/2})$.
This common rate is stronger than some component-specific product
conditions. For a regression of smoothness $s$ in $d$ continuous
coordinates, the usual rate $n^{-s/(2s+d)}$ is $o(n^{-1/4})$ when $2s>d$,
provided the weighted, boundary, and design conditions also hold;
bandwidth orders alone do not verify the theorems. The dimension is that
of each regression: fitting $m$ and the outcome means involves the full
$X$, so with second-order smoothers ($s=2$) the condition allows at most
three continuous coordinates in $X$, a point that matters for the
treatment of AGE in Section~\ref{sec:app}.

Let $\eta_0(\gamma)$ denote the population nuisance profile, and define
\[
\mathcal J=\left.\frac d{d\gamma}
E\{\Phi_\perp(W;\gamma,\eta_0(\gamma))\}\right|_{\gamma_0},
\qquad
G_{\mathrm{MI},\gamma}
=E\bigl[\delta O^B Y\{Y-\mu(X)-a_g(X_1)\}\bigr],
\]
where functions without a $\gamma$ argument are evaluated at $\gamma_0$.
Write $\Phi_{\perp,i}=\Phi_\perp(W_i;\gamma_0,\eta_0)$ and
$\pi_i^B=\pi^B(X_{1i},Y_i;\gamma_0)$. The profile identities of
Section~\ref{sec:orth} give $\nabla_\gamma m(X;\gamma_0)=\mu_0(X;\gamma_0)$
and $\nabla_\gamma g(X_1;\gamma_0)=b(X_1)$, where
$b(X_1)=E(Y\mid X_1,\delta=0)$, and
$c(X;\gamma_0)=\{\mu_0(X;\gamma_0)-b(X_1)\}\{1-p(X)\}/p(X)$, so that
\begin{equation}\label{eq:J-interpret}
\mathcal J=E\bigl[\{1-p(X)\}^2\{\mu_0(X;\gamma_0)-b(X_1)\}^2\bigr].
\end{equation}
The information about $\gamma$ is thus the squared variation of the tilted
conditional mean of $Y$ across values of the instrument $X_2$ at fixed
$X_1$, weighted by the non-selection probability. Instruments that move the
non-respondent conditional mean little given $X_1$ make $\mathcal J$ small
and the score flat; this is the mechanism behind the loss of precision
under thin instrument sets in Section~\ref{sec:app-iv}.

\begin{theorem}[Joint inference for the tilting parameter and mean]
\label{thm:gamma}\label{cor:mi-orth}\label{thm:joint}
Under Conditions~(R1)--(R6) and (R8)--(R10), the cross-fitted estimators
\eqref{eq:score-orth} and \eqref{eq:mi-orth} satisfy
\begin{equation}\label{eq:gamma-expansion}
\hat\gamma-\gamma_0=-\frac1N\sum_{i=1}^N\mathcal J^{-1}\Phi_{\perp,i}
+o_p(N^{-1/2}),
\end{equation}
and
\begin{equation}\label{eq:mi-main-expansion}
\hat\theta_{\mathrm{MI},\perp}-\theta_N
=\frac1N\sum_{i=1}^N\varphi^F_{\mathrm{MI},\perp,i}+o_p(N^{-1/2}),
\end{equation}
where $\mathcal J\ne0$ and
\begin{equation}\label{eq:mi-main-if}
\begin{split}
\varphi^F_{\mathrm{MI},\perp,i}
={}&(I_i^Ad_i^A-1)\{\mu(X_i)+a_g(X_{1i})\}\\
&+\left(\frac{\delta_i}{\pi_i^B}-1\right)
\{Y_i-\mu(X_i)-a_g(X_{1i})\}
-G_{\mathrm{MI},\gamma}\mathcal J^{-1}\Phi_{\perp,i}.
\end{split}
\end{equation}
Consequently,
$\sqrt N(\hat\gamma-\gamma_0)\Rightarrow
\mathcal N(0,\mathcal J^{-1}\Omega_\gamma\mathcal J^{-1})$ and
$\sqrt N(\hat\theta_{\mathrm{MI},\perp}-\theta_N)
\Rightarrow\mathcal N(0,V^F_{\mathrm{MI},\perp})$,
with $\Omega_\gamma$ and $V^F_{\mathrm{MI},\perp}$ the limiting variances
of $N^{-1/2}\sum_i\Phi_{\perp,i}$ and
$N^{-1/2}\sum_i\varphi^F_{\mathrm{MI},\perp,i}$, respectively.
\end{theorem}

The last term of \eqref{eq:mi-main-if} propagates estimation of $\gamma$.
It remains relevant when $\gamma_0=0$ unless MAR is imposed a priori and
$\gamma$ is fixed. The mean score is orthogonal to the infinite-dimensional
nuisances with $\gamma$ held fixed; joint estimation accounts for its
finite-dimensional variation. The proof is in Supplementary
Section~\ref{appendix:proofs}.

The finite-population influence function has a sum of centered reference-
sample and Bernoulli-selection contributions. Their conditional variances
add under independent sampling. Supplementary
Section~\ref{appendix:variance} gives the sample-separated plug-in formula
$\widehat V^F_{\mathrm{MI},\perp}$, including the SRSWOR reference-design
specialization, with $N\widehat V^F_{\mathrm{MI},\perp}\to_p
V^F_{\mathrm{MI},\perp}$. The Wald standard error is
$\sqrt{\widehat V^F_{\mathrm{MI},\perp}}$. Other reference designs require
their actual pairwise inclusion probabilities or a consistent
design-specific variance estimator.

For clarity, write the population tilt score as
$\Phi_{\perp,i}=I_i^Ad_i^A A_{\gamma,i}+\delta_i B_{\gamma,i}$.
The influence function \eqref{eq:mi-main-if} then has the form
\[
\varphi^F_{\mathrm{MI},\perp,i}
=(I_i^Ad_i^A-1)u_{A,i}+(\delta_i-\pi_i^B)u_{B,i},
\]
where
\begin{align*}
u_{A,i}&=\mu(X_i)+a_g(X_{1i})
-G_{\mathrm{MI},\gamma}\mathcal J^{-1}A_{\gamma,i},\\
u_{B,i}&=\frac{Y_i-\mu(X_i)-a_g(X_{1i})}{\pi_i^B}
-G_{\mathrm{MI},\gamma}\mathcal J^{-1}B_{\gamma,i}.
\end{align*}
The identity $A_{\gamma,i}+\pi_i^B B_{\gamma,i}=0$ at the truth
justifies this conditional centering (Supplementary
Lemma~\ref{lem:mi-common-if}). Under SRSWOR for $A$, a plug-in variance
estimator is
\[
\widehat V^F_{\mathrm{MI},\perp}
=\left(1-\frac{n_A}{N}\right)\frac{s_{\hat u_A,A}^2}{n_A}
+\frac1{N^2}\sum_{i\in B}(1-\hat\pi_i^B)\hat u_{B,i}^2,
\]
where $s_{\hat u_A,A}^2$ is the sample variance of the fitted reference
coefficients. The separate sample contributions make the finite-population
target explicit and allow variance computation without matching
overlapping records. Validity of the nuisance fits still requires (R6).
Replacing this calculation by an empirical second moment of a
superpopulation influence function would generally target a different
variance; Supplementary Section~\ref{appendix:variance} distinguishes the
two settings.

The plain IPW and MI estimators can retain first-order smoothing bias;
Supplementary Theorems~\ref{thm:ipw}--\ref{thm:mi} give their bias-centered
limits. Under the additional integrated-linearization condition (R7),
the nonparametric fits in Supplementary Section~\ref{app:nuisance-kernel}
satisfy
\begin{equation}\label{eq:mi-bias-removal}
\hat\theta_{\mathrm{MI}}-\hat\theta_{\mathrm{MI},\perp}
=\mathcal B_{\mathrm{MI},N}+o_p(N^{-1/2}).
\end{equation}
The bias-centered limits of the plain estimators have the form
\[
\sqrt N\{\hat\theta_j-\theta_N-\mathcal B_{j,N}\}
\Rightarrow\mathcal N(0,V_j),\qquad j\in\{\mathrm{IPW},\mathrm{MI}\}.
\]
Although $\mathcal B_{j,N}\to0$, it need not be $o(N^{-1/2})$.
If $\sqrt N\mathcal B_{j,N}\to b_j$, the uncorrected root-$N$ error
has limiting mean $b_j$. Centered inference then requires suitable
undersmoothing or an explicit bias correction, with any first-order
uncertainty in the estimated correction included in the variance.
Cross-fitting alone does not remove this centering error.

Thus, under (R7) and for the kernel nuisance estimators of Supplementary
Section~\ref{app:nuisance-kernel}, the two MI estimators share their
first-order stochastic fluctuation when they use the same fits, while
orthogonalization removes the leading smoothing bias (Supplementary
Lemma~\ref{lem:mi-common-if}). Orthogonality alone does not give this
equivalence; it need not hold for other nuisance learners or for a
parametric respondent model, and it does not imply equal finite-sample
variances. Theorem~\ref{thm:joint} does not require (R7).
Its weaker nuisance-rate requirement motivates our recommendation of
\eqref{eq:mi-orth}; neither orthogonality nor cross-fitting removes bias
from invalid instruments, an incorrect tilt, or an active fixed propensity
floor.

\section{Simulation study}\label{sec:sim}

We first examine misspecification of the covariate component with continuous
$X_1$, then violations of the identifying assumptions with categorical
$X_1$, where $g$ can be estimated by exact cell matching or saturated
parametrically. The two studies use different finite populations, so their
magnitudes are not directly comparable. Each uses 500 Monte Carlo
replicates.

\subsection{Designs and competing estimators}\label{sec:sim-designs}

The first study has $N=20{,}000$, $X_{1i}\sim\mathrm{Unif}(-1.5,1.5)$,
and $X_{2i}\sim\mathcal N(0,1)$. Sample $A$ is selected by simple random
sampling without replacement (SRSWOR), with $n_A=2{,}000$. Independent
Bernoulli selection generates $B$ under \eqref{eq:semi-ps}, with
$\gamma=0.6$ and the intercept calibrated to give $E(n_B)=1{,}500$.
The respondent law is $Y_i\mid X_i,\delta_i=1\sim\mathcal N\{m_1(X_i),1\}$,
so that by \eqref{eq:tilt} the non-respondent law is
$\mathcal N\{m_1(X_i)+\gamma,1\}$ and the population law of $Y_i$ given
$X_i$ is the two-component mixture with weights $p(X_i)$ and $1-p(X_i)$;
the finite population is generated from this mixture together with the
selection indicator, so that the stated respondent law holds exactly and
the Gaussian working model of the parametric competitor is correctly
specified under MNAR-I. \dlfill{The population is generated by computing $m_1(X_i)$,
forming $p(X_i)$, drawing $\delta_i$, and drawing $Y_i$ last, from
$\mathcal N\{m_1(X_i),1\}$ if $\delta_i=1$ and from
$\mathcal N\{m_1(X_i)+\gamma,1\}$ if $\delta_i=0$. The same construction is
used for the MAR, GMAR and INV-IV designs of
Section~\ref{sec:sim-sensitivity}; NONLIN-Y is generated in the opposite
order, as noted there.}
Here
\begin{align*}
m_1(x)&=0.5-0.6x_1+0.6x_2+\beta_4\sin(\pi x_1),\\
g(x_1)&=\phi_0+\phi_1x_1+\phi_2\sin(\pi x_1).
\end{align*}
Here $m_1$ is the respondent mean, not the log tilted moment $m(x;\gamma)$.
The three designs are
\begin{itemize}
\item \textbf{MNAR-I:} $(\phi_1,\phi_2,\beta_4)=(-0.45,0,0)$; both
the selection function and respondent mean are linear in $x_1$.
\item \textbf{MNAR-II:} $(\phi_1,\phi_2,\beta_4)=(-0.20,0.5,0)$;
only the selection function is nonlinear.
\item \textbf{MNAR-III:} $(\phi_1,\phi_2,\beta_4)=(-0.20,0.5,0.5)$;
both functions contain the same nonlinear term.
\end{itemize}
MNAR-II and MNAR-III share the covariate-dependent component of $g$;
their intercepts $\phi_0$ are calibrated separately, because changing the
respondent mean changes the intercept that gives $E(n_B)=1{,}500$. The
   standard deviations of $g(X_1)$ are similar across designs: $0.390$ in MNAR-I and
   $0.410$ in the other two. The residual from projecting $g(X_1)$ on $X_1$
   has standard deviation $0.351$ in MNAR-II and MNAR-III, but its correlation
   with $m_1(X)$ changes from zero to $0.394$. These diagnostics help interpret
how covariate misspecification affects mean estimation; they are not a
general characterization of estimator bias. Supplementary
Table~\ref{tab:sim-designs} records them.

Competitors are the naive mean $\hat\theta_B$, the MAR-based estimator
$\hat\theta_{\mathrm{CLW}}$ of \citet{chen2020doubly}, and the parametric
pseudo-likelihood estimator $\hat\theta_{\mathrm{PML}}$ of
\citet{liu2024statistical}, which uses a linear $g$. We compare these with
\eqref{eq:ipw}, \eqref{eq:mi}, and \eqref{eq:mi-orth}, denoted
\estlab{}.ipw, \estlab{}.mi, and \estlab{}.mi$\perp$ in the tables. Implementation
details are in Supplementary Section~\ref{supp:impl}.

\subsection{Results}\label{sec:sim-main}

Table~\ref{tab:sim-mean} reports bias, standard deviation and root mean
squared error (RMSE), and Table~\ref{tab:sim-var} reports the performance
of the linearization variance estimator and the coverage of nominal $95\%$
Wald intervals for the proposed estimators. Under MNAR-I, PML is correctly specified
and has bias $0.0057$. The recommended estimator has bias $0.0065$ and
RMSE $0.0923$, compared with $0.0896$ for PML, a $3.0\%$ increase.

Under MNAR-II, a linear model poorly approximates $g$, yet PML has bias
$0.0006$. This is consistent with the projection residual being
uncorrelated with the respondent mean. Adding the same nonlinearity to
the outcome in MNAR-III raises the PML bias to $0.1147$, while the bias
of \estlab{}.mi$\perp$ remains $0.0094$. Thus the consequences of a
misspecified covariate selection function depend on its relationship to
the outcome, not only on the size of the approximation error.

The proposed $\hat\gamma$ averages $0.602$, $0.602$, and $0.603$, with
standard deviations approximately $0.084$--$0.085$, against a true value
of $0.6$.
\dlfill{The tilting parameter is obtained by evaluating the orthogonal
score on a grid over $\Gamma$ and refining each sign-change bracket. A
replicate in which no sign change is found returns the minimizer of the
squared score, is recorded as a boundary outcome, and is retained in the
point estimates; where several roots are found the root-selection rule
applies, which occurred in no replicate of these three designs. Boundary outcomes occurred in 10, 5 and 3 of the 500 replicates under MNAR-I, MNAR-II and MNAR-III.} These
results support the stability of the proposed tilting-parameter estimator;
they do not by themselves isolate the source of the PML mean bias.

Among the proposed means, plain IPW has the largest RMSE in all three
designs, $0.0948$, $0.0975$ and $0.1025$, and carries the largest bias
under MNAR-III at $0.0318$. Replacing the propensity-weighted outcome by
the smoother quantity $\hat\mu(X)$ reduces variability when some fitted
propensities are small; the smallest fitted $\hat\pi^B$ across all
replicates is $1.8\times10^{-3}$.

The two mass-imputation estimators differ in the nuisance rates they
require: \eqref{eq:mi} needs the smoothing bias of $\hat\mu$ to be
$o(N^{-1/2})$, which the bandwidths of Section~\ref{sec:implementation}
do not deliver, whereas \eqref{eq:mi-orth} needs only the product-rate
condition of Section~\ref{sec:theory}. They do not differ in their
first-order variance: by Supplementary Lemma~\ref{lem:mi-common-if} the
two share the same linearized variables at the truth, so one variance
routine serves both, and Table~\ref{tab:sim-var} reports a common
estimated standard error. Their first-order difference is therefore the
smoothing bias. In these designs that difference is small and its sign
depends on the scenario. The bias of \eqref{eq:mi} is $-0.0107$,
$-0.0091$ and $0.0281$ across the three designs, against $0.0065$,
$0.0075$ and $0.0094$ for \eqref{eq:mi-orth}. Because the two estimators
are correlated at approximately $0.99$, the paired difference in absolute
error is estimated more precisely than either bias: it favors
\eqref{eq:mi} under MNAR-I and MNAR-II, by $0.0028$ and $0.0036$ with
$t=-3.1$ and $-3.9$, and \eqref{eq:mi-orth} under MNAR-III, by $0.0011$
with $t=1.2$, which is not resolved at this replicate count. What
distinguishes \eqref{eq:mi-orth} in these designs is stability rather than
uniform superiority: its biases of $0.0065$, $0.0075$ and $0.0094$ have
Monte Carlo standard errors of about $0.004$ and are statistically
indistinguishable from zero and from one another, whereas the bias of
\eqref{eq:mi} rises to $0.0281$ under MNAR-III. This is an observed
property of three designs, not a general invariance. The recommendation of
the orthogonalized estimator rests on its weaker rate requirement and this
observed stability, not on a uniform finite-sample variance advantage; Supplementary
Section~\ref{supp:mi-comparison} discusses why the gap is narrow when
$X_1$ is continuous.

Table~\ref{tab:sim-var} shows that the estimated standard errors and the
interval coverage are reasonably accurate in these designs. Relative bias
of the average estimated standard error ranges from $-6.1\%$ to $9.4\%$
across the nine estimator-by-design combinations; the largest values,
$9.4\%$ and $7.0\%$, occur for plain MI under MNAR-I and MNAR-II, where the
common standard error of Supplementary Lemma~\ref{lem:mi-common-if} is
used and the Monte Carlo standard deviation of plain MI is smaller than
that of its orthogonalized version. Coverage for the proposed means ranges
from $93.6\%$ to $96.4\%$; the Monte Carlo standard error of a coverage
rate near $95\%$ with 500 replicates is about one percentage point.
Under MNAR-I, the reference- and non-probability-sample components of the
standard error of \estlab{}.mi$\perp$ average $0.054$ and $0.077$, giving a
total of $0.094$; both include the effect of estimating $\gamma$. For the
plain estimators, these coverage results do not remove the bias requirement
for centered asymptotic inference.

\begin{table}[tbp]
\caption{Bias, standard deviation (SD) and root mean squared error (RMSE)
of estimators of the finite population mean, based on 500 Monte Carlo
replicates. Coverage of the proposed estimators is reported in
Table~\ref{tab:sim-var}.}
\label{tab:sim-mean}
\begin{center}
\begin{tabular}{llrrr}
\toprule
Design & Estimator & Bias & SD & RMSE \\
\midrule
MNAR-I
 & $\hat{\theta}_B$                       & $-0.7148$ & $0.0325$ & $0.7155$ \\
 & $\hat{\theta}_{\mathrm{CLW}}$          & $-0.5524$ & $0.0339$ & $0.5534$ \\
 & $\hat{\theta}_{\mathrm{PML}}$          & $\phantom{-}0.0057$ & $0.0895$ & $0.0896$ \\
 & $\hat{\theta}_{\mathrm{\estlab{}.ipw}}$     & $\phantom{-}0.0052$ & $0.0948$ & $0.0948$ \\
 & $\hat{\theta}_{\mathrm{\estlab{}.mi}}$      & $-0.0107$ & $0.0864$ & $0.0870$ \\
 & $\hat{\theta}_{\mathrm{\estlab{}.mi}\perp}$ & $\phantom{-}0.0065$ & $0.0922$ & $0.0923$ \\
\midrule
MNAR-II
 & $\hat{\theta}_B$                       & $-0.7980$ & $0.0321$ & $0.7987$ \\
 & $\hat{\theta}_{\mathrm{CLW}}$          & $-0.5537$ & $0.0342$ & $0.5547$ \\
 & $\hat{\theta}_{\mathrm{PML}}$          & $\phantom{-}0.0006$ & $0.0877$ & $0.0876$ \\
 & $\hat{\theta}_{\mathrm{\estlab{}.ipw}}$     & $\phantom{-}0.0091$ & $0.0971$ & $0.0975$ \\
 & $\hat{\theta}_{\mathrm{\estlab{}.mi}}$      & $-0.0091$ & $0.0888$ & $0.0892$ \\
 & $\hat{\theta}_{\mathrm{\estlab{}.mi}\perp}$ & $\phantom{-}0.0075$ & $0.0959$ & $0.0961$ \\
\midrule
MNAR-III
 & $\hat{\theta}_B$                       & $-0.7694$ & $0.0345$ & $0.7701$ \\
 & $\hat{\theta}_{\mathrm{CLW}}$          & $-0.5078$ & $0.0367$ & $0.5091$ \\
 & $\hat{\theta}_{\mathrm{PML}}$          & $\phantom{-}0.1147$ & $0.0962$ & $0.1496$ \\
 & $\hat{\theta}_{\mathrm{\estlab{}.ipw}}$     & $\phantom{-}0.0318$ & $0.0975$ & $0.1025$ \\
 & $\hat{\theta}_{\mathrm{\estlab{}.mi}}$      & $\phantom{-}0.0281$ & $0.0914$ & $0.0956$ \\
 & $\hat{\theta}_{\mathrm{\estlab{}.mi}\perp}$ & $\phantom{-}0.0094$ & $0.0936$ & $0.0940$ \\
\bottomrule
\end{tabular}
\end{center}
\end{table}

\begin{table}[tbp]
\caption{Monte Carlo standard deviation, average estimated
standard error, relative bias of the estimated standard error, and coverage of the
nominal 95\% Wald interval, 500 replicates. The two mass-imputation estimators
use the same estimated linearized variables, consistent with Supplementary Lemma~\ref{lem:mi-common-if}.}
\label{tab:sim-var}
\begin{center}
\begin{tabular}{llcccc}
\toprule
Design & Estimator & MC SD & SE & RB (\%) & CR (\%) \\
\midrule
MNAR-I
 & $\hat{\gamma}$                         & 0.0842 & 0.0822 & $-2.4$ & --- \\
 & $\hat{\theta}_{\mathrm{\estlab{}.ipw}}$     & 0.0948 & 0.0944 & $-0.4$ & 95.2 \\
 & $\hat{\theta}_{\mathrm{\estlab{}.mi}}$      & 0.0864 & 0.0945 & $\phantom{-}9.4$ & 96.4 \\
 & $\hat{\theta}_{\mathrm{\estlab{}.mi}\perp}$ & 0.0922 & 0.0945 & $\phantom{-}2.5$ & 96.2 \\
\midrule
MNAR-II
 & $\hat{\gamma}$                         & 0.0845 & 0.0827 & $-2.1$ & --- \\
 & $\hat{\theta}_{\mathrm{\estlab{}.ipw}}$     & 0.0971 & 0.0942 & $-3.0$ & 95.6 \\
 & $\hat{\theta}_{\mathrm{\estlab{}.mi}}$      & 0.0888 & 0.0950 & $\phantom{-}7.0$ & 95.8 \\
 & $\hat{\theta}_{\mathrm{\estlab{}.mi}\perp}$ & 0.0959 & 0.0950 & $-0.9$ & 95.4 \\
\midrule
MNAR-III
 & $\hat{\gamma}$                         & 0.0836 & 0.0801 & $-4.2$ & --- \\
 & $\hat{\theta}_{\mathrm{\estlab{}.ipw}}$     & 0.0975 & 0.0916 & $-6.1$ & 93.6 \\
 & $\hat{\theta}_{\mathrm{\estlab{}.mi}}$      & 0.0914 & 0.0922 & $\phantom{-}0.9$ & 95.6 \\
 & $\hat{\theta}_{\mathrm{\estlab{}.mi}\perp}$ & 0.0936 & 0.0922 & $-1.5$ & 95.6 \\
\bottomrule
\end{tabular}
\end{center}
\vspace{0.2cm}
{\footnotesize NOTE: RB is the relative bias of the average estimated standard
error against the Monte Carlo standard deviation. CR is the empirical coverage
of the nominal 95\% Wald interval; its Monte Carlo standard error at 500
replicates is 1.0 percentage point. A Wald interval for a plain estimator
additionally requires negligible bias, so the coverage reported for
\eqref{eq:mi} and \eqref{eq:ipw} is a diagnostic rather than a guaranteed
level. \dlfill{Every coverage rate in this table is a count out of 500: the
linearization variance is defined in every replicate of these designs.}}
\end{table}

\subsection{Sensitivity to model assumptions}\label{sec:sim-sensitivity}

The second study uses a four-level categorical $X_1$, so $g$ is estimated
by exact cell matching. The reference design and sample sizes are as
above, with a newly generated finite population. In addition to the
previous competitors, PMLd enters $X_1$ through $C-1$ dummy variables,
saturating the covariate component. We also examine an oracle adaptation
of \citet{shetty2025robust}, denoted SMZ$^*$, supplied with the true $g$.
Our implementation uses Gaussian working moments; its construction and
scope are described in Supplementary Section~\ref{supp:competitors}.

Four selection mechanisms are considered, with intercepts calibrated to
the target non-probability sample size:
\begin{itemize}
\item \textbf{MAR:} $\operatorname{logit}(\pi_i^B)=\phi_0-0.5X_{1i}$.
This is the restricted MAR submodel, with $\gamma=0$.
\item \textbf{GMAR:} $\operatorname{logit}(\pi_i^B)=\phi_0-0.5X_{1i}+0.2X_{2i}$.
Selection is MAR given $X$, but the IV exclusion restriction fails.
\item \textbf{INV-IV:} $\operatorname{logit}(\pi_i^B)=\phi_0-0.5X_{1i}+0.3X_{2i}-0.6Y_i$.
Both outcome dependence and a direct instrument effect are present.
\item \textbf{NONLIN-Y:} $\operatorname{logit}(\pi_i^B)=\phi_0-0.5X_{1i}-0.3Y_i^2$.
Exclusion holds, but the linear outcome tilt is misspecified. \dlfill{In this design
$Y_i$ is drawn for every population unit before selection, unlike the others,
so that the respondent law carries the quadratic tilt.}
\end{itemize}
Table~\ref{tab:sensitivity} gives the complete point-estimation comparisons.
Competitor implementation details are in Supplementary
Section~\ref{supp:competitors}. \dlfill{Boundary outcomes of the root search
occurred in two replicates under GMAR and three under NONLIN-Y and in none
under MAR or INV-IV; the root-selection rule was invoked in two replicates,
both under NONLIN-Y.}

With a fixed number of levels, dummy coding removes functional-form
restrictions on $g$, so comparing PML and PMLd tests whether the covariate
restriction explains the observed behavior; it does not address a
misspecified tilt or respondent law. SMZ$^*$ supplies the true $g$, but
its working moments and two-sample adaptation still matter.

Under MAR, the proposed means and both parametric estimators have small
bias. \estlab{}.mi$\perp$ has RMSE $0.0816$, compared with $0.0788$ for PMLd
and $0.0380$ for CLW. Thus allowing for outcome-dependent selection costs
little relative to PMLd in this design, but appreciably more than imposing
the correct MAR assumption. SMZ$^*$ has RMSE $0.1068$. This finite-sample
comparison does not establish an efficiency ordering in the two-sample
experiment. At $\gamma_0=0$, $a_g(X_1)=0$, but the variance contribution
from estimating $\gamma$ remains; see Supplementary
Remark~\ref{rmk:mar-var}.

Two diagnostics clarify the MAR result. The fitted correction
$\hat a_g$ averages $-0.0003$, consistent with its population value zero.
However, the estimated derivative of the mean with respect to $\gamma$
averages $0.96$, so uncertainty in the estimated tilt is not negligible
when the true tilt is zero. Removing its variance contribution merely
because $|\hat\gamma|$ is small would discard an important source of
sampling variability in this design.

Under GMAR, exclusion-based estimators are biased, while CLW remains
nearly unbiased because selection is ignorable given the full $X$.
Under INV-IV, both MAR-based and exclusion-based methods are biased.
Flexibility in $g$ cannot repair these violations. Variables with a
plausible direct effect on selection should be included in $X_1$, and
the exclusion of variables assigned to $X_2$ requires substantive support.

Under NONLIN-Y, \estlab{}.mi$\perp$ has bias $-0.0080$, compared with
$0.1641$ for PML and $0.1549$ for PMLd. Saturating $g$ therefore does not
resolve the parametric bias. Quadratic outcome selection also makes the
respondent law non-Gaussian in this design, affecting the Gaussian moment
calculations used by PML and the implemented SMZ$^*$ variant. The proposed
method estimates these moments nonparametrically, but still assumes a
linear tilt. Its mean $\hat\gamma=-0.262$ has no true linear-tilt value to
recover. The small mean bias is empirical evidence for this design, not
a consistency guarantee under tilt misspecification. The kernel variant
of \citet{shetty2025robust} avoids Gaussian working moments, but its
performance under a quadratic tilt is not established by this comparison.

Together with MNAR-III in Section~\ref{sec:sim-main}, this design
illustrates two distinct departures from a parametric specification,
although the two studies use different finite populations and their
magnitudes are not directly comparable. Under MNAR-III only the covariate
component of the parametric selection model is misspecified, and the
parametric mean is biased while the proposed mean is not; the reported
tilting estimates belong to the proposed method, so these results do not
decompose the PML bias into a tilt component and a covariate component.
Under NONLIN-Y the covariate component is saturated and two things change
at once: the linear tilt is wrong for every method, and the respondent law
is no longer Gaussian, which additionally affects the Gaussian working
moments of PML, PMLd and the implemented SMZ$^*$. Comparing like with like
within each design, the proposed estimator's RMSE is $0.0816$ against
$0.0788$ for PMLd under MAR, and $0.0991$ against $0.1698$ for PMLd under
NONLIN-Y.

\begin{table}[tbp]
\centering
\caption{Bias, standard deviation (SD) and root mean squared error (RMSE) for
estimators of the finite population mean $\theta$ under the MAR submodel and
three departures from the model assumptions, based on 500 Monte Carlo
replicates; $X_1$ is categorical with four levels. SMZ$^*$ is the
implemented oracle comparator; it is not evaluated under GMAR and INV-IV
because its model class requires the selection logit $g^*(X_1)-\gamma Y$,
which fails when $X_2$ enters selection directly.}
\label{tab:sensitivity}
\footnotesize
\setlength{\tabcolsep}{2pt}
\begin{tabular}{lrrrrrrrrrrrr}
\toprule
& \multicolumn{3}{c}{MAR} & \multicolumn{3}{c}{GMAR}
& \multicolumn{3}{c}{INV-IV} & \multicolumn{3}{c}{NONLIN-Y} \\
\cmidrule(lr){2-4}\cmidrule(lr){5-7}\cmidrule(lr){8-10}\cmidrule(lr){11-13}
Estimator & Bias & SD & RMSE & Bias & SD & RMSE & Bias & SD & RMSE & Bias & SD & RMSE \\
\midrule
Naive         &  0.3267 & 0.0339 & 0.3285 &  0.4340 & 0.0350 & 0.4354 & $-$0.4911 & 0.0348 & 0.4923 &  0.3773 & 0.0236 & 0.3780 \\
CLW           &  0.0005 & 0.0381 & 0.0380 & $-$0.0001 & 0.0411 & 0.0411 & $-$0.5551 & 0.0330 & 0.5561 &  0.1825 & 0.0273 & 0.1845 \\
PML           & $-$0.0003 & 0.0797 & 0.0797 & $-$0.3108 & 0.0874 & 0.3229 & {$-$0.4630} & {0.0735} & {0.4688} &  0.1641 & 0.0695 & 0.1782 \\
PMLd          & $-$0.0008 & 0.0789 & 0.0788 & $-$0.3117 & 0.0870 & 0.3236 & {$-$0.4629} & {0.0734} & {0.4686} &  0.1549 & 0.0697 & 0.1698 \\
SMZ$^*$       & {$-$0.0005} & {0.1069} & {0.1068} & --- & --- & --- & --- & --- & --- & {0.1724} & {0.0718} & {0.1867} \\
OP.ipw        & $-$0.0049 & 0.0816 & 0.0816 & $-$0.3180 & 0.0958 & 0.3321 & {$-$0.4642} & {0.0759} & {0.4703} & $-$0.0110 & 0.1006 & 0.1011 \\
OP.mi         & $-$0.0032 & 0.0789 & 0.0789 & $-$0.3050 & 0.0898 & 0.3179 & {$-$0.4662} & {0.0740} & {0.4720} & $-$0.0019 & 0.0940 & 0.0939 \\
OP.mi$^\perp$ & $-$0.0034 & 0.0816 & 0.0816 & $-$0.3181 & 0.0947 & 0.3319 & {$-$0.4634} & {0.0761} & {0.4696} & $-$0.0080 & 0.0988 & 0.0991 \\
\bottomrule
\end{tabular}
\end{table}

\section{A simulation based on national health-screening data}\label{sec:app}
\subsection{Finite population and sampling mechanisms}
\label{sec:app-data}\label{sec:app-design}

We use an anonymized extract of the Republic of Korea's 2014 National
Health Screening Program, provided by the National Health Insurance
Sharing Service (NHISS).\footnote{\url{https://nhiss.nhis.or.kr/bd/ab/bdabf006cv.do}}
The extract contains sex, age bands, hemoglobin (HGB), total cholesterol
(TCHOL), triglycerides (TG), and high-density lipoprotein cholesterol
(HDL). We take $N=50{,}000$ records as a finite population and estimate
the mean of $y=\log(\mathrm{TG})$, whose population value is $4.687$.
This study preserves the empirical joint distribution of $(X,Y)$ while
imposing the sampling and selection mechanisms. It is a simulation using
real records, not an analysis identifying the screening program's actual
participation mechanism.

Sample $A$ is SRSWOR with $n_A=1{,}000$ and $d_i^A=N/n_A$.
Independently, sample $B$ is selected by Bernoulli draws with
\begin{equation}\label{eq:app-pib}
\pi_i^B=\operatorname{expit}\{\alpha+g(x_{1i})-\gamma(y_i-\bar y)\},
\end{equation}
where $\alpha$ is calibrated to $E(n_B)=2{,}500$. Centering $y$ is an
intercept reparameterization and does not change the tilt. Set
$\gamma=1.047$, so $\gamma\operatorname{sd}(y)=0.6$, and consider
\begin{align*}
\mathrm{M}_1:\quad g(x_{1i})&=-1.5-0.2\tilde h_i-0.5\sin(\pi\tilde h_i),\\
\mathrm{M}_2:\quad g(x_{1i})&=-1.5-0.2\tilde h_i-0.5\sin(\pi\tilde h_i)
+0.4\sin(\pi\tilde a_i),
\end{align*}
where $\tilde h_i$ and $\tilde a_i$ are standardized HDL and age.
\dlfill{In the $X$-regressions SEX is matched exactly, while AGE enters as its
standardized band code with a bandwidth of its own. The smoothing dimension is
therefore $d=3$ under M1, where the smoothed coordinates are HDL, TCHOL and
HGB, and $d=4$ under M2, which adds AGE; the exponent $-1/(d+4)$ uses those
values. Band codes and band midpoints are an affine transformation of one
another and give the same kernel after standardization. With $s=2$ the
sufficient condition $2s>d$ of Section~\ref{sec:theory} admits at most $d=3$,
so M1 satisfies it and M2 lies outside it by one dimension. The condition is
sufficient and not necessary, and the coverage reported below under M2 is an
empirical diagnostic at a dimension it does not cover.}
The selection covariates are $X_1=\mathrm{HDL}$ under M1 and
$X_1=(\mathrm{HDL},\mathrm{AGE})$ under M2. The baseline instrument set
is $X_2=(\mathrm{SEX},\mathrm{TCHOL},\mathrm{HGB})$; exclusion holds by
construction. A linear approximation to $g$ explains about
$27\%$ of its variation, making the linear working selection model
misspecified.

Estimation uses five-fold cross-fitting, a bandwidth multiplier of three,
and a fitted-propensity floor of $5\times10^{-3}$. Supplementary
Section~\ref{supp:constants} gives the bandwidth and search settings.
The multiplier was chosen to improve numerical stability in this study:
at multiplier one, boundary outcomes were reported in \dlnum{$20.2\%$} of
replicates, versus none at multiplier three. Supplementary
Section~\ref{supp:bw} reports the full comparison. There are 500
replicates, with matched samples across the instrument specifications
within each mechanism. An active fixed propensity floor is not covered
by the negligibility requirement in (R3), so the reported coverage is
an empirical diagnostic rather than a direct verification of the theorem.
\dlfill{Over the 500 replicates the floor binds at $222$ $B$-records under M1
and $77$ under M2, fewer than one per replicate out of $n_B\approx2{,}500$,
rising to $1{,}546$ under the AGE-only specification of
Section~\ref{sec:app-iv}. Supplementary Section~\ref{supp:constants} lists
the quantities it is applied to, and Supplementary Section~\ref{supp:floor}
reports a rerun at floors $10^{-3}$ and $10^{-2}$: the bias of the
inverse-probability-weighting estimator moves monotonically with the floor,
from $-0.168$ to $-0.190$, while the orthogonalized mass-imputation estimate
moves by less than a tenth of its Monte Carlo standard deviation and keeps
coverage $94.4\%$ at all three floors.}

\subsection{Estimation of the population mean}\label{sec:app-main}

Table~\ref{tab:app-theta} reports the six mean estimators. The naive and
MAR-based estimates remain substantially biased: under M1 their biases
are $-0.248$ and $-0.200$. \estlab{}.mi$\perp$ reduces the bias to \dlnum{$-0.0067$}
under M1 and \dlnum{$-0.0217$} under M2. Its nominal $95\%$ intervals have
coverage \dlnum{$94.4\%$} and \dlnum{$89.6\%$}, respectively.
\dlfill{Table~\ref{tab:app-cov} gives the variance estimator behind those
intervals: the average estimated standard error falls below the Monte Carlo
standard deviation by $5.2\%$ under M1 and by $2.3\%$ under M2.} The M2
bias, \dlnum{$-0.0217$ against a standard deviation of $0.0496$}, is about
\dlnum{$44\%$} of that standard deviation and contributes to
the coverage shortfall.

Orthogonalization improves RMSE relative to plain MI, from \dlnum{$0.0646$} to
\dlnum{$0.0559$} under M1 and from \dlnum{$0.0551$} to \dlnum{$0.0541$} under M2, but not through
bias: under M2 the orthogonalized estimator has the larger bias,
\dlnum{$-0.0217$} against \dlnum{$-0.0047$} for plain MI, and its RMSE is smaller only
because its standard deviation is smaller (\dlnum{$0.0496$} against \dlnum{$0.0549$}).
This is not in conflict with Theorem~\ref{thm:joint}, which concerns
first-order centering, but it does mean that orthogonalization does not
uniformly reduce empirical bias at these sample sizes, and the
\dlnum{$89.6\%$} coverage under M2 reflects that bias. The paired absolute-error
comparison favors the orthogonalized estimator under M1 with \dlnum{$t=4.3$}; the
difference under M2 is not resolved at this replicate count. In contrast, PML has smaller RMSE than \estlab{}.mi$\perp$ under both
mechanisms: $0.0479$ versus \dlnum{$0.0559$} and $0.0524$ versus \dlnum{$0.0541$}.
The more flexible method reduces bias here at a cost in sampling variability.

With 500 replicates the Monte Carlo standard error of the PML mean is
about $0.002$, so its biases of $-0.0158$ and $-0.0269$ are detectable,
yet its smaller sampling standard deviation yields the lower RMSE.

The relatively small PML biases are consistent with the weak correlation,
$-0.013$, between the projection residual of $g$ and $y$: as in MNAR-II,
poor approximation of the selection function need not produce a large mean
bias. This diagnostic requires the known selection mechanism of the
simulation.

Plain IPW is more sensitive, with biases \dlnum{$-0.1730$} and \dlnum{$-0.1842$}.
Smoothing of $g$ and truncation of large inverse-propensity weights can
both contribute. Some fitted propensities reach the floor for units with
large outcomes. Mean augmentation reduces the observed bias in this study,
but the orthogonalized estimator still uses inverse-propensity residuals
and is not immune to truncation. The normalization diagnostic
$N^{-1}\sum_i\delta_i/\hat\pi_i^B$ averages $0.964$ and $0.965$;
its deviation from one can reflect smoothing, truncation, and estimation
of $\gamma$, rather than isolating a single source of bias.

The proposed tilting estimates average \dlnum{$1.020$} and \dlnum{$0.949$}, compared
with $0.978$ and $0.931$ for PML and a true value of $1.047$.
Table~\ref{tab:app-var} gives their bias and variability: the proposed
estimator is less biased but more variable than PML under both
mechanisms. No boundary outcomes are reported under the baseline
instrument set. The next
comparison examines sensitivity to using less informative instruments.

\begin{table}[tbp]
\caption{Bias, standard deviation and root mean squared error of estimators of
the finite population mean $\theta$, based on 500 Monte Carlo replicates from
the NHISS finite population. True values: $\theta = 4.687$, $\gamma = 1.047$.}
\label{tab:app-theta}
\begin{center}
\begin{tabular}{llrrr}
\toprule
Model & Estimator & Bias & SD & RMSE \\
\midrule
$\mathrm{M}_1$
 & $\hat{\theta}_B$                        & $-0.2479$ & $0.0103$ & $0.2481$ \\
 & $\hat{\theta}_{\mathrm{CLW}}$           & $-0.2000$ & $0.0131$ & $0.2004$ \\
 & $\hat{\theta}_{\mathrm{PML}}$           & $-0.0158$ & $0.0452$ & $0.0479$ \\
 & $\hat{\theta}_{\mathrm{\estlab{}.ipw}}$      & \dlnum{$-0.1730$} & \dlnum{$0.0576$} & \dlnum{$0.1823$} \\
 & $\hat{\theta}_{\mathrm{\estlab{}.mi}}$       & \dlnum{$\phantom{-}0.0061$} & \dlnum{$0.0643$} & \dlnum{$0.0646$} \\
 & $\hat{\theta}_{\mathrm{\estlab{}.mi}\perp}$  & \dlnum{$-0.0067$} & \dlnum{$0.0555$} & \dlnum{$0.0559$} \\
\midrule
$\mathrm{M}_2$
 & $\hat{\theta}_B$                        & $-0.2529$ & $0.0104$ & $0.2531$ \\
 & $\hat{\theta}_{\mathrm{CLW}}$           & $-0.1993$ & $0.0131$ & $0.1998$ \\
 & $\hat{\theta}_{\mathrm{PML}}$           & $-0.0269$ & $0.0451$ & $0.0524$ \\
 & $\hat{\theta}_{\mathrm{\estlab{}.ipw}}$      & \dlnum{$-0.1842$} & \dlnum{$0.0564$} & \dlnum{$0.1926$} \\
 & $\hat{\theta}_{\mathrm{\estlab{}.mi}}$       & \dlnum{$-0.0047$} & \dlnum{$0.0549$} & \dlnum{$0.0551$} \\
 & $\hat{\theta}_{\mathrm{\estlab{}.mi}\perp}$  & \dlnum{$-0.0217$} & \dlnum{$0.0496$} & \dlnum{$0.0541$} \\
\bottomrule
\end{tabular}
\end{center}
\end{table}

\begin{table}[tbp]
\caption{Monte Carlo bias and standard deviation of the tilting-parameter
estimators, based on 500 replicates. True $\gamma = 1.047$.}
\label{tab:app-var}
\begin{center}
\begin{tabular}{lrrrr}
\toprule
 & \multicolumn{2}{c}{$\hat{\gamma}_{\mathrm{\estlab{}}}$}
 & \multicolumn{2}{c}{$\hat{\gamma}_{\mathrm{PML}}$} \\
\cmidrule(lr){2-3}\cmidrule(lr){4-5}
Model & Bias & SD & Bias & SD \\
\midrule
$\mathrm{M}_1$ & \dlnum{$-0.027$} & \dlnum{$0.248$} & $-0.069$ & $0.180$ \\
$\mathrm{M}_2$ & \dlnum{$-0.098$} & \dlnum{$0.209$} & $-0.116$ & $0.184$ \\
\bottomrule
\end{tabular}
\end{center}
\end{table}

\begingroup\ifdlmarks\color{dlblue}\fi
\begin{table}[tbp]
\ifdlmarks\color{dlblue}\fi
\caption{\ifdlmarks\color{dlblue}\fi Linearization variance and interval coverage of
$\hat{\theta}_{\mathrm{\estlab{}.mi}\perp}$, based on 500 replicates.
True $\theta = 4.687$.}
\label{tab:app-cov}
\begin{center}
\begin{tabular}{lcccc}
\toprule
Model & MC SD & SE & RB (\%) & CR (\%) \\
\midrule
$\mathrm{M}_1$ & $0.0555$ & $0.0526$ & $-5.2$ & $94.4$ \\
$\mathrm{M}_2$ & $0.0496$ & $0.0485$ & $-2.3$ & $89.6$ \\
\bottomrule
\end{tabular}
\end{center}
\vspace{0.2cm}
{\footnotesize NOTE: MC SD is the Monte Carlo standard
deviation, SE the average estimated standard error, RB the relative bias of SE
against MC SD, and CR the empirical coverage of the nominal $95\%$ Wald
interval, a count out of 500 in both designs.}
\end{table}
\endgroup

\subsection{Sensitivity to the instrument set}\label{sec:app-iv}

We repeat M1 holding $X_1=\mathrm{HDL}$ fixed and varying the included
instruments. All four sets in Table~\ref{tab:app-iv} satisfy exclusion
by construction. The same realized samples are used across specifications,
so comparisons are paired. Age could not serve as an excluded instrument
under M2 because it enters that mechanism directly.

Table~\ref{tab:app-iv} lists the four specifications; the partial
correlation of each instrument with $y$ given HDL is $-0.15$ for SEX,
$0.42$ for TCHOL, $0.21$ for HGB, and $0.09$ for AGE. These correlations
describe relevance but do not verify completeness.
The AGE-only set is discrete and cannot satisfy full $L^2$ completeness
for a continuously modeled outcome; it is a sensitivity analysis outside
the sufficient identification framework, which does not itself prove
nonidentification of the scalar tilt (Supplementary
Section~\ref{supp:completeness}).

\begin{table}[tbp]
\caption{Sensitivity to the instrument set under
$\mathrm{M}_1$, based on 500 Monte Carlo replicates per specification with
identical samples across specifications. Entries for $\hat{\theta}$ are bias;
$x_1 = \mathrm{HDL}$ throughout. True $\theta = 4.687$, $\gamma = 1.047$.}
\label{tab:app-iv}
\begin{center}
\footnotesize
\setlength{\tabcolsep}{3.5pt}
\begin{tabular}{llrrrrrrrr}
\toprule
& & & \multicolumn{4}{c}{Bias of $\hat{\theta}$}
& \multicolumn{2}{c}{$\hat{\gamma}$} & \\
\cmidrule(lr){4-7}\cmidrule(lr){8-9}
Spec & $x_2$ & Partial corr. & $\hat{\theta}_B$ & PML & MI & MI$\perp$
& PML & \estlab{} & CR(\%) \\
\midrule
$\mathrm{S}_1$ & SEX, TCHOL, HGB & $-0.15$, $0.42$, $0.21$ & $-0.248$ & $-0.016$ & \dlnum{$\phantom{-}0.006$} & \dlnum{$-0.007$}
               & $0.978$ & \dlnum{$1.020$} & \dlnum{$94.4$} \\
$\mathrm{S}_2$ & TCHOL           & $0.42$                  & $-0.248$ & $-0.016$ & \dlnum{$\phantom{-}0.014$} & \dlnum{$\phantom{-}0.000$}
               & $0.979$ & \dlnum{$1.048$} & \dlnum{$95.4$} \\
$\mathrm{S}_3$ & HGB, SEX        & $0.21$, $-0.15$         & $-0.248$ & $-0.120$ & \dlnum{$\phantom{-}0.006$} & \dlnum{$-0.008$}
               & $0.546$ & \dlnum{$1.016$} & \dlnum{$95.5$} \\
$\mathrm{S}_4$ & AGE             & $0.09$                  & $-0.248$ & $-0.177$ & \dlnum{$\phantom{-}0.069$} & \dlnum{$\phantom{-}0.029$}
               & $0.302$ & \dlnum{$1.180$} & \dlnum{$91.3$} \\
\midrule
\multicolumn{3}{l}{Range across specifications}
               & $0.000$ & $0.161$ & \dlnum{$0.063$} & \dlnum{$0.037$}
               & $0.677$ & \dlnum{$0.164$} & \\
\bottomrule
\end{tabular}
\end{center}
\vspace{0.2cm}
{\footnotesize NOTE: $x_1=\mathrm{HDL}$ throughout. ``Partial corr.'' is
the partial correlation between each instrument and $y$ after $x_1$ is
partialled out, in the order the instruments are listed; it describes
relevance rather than verifying conditional completeness. CR is the empirical coverage of the
nominal $95\%$ Wald interval for $\hat{\theta}_{\mathrm{\estlab{}.mi}\perp}$.
The Monte Carlo standard error of each specification mean is \dlnum{$0.002$--$0.007$}
for $\hat{\theta}_{\mathrm{\estlab{}.mi}\perp}$ and $0.002$--$0.003$ for
$\hat{\theta}_{\mathrm{PML}}$. \dlfill{Coverage is computed over the
replicates for which the linearization variance is defined, which is $500$,
$500$, $491$ and $449$ for $\mathrm{S}_1$ to $\mathrm{S}_4$; the biases are
computed over all 500.}}
\end{table}

The PML bias ranges from $-0.016$ to $-0.177$, a spread of $0.161$.
The corresponding range for \estlab{}.mi$\perp$ is \dlnum{$0.037$}, \dlnum{about a
quarter} as large, calculated from the displayed biases. The proposed
tilting estimate stays between \dlnum{$1.016$} and \dlnum{$1.180$}, while PML ranges
from $0.302$ to $0.979$. Thus the instability of the PML mean coincides
with substantial changes in its fitted tilt.

The naive mean is unchanged across instrument sets, with bias $-0.248$.
Plain MI has a bias range of \dlnum{$0.063$}, between the ranges for PML and its
orthogonalized counterpart. The variation across specifications is
substantively larger than the Monte Carlo uncertainty of the individual
means given in the table note; because the comparisons reuse the same
samples, their paired uncertainty is not obtained by treating the rows as
independent.

Retaining TCHOL alone gives results similar to the baseline set, whereas
retaining HGB and SEX without TCHOL produces greater PML bias. This
pattern is consistent with the relative informativeness of the instruments
and does not suggest that simply adding more variables improves estimation.
The proposed estimator is less sensitive in bias, but loses precision:
its Monte Carlo standard deviation increases from \dlnum{$0.056$} with the
baseline set to $0.158$ with AGE alone; the standard deviation of
$\hat\gamma$ increases from \dlnum{$0.248$} to \dlnum{$0.692$}.

Coverage is \dlnum{$94.4\%$}, \dlnum{$95.4\%$} and \dlnum{$95.5\%$} for the first
three sets and \dlnum{$91.3\%$} for AGE alone, where boundary outcomes are reported in \dlnum{$1.6\%$}
of replicates. \dlfill{Thin instrument sets also make the estimated Jacobian
small: the linearization variance is undefined in $9$ replicates under
$\mathrm{S}_3$ and $51$ under $\mathrm{S}_4$, so the coverage for those sets
is computed over the remaining replicates, as the table notes state.} These results concern sensitivity to the included instrument
set under imposed exclusion. They do not establish robustness to invalid
instruments or verify the sufficient completeness condition in the
empirical population.

\section{Concluding remarks}\label{sec:conclusion}

The proposed approach combines probability and non-probability samples
under a semiparametric outcome-dependent selection model. Profiling avoids
a parametric specification of the covariate component, while separate
orthogonalization of the tilt and mean equations permits centered
root-$N$ inference under the stated nuisance-rate and sampling conditions.
The recommended mass-imputation estimator retains the contribution from
estimating the tilt in its variance, including under the restricted MAR
submodel.

The numerical studies illustrate the value and cost of this flexibility.
It can reduce bias from covariate misspecification and sensitivity to the
instrument set, but does not uniformly improve RMSE. Valid exclusion,
sufficient identifying information, and stable nuisance estimation remain
essential. The small bias observed under a quadratic tilt is
design-specific, and fixed propensity truncation can affect inference.

Further work could investigate calibration through empirical likelihood,
alternative models for outcome-dependent selection, and more flexible
nuisance learners. Such extensions require appropriate identification,
weighted convergence rates, and sampling-design conditions. Practical
procedures for bandwidth choice and inference with weak instruments are
also important.

\section*{Supplementary Material}

The Supplementary Material contains the nuisance estimators and variance
formulas (Section~\ref{appendix:A1}), implementation details
(Section~\ref{supp:impl}), the regularity conditions
(Section~\ref{appendix:regularity}), all proofs
(Sections~\ref{appendix:proofs}--\ref{appendix:identification}), and
additional numerical results (Section~\ref{supp:additional}).



\section*{Data Availability Statement}

\ifanon
\else
The R and C++ code that reproduces all numerical results, the data and the
saved Monte Carlo output are openly available in Zenodo at
\url{https://doi.org/10.5281/zenodo.23513896}. The simulations of
Section~\ref{sec:sim} use generated data only. Section~\ref{sec:app} uses a
50,000-record subset of the 2014 health-screening public data of the National
Health Insurance Service, Republic of Korea, distributed through the National
Health Insurance Sharing Service.

\fi
\section*{AI disclosure statement} 

The authors used Claude (Anthropic; Opus 5.5) to improve the presentation of the manuscript and the Supplementary Material, including organization, wording, length, and the consistency of notation, cross-references and reported figures. The authors also used the tool to check and refine the theoretical development, including the precision of the regularity conditions and the rigor and completeness of the proofs. The proposed methodology, the main results and the conclusions are the authors’ own, and the tool was not used to generate data or to produce numerical results. The authors  take full responsibility for the originality, accuracy and integrity of the entire content, including the references.

\bibliography{ref}

\clearpage


\section*{Supplementary material}

# Supplementary Material for "A Semiparametric Model Approach to Data Integration under Missing Not at Random"

This supplement is organized as follows. Section S1 specifies the kernel nuisance estimators, the cross-fitting algorithm, and the linearization variance estimators used in the main text. Section S2 gives the implementation constants for each numerical study, the likelihood constructions, and competitor details. Section S3 states the regularity conditions (R1)–(R10). Section S4 proves Proposition 1 and Theorem 1 of the main text, and Supplementary Theorems S1–S2 and derives the bias and influence representations of the plain estimators. Section S5 collects the verification lemmas, and Section S6 proves identification under conditional completeness. Section S7 collects additional numerical results: full simulation tables, a discussion of the comparison between the two mass-imputation estimators, a discussion of completeness with discrete instruments, and a bandwidth sensitivity study. Equation, section, theorem, lemma, and table numbers prefixed by "S" refer to this supplement; all other numbers refer to the main text.

## S1 Nuisance estimation, cross-fitting, and variance estimation

This section specifies the kernel-based nuisance estimators used throughout the paper, the cross-fitting algorithm, and the linearization variance estimators. For simplicity throughout, we assume that the finite population size $N$ is known; this entails no loss of generality for the first-order results, since replacing $N$ by a consistent estimator only adds the usual ratio-linearization term.

### S1.1 Kernel nuisance estimators

Let $X \in \mathbb{R}^d$ and $X_1 \in \mathbb{R}^{d_1}$. For a kernel $K$ and bandwidths $h, h_1 > 0$, define

$$K_h(u) = h^{-d}K(u/h), \qquad K_{h_1}(u) = h_1^{-d_1}K(u/h_1).$$

For fixed $\gamma$, the profile nuisance functions are estimated by

$$\exp\{\hat{m}(x;\gamma)\} = \frac{\sum_{j\in B} e^{\gamma Y_j} K_h(x - X_j)}{\sum_{j\in B} K_h(x - X_j)}, \tag{S1}$$

The two conditional means entering (15) are estimated by

$$\hat{\mu}_1(x) = \frac{\sum_{j\in B} Y_j K_h(x - X_j)}{\sum_{j\in B} K_h(x - X_j)}, \qquad \hat{\mu}_0(x;\gamma) = \frac{\sum_{j\in B} Y_j e^{\gamma Y_j} K_h(x - X_j)}{\sum_{j\in B} e^{\gamma Y_j} K_h(x - X_j)}.$$

Thus $\hat{\mu}(x;\gamma) = \hat{\pi}_m^B(x;\gamma)\hat{\mu}_1(x) + \{1 - \hat{\pi}_m^B(x;\gamma)\}\hat{\mu}_0(x;\gamma)$.

$$\exp\{\hat{g}(x_1;\gamma)\} = \frac{\sum_{j\in B} e^{\gamma Y_j} K_{h_1}(x_1 - X_{1j})}{\sum_{j\in A} d_j^A K_{h_1}(x_1 - X_{1j}) - \sum_{j\in B} K_{h_1}(x_1 - X_{1j})}. \tag{S2}$$

The denominator in (S2) estimates the design-weighted number of units not selected into the non-probability sample in a neighborhood of $x_1$. Where the reference sample is locally sparse this estimated count can be non-positive. It is then truncated at machine epsilon and carried through; the evaluation point is retained in the kernel sums, and the occurrences are counted and reported in Section S7.5.

For the orthogonal score, define the four conditional expectations

$$\pi_{m,0}^B(x) = E(\delta \mid X = x), \qquad \varpi_{g,0}^B(x_1) = E(\delta \mid X_1 = x_1),$$

$$r_m^{(k)}(x;\gamma) = E\{\delta Y^{\mathbb{I}\{k=1\}} e^{\gamma Y} \mid X = x\}, \quad r_g^{(k)}(x_1;\gamma) = E\{\delta Y^{\mathbb{I}\{k=1\}} e^{\gamma Y} \mid X_1 = x_1\}$$

for $k = 0, 1$, and their kernel estimators

$$\hat{\pi}_{m,0}^B(x) = \frac{\sum_{j\in B} K_h(x - X_j)}{\sum_{j\in A} d_j^A K_h(x - X_j)}, \quad \hat{\varpi}_{g,0}^B(x_1) = \frac{\sum_{j\in B} K_{h_1}(x_1 - X_{1j})}{\sum_{j\in A} d_j^A K_{h_1}(x_1 - X_{1j})},$$

$$\hat{r}_m^{(k)}(x;\gamma) = \frac{\sum_{j\in B} Y_j^{I\{k=1\}} e^{\gamma Y_j} K_h(x - X_j)}{\sum_{j\in A} d_j^A K_h(x - X_j)}, \quad \hat{r}_g^{(k)}(x_1;\gamma) = \frac{\sum_{j\in B} Y_j^{I\{k=1\}} e^{\gamma Y_j} K_{h_1}(x_1 - X_{1j})}{\sum_{j\in A} d_j^A K_{h_1}(x_1 - X_{1j})}.$$

The population profile components satisfy

$$e^{m(x;\gamma)} = \frac{r_m^{(0)}(x;\gamma)}{\pi_{m,0}^B(x)}, \qquad e^{g(x_1;\gamma)} = \frac{r_g^{(0)}(x_1;\gamma)}{1 - \varpi_{g,0}^B(x_1)}.$$

Together with the derivative ratios below, these identities express the score's nuisance functions through the primitive conditional moments.

The derivatives of the profile functions are then computed as

$$\nabla_\gamma \hat{m}(x;\gamma) = \frac{\hat{r}_m^{(1)}(x;\gamma)}{\hat{r}_m^{(0)}(x;\gamma)}, \qquad \nabla_\gamma \hat{g}(x_1;\gamma) = \frac{\hat{r}_g^{(1)}(x_1;\gamma)}{\hat{r}_g^{(0)}(x_1;\gamma)}.$$

The function $\xi(x_1;\gamma)$ in Section 3.2 is a population conditional expectation over the distribution of $X$ given $X_1$. It is therefore estimated from the probability sample, using its design weights:

$$\hat{\xi}(x_1;\gamma) = \frac{\sum_{j\in A} d_j^A K_{h_1}(x_1 - X_{1j})\hat{c}_j(\gamma)\hat{\pi}_m^B(X_j;\gamma)\{1-\hat{\pi}_m^B(X_j;\gamma)\}}{\sum_{j\in A} d_j^A K_{h_1}(x_1 - X_{1j})}.$$

Within cross-fitting, both sums are restricted to the training portion of sample $A$. An unweighted smoother over $A\cup B$ generally converges to a selected mixture distribution and does not estimate the coefficient required for the cancellation in (S11).

The variance formulas below also use

$$\hat{\kappa}_g(x_1;\gamma) = \frac{\sum_{j\in B} Y_j e^{\gamma Y_j} K_{h_1}(x_1 - X_{1j})}{\sum_{j\in B} e^{\gamma Y_j} K_{h_1}(x_1 - X_{1j})},$$

and the tilted conditional variance

$$\widehat{V}_0(x;\gamma) = \frac{\sum_{j\in B}\{Y_j - \hat{\mu}_0(x;\gamma)\}^2 e^{\gamma Y_j} K_h(x - X_j)}{\sum_{j\in B} e^{\gamma Y_j} K_h(x - X_j)}.$$

All such quantities are evaluated out of fold.

The orthogonalized MI estimator additionally requires $a_g$ from (16). For each outer fold $k$ and fixed candidate $\gamma$, define

$$\hat{O}_j^{B,[-k]}(\gamma) = \exp\{\gamma Y_j - \hat{g}^{[-k]}(X_{1j};\gamma)\}, \qquad \hat{\mu}_j^{[-k]}(\gamma) = \hat{\mu}^{[-k]}(X_j;\gamma),$$

where both functions are fitted using only the outer training records $D_{-k}$. For $i \in U^{(k)}$, set

$$\hat{a}_{g,i}(\gamma) = \frac{\sum\limits_{j\in B\cap D_{-k}} \hat{O}_j^{B,[-k]}(\gamma)\{Y_j - \hat{\mu}_j^{[-k]}(\gamma)\}K_{h_1}(X_{1i} - X_{1j})}{\sum\limits_{j\in B\cap D_{-k}} \hat{O}_j^{B,[-k]}(\gamma)K_{h_1}(X_{1i} - X_{1j})}. \tag{S3}$$

The final prediction is $\hat{a}_{g,i} = \hat{a}_{g,i}(\hat{\gamma})$. Using the usual out-of-fold prediction for each training record $j$ is not sufficient: that prediction may have been trained on the outer evaluation fold $k$. Any inner cross-fitting or tuning is confined to $D_{-k}$. The pooled finite-dimensional estimate $\hat{\gamma}$ is substituted only after these fold-specific functions of $\gamma$ have been constructed.

## S1.2 Cross-fitting algorithm

We use $K$ outer folds. For theoretical normalization, $U^{(k)}$ denotes a fold of population indices and $D_{-k}$ its observed training records from $A$ and $B$. Folds and weights must

satisfy (R6); linked records from the same population unit share a fold. The steps below expand Algorithm 1 of the main text, which uses a single pooled root (DML2).

1. Assign outer folds as in (R6), keeping linked records together.
2. For each candidate $\gamma$ and each fold $k$, compute all kernel nuisance fits in Section S1.1 using $D_{-k}$ only.
3. Evaluate $\hat{\pi}_m^B(x_i; \gamma)$ and $\hat{\Phi}_{\perp,i}(\gamma)$ on the held-out fold $U^{(k)}$ and pool the score contributions across folds.
4. Refine every sign-change bracket for $N^{-1} \sum_i \hat{\Phi}_{\perp,i}(\gamma) = 0$ and select the root by the rule in Section 3.4 of the main text.
5. Evaluate $\hat{\mu}_i$ and $\hat{a}_{g,i}(\hat{\gamma})$ out of fold. Every intermediate regression entering (S3) uses the same outer training set; inner tuning or splitting stays within $D_{-k}$.
6. Assemble the three mean estimators and their linearization variances from Section S1.3.

Nuisance fits must be recomputed at each candidate $\gamma$ because the profile depends on the tilted moments. Substituting the pooled estimate $\hat{\gamma}$ occurs only after these fold-specific functions have been constructed. The dominant direct-computation cost is $O\{KG(n_A + n_B)^2\}$ for $G$ grid candidates, before local refinement.

## S1.3 Linearization variance estimators

Let $k(i)$ denote the fold containing observation $i$, and evaluate all nuisance functions out of fold. The formulas in this subsection estimate the variance of the stochastic linear term. For the plain estimators they do not remove the first-order centering terms $\mathcal{B}_{\mathrm{IPW},N}$ and $\mathcal{B}_{\mathrm{MI},N}$. The sample-separated formulas are the primary formulas for inference on the finite-population target $\theta_N$. The unit-level full influence contributions displayed below instead correspond to the i.i.d.-Poisson superpopulation specialization and are included only to connect with the usual empirical-second-moment implementation.

It is useful first to decompose the orthogonal score as

$$\hat{\Phi}_{\perp,i}(\gamma) = I_i^A d_i^A \hat{A}_{\gamma,i}(\gamma) + \delta_i \hat{B}_{\gamma,i}(\gamma),$$

where, with $\hat{q}_i = 1 - \hat{\varpi}^B_{g,0}(X_{1i})$,

$$\hat{A}_{\gamma,i} = -\hat{\pi}^B_m(X_i)\hat{c}_i + \frac{\hat{\xi}(X_{1i})}{\hat{q}_i},$$

$$\begin{aligned}\hat{B}_{\gamma,i} = \hat{c}_i - &\frac{\hat{\xi}(X_{1i})}{\hat{q}_i}\{1 + e^{\hat{\gamma}Y_i - \hat{g}_i}\} \\ &+ \frac{\hat{c}_i\{\hat{\pi}^B_m(X_i)\}^2}{e^{\hat{g}_i}\hat{\pi}^B_{m,0}(X_i)}\{e^{\hat{\gamma}Y_i} - e^{\hat{m}_i}\}.\end{aligned}$$

The Jacobian is

$$\hat{\mathcal{J}}_\gamma = \frac{1}{\hat{N}^A}\sum_{i=1}^N I_i^A d_i^A \hat{\pi}^B_m(X_i)\{1 - \hat{\pi}^B_m(X_i)\}\hat{c}_i\{\nabla_\gamma \hat{m}_i - \nabla_\gamma \hat{g}_i\}, \qquad \hat{N}^A = \sum_i I_i^A d_i^A. \tag{S4}$$

Under independent Poisson reference sampling and unit-level linkage, one may estimate a superpopulation variance by the empirical second moment of the corresponding full influence contribution. For a general probability design, or when overlap between $A$ and $B$ is not linked, the finite-population variance is more transparently estimated by separating the two independent sample contributions. For a scalar array $u_i$ observed in sample $A$, define

$$\widehat{\mathcal{V}}_A(u) = \frac{1}{N^2}\sum_{i,j\in A}\frac{\pi^A_{ij} - \pi^A_i\pi^A_j}{\pi^A_{ij}}\frac{u_i}{\pi^A_i}\frac{u_j}{\pi^A_j},$$

with $\pi^A_{ii} = \pi^A_i$, provided the required pairwise inclusion probabilities are positive. Terms with $\pi^A_{ij} = 0$ cannot simply be omitted while retaining an unbiased-variance claim. For SRSWOR this is

$$\widehat{\mathcal{V}}_A(u) = \left(1 - \frac{n_A}{N}\right)\frac{s^2_{u,A}}{n_A}, \qquad s^2_{u,A} = \frac{1}{n_A - 1}\sum_{i\in A}(u_i - \bar{u}_A)^2.$$

For independent Poisson sampling it is $N^{-2}\sum_{i\in A}(1 - \pi^A_i)u_i^2/(\pi^A_i)^2$. Other designs require their actual pairwise probabilities or a consistent design-specific variance procedure. A systematic sample with a single random start is not automatically covered. The ordering randomization, pairwise-probability calculation, or replication scheme for such a design must be specified and satisfy (R10); the SRS or Poisson formula is not interchangeable with it. For a sample-$B$ contribution $v_i$ under conditionally independent Bernoulli selection, define

$$\widehat{\mathcal{V}}_B(v) = \frac{1}{N^2}\sum_{i\in B}\{1 - \hat{\pi}^B_i\}v_i^2.$$

For the variance of $\hat{\gamma}$, set

$$\hat{u}_{\gamma,A,i} = -\hat{\mathcal{J}}_\gamma^{-1}\hat{A}_{\gamma,i}, \qquad \hat{u}_{\gamma,B,i} = -\hat{\mathcal{J}}_\gamma^{-1}\hat{B}_{\gamma,i}.$$

The design-aware linearization variance is

$$\widehat{\mathrm{Var}}(\hat{\gamma}) = \widehat{\mathcal{V}}_A(\hat{u}_{\gamma,A}) + \widehat{\mathcal{V}}_B(\hat{u}_{\gamma,B}).$$

Under the i.i.d.-Poisson specialization this agrees asymptotically with the quantity $N^{-2}\sum_i(-\hat{\mathcal{J}}_\gamma^{-1}\hat{\Phi}_{\perp,i})^2$. In particular, $N\widehat{\mathrm{Var}}(\hat{\gamma}) \to_p \mathcal{J}^{-1}\Omega_\gamma\mathcal{J}^{-1}$.

For the plain IPW estimator, estimate

$$\hat{H} = \frac{1}{N}\sum_{i\in B}\hat{O}_i^B\{Y_i - \hat{\kappa}_g(X_{1i})\}^2.$$

For the i.i.d.-Poisson superpopulation specialization, the full plug-in influence contribution is

$$\widehat{\mathrm{IF}}_{\theta,i}^{\mathrm{IPW}} = I_i^A d_i^A \hat{\kappa}_g(X_{1i}) + \frac{\delta_i}{\hat{\pi}_i^B}\{Y_i - \hat{\kappa}_g(X_{1i})\} - \hat{\theta}_{\mathrm{IPW}} - \hat{H}\hat{\mathcal{J}}_\gamma^{-1}\hat{\Phi}_{\perp,i}.$$

This expression contains the $g$-profiling term omitted by the unaugmented formula. For design-separated finite-population variance estimation use

$$\hat{u}_{A,i}^{\mathrm{IPW}} = \hat{\kappa}_g(X_{1i}) - \hat{H}\hat{\mathcal{J}}_\gamma^{-1}\hat{A}_{\gamma,i},$$

$$\hat{u}_{B,i}^{\mathrm{IPW}} = \frac{Y_i - \hat{\kappa}_g(X_{1i})}{\hat{\pi}_i^B} - \hat{H}\hat{\mathcal{J}}_\gamma^{-1}\hat{B}_{\gamma,i},$$

and set

$$\widehat{V}_{\mathrm{IPW}} = \widehat{\mathcal{V}}_A(\hat{u}_A^{\mathrm{IPW}}) + \widehat{\mathcal{V}}_B(\hat{u}_B^{\mathrm{IPW}}).$$

This estimates the variance of the estimator, not the $O(1)$ variance of its root-$N$ limit: $N\widehat{V}_{\mathrm{IPW}} \to_p V_{\mathrm{IPW}}$. A Wald interval for $\theta_N$ additionally requires negligible bias or a bias correction satisfying (R10).

For the plain MI estimator, let

$$\hat{k}_i = \{\hat{\mu}_{1i} - \hat{\mu}_{0i}\}\hat{\pi}_m^B(X_i)\{1 - \hat{\pi}_m^B(X_i)\}, \qquad \hat{c}_{g,i} = \hat{a}_{g,i},$$

and, for $i \in B$, define

$$\hat{\lambda}_{f,i} = \frac{Y_i - \hat{\mu}_i}{\hat{\pi}_i^B}.$$

This uses the identity $\Lambda_f = \delta(Y - \mu)/\pi^B$ proved in Lemma S4. It is a consistent plug-in representation; separately estimated marginal propensities need not satisfy the same algebraic identities at finite sample size. An analytic plug-in estimator of $H_\gamma$ is

$$\hat{H}_\gamma = \frac{1}{\hat{N}^A} \sum_{i=1}^{N} I_i^A d_i^A \left[ \hat{k}_i \{ \nabla_\gamma \hat{g}_i - \nabla_\gamma \hat{m}_i \} + \{ 1 - \hat{\pi}_m^B(X_i) \} \widehat{V}_0(X_i; \hat{\gamma}) \right].$$

For the i.i.d.-Poisson superpopulation specialization, the full plug-in influence contribution is

$$\widehat{\mathrm{IF}}_{\theta,i}^{\mathrm{MI}} = I_i^A d_i^A \hat{\mu}_i + \delta_i \hat{\lambda}_{f,i} + \hat{c}_{g,i} \left( I_i^A d_i^A - \frac{\delta_i}{\hat{\pi}_i^B} \right) - \hat{\theta}_{\mathrm{MI}}$$
$$- \hat{H}_\gamma \hat{\mathcal{J}}_\gamma^{-1} \hat{\Phi}_{\perp,i}.$$

For a design-separated variance, use

$$\hat{u}_{A,i}^{\mathrm{MI}} = \hat{\mu}_i + \hat{c}_{g,i} - \hat{H}_\gamma \hat{\mathcal{J}}_\gamma^{-1} \hat{A}_{\gamma,i},$$

$$\hat{u}_{B,i}^{\mathrm{MI}} = \hat{\lambda}_{f,i} - \frac{\hat{c}_{g,i}}{\hat{\pi}_i^B} - \hat{H}_\gamma \hat{\mathcal{J}}_\gamma^{-1} \hat{B}_{\gamma,i},$$

and set

$$\widehat{V}_{\mathrm{MI}} = \widehat{\mathcal{V}}_A(\hat{u}_A^{\mathrm{MI}}) + \widehat{\mathcal{V}}_B(\hat{u}_B^{\mathrm{MI}}).$$

Under (R10), $N\widehat{V}_{\mathrm{MI}} \to_p V_{\mathrm{MI}}$. In the nonparametric outcome-law case, one can alternatively use $\hat{H}_\gamma = \hat{G}_{\mathrm{MI},\gamma}$ below; the two MI variance formulas then use the same estimated linearized variables. With different consistent derivative estimates they need only be asymptotically equivalent.

Under a parametric respondent model, let $\hat{u}_{\xi,i} = -\hat{A}_\xi^{-1} s_\xi(Y_i \mid X_i; \hat{\xi})$ for $i \in B$. In the sample-$B$ linearized variable, replace $\hat{\lambda}_{f,i}$ by $\hat{\Gamma}_\xi^\top \hat{u}_{\xi,i}$; the factor $\delta_i$ is supplied by the sample-$B$ total itself. Unlike the unrestricted outcome-law linearization, this parametric projection does not in general satisfy the pointwise identity $\mu(X) - Y + \pi^B(X_1, Y)\lambda_f(W) = 0$. Under the combined superpopulation–sampling law, define

$$\hat{r}_{\xi,i} = \hat{\mu}_i - Y_i + \hat{\pi}_i^B \hat{\Gamma}_\xi^\top \hat{u}_{\xi,i}, \qquad i \in B,$$

$$\bar{\hat{r}}_\xi = \frac{1}{N} \sum_{i \in B} \frac{\hat{r}_{\xi,i}}{\hat{\pi}_i^B}, \qquad \widehat{\mathcal{V}}_{\mathrm{mod}} = \frac{1}{N^2} \sum_{i \in B} \frac{(\hat{r}_{\xi,i} - \bar{\hat{r}}_\xi)^2}{\hat{\pi}_i^B}.$$

The parametric-$f_1$ variance estimator is then $\widehat{\mathcal{V}}_A(\hat{u}_A^{\mathrm{MI}}) + \widehat{\mathcal{V}}_B(\hat{u}_B^{\mathrm{MI}}) + \widehat{\mathcal{V}}_{\mathrm{mod}}$. The last term estimates the superpopulation variation of the conditional centering component. If inference

is instead conditional on a fixed finite population, $N^{-1}\sum_i r_{\xi,i}$ is a local model-centering term and must be retained; the two sample-design variances alone are not sufficient. As before, none of these variance terms removes the deterministic smoothing bias $\mathcal{B}_{\mathrm{MI},N}$.

For the orthogonalized MI estimator, estimate

$$\hat{G}_{\mathrm{MI},\gamma} = \frac{1}{N}\sum_{i\in B}\hat{O}_i^B Y_i\{Y_i - \hat{\mu}_i - \hat{a}_{g,i}\}.$$

For the i.i.d.-Poisson superpopulation specialization, the full plug-in influence contribution is

$$\begin{aligned}\widehat{\mathrm{IF}}_{\theta,i}^{\mathrm{MI},\perp} &= I_i^A d_i^A(\hat{\mu}_i + \hat{a}_{g,i}) + \frac{\delta_i}{\hat{\pi}_i^B}(Y_i - \hat{\mu}_i - \hat{a}_{g,i}) \\ &\quad - \hat{\theta}_{\mathrm{MI},\perp} - \hat{G}_{\mathrm{MI},\gamma}\hat{\mathcal{J}}_\gamma^{-1}\hat{\Phi}_{\perp,i}.\end{aligned}$$

For a general reference design, use

$$\hat{u}_{A,i}^{\mathrm{MI},\perp} = \hat{\mu}_i + \hat{a}_{g,i} - \hat{G}_{\mathrm{MI},\gamma}\hat{\mathcal{J}}_\gamma^{-1}\hat{A}_{\gamma,i},$$

$$\hat{u}_{B,i}^{\mathrm{MI},\perp} = \frac{Y_i - \hat{\mu}_i - \hat{a}_{g,i}}{\hat{\pi}_i^B} - \hat{G}_{\mathrm{MI},\gamma}\hat{\mathcal{J}}_\gamma^{-1}\hat{B}_{\gamma,i},$$

and

$$\widehat{V}_{\mathrm{MI},\perp}^F = \widehat{\mathcal{V}}_A(\hat{u}_A^{\mathrm{MI},\perp}) + \widehat{\mathcal{V}}_B(\hat{u}_B^{\mathrm{MI},\perp}).$$

Under (R10), $N\widehat{V}_{\mathrm{MI},\perp}^F \to_p V_{\mathrm{MI},\perp}^F$. The standard error used in a Wald interval is $\sqrt{\widehat{V}_{\mathrm{MI},\perp}^F}$, not that quantity divided by another $\sqrt{N}$. The variance formula itself needs no record linkage, but the validity of the fitted nuisance construction still requires the overlap and conditional-evaluation conditions in (R6). For bias-corrected plain estimates, an error of order $N^{-1/2}$ in the bias estimate must be propagated jointly, rather than ignored.

*Remark* S1 (The $\gamma$-correction under the MAR submodel). Even when $\gamma_0 = 0$, the parameter is estimated under the unrestricted model and its root-$N$ variation remains in the mean influence function. The $\gamma$ correction may be omitted only if the MAR submodel is imposed a priori and $\gamma$ is fixed at zero rather than estimated. We retain the correction in every replicate reported here.

## S2 Implementation details

### S2.1 Bandwidth rates

For bandwidth selection, let $s > 0$ denote the Hölder smoothness of a nuisance regression on $d$ continuous coordinates, and use a local-polynomial or higher-order kernel smoother that exploits this smoothness. Under the corresponding density, moment, and boundary conditions, its bias is $O(h^s)$ and its stochastic $L^2$ error is $O_p\{(nh^d)^{-1/2}\}$, where $n$ is the relevant training-sample size (Fan and Gijbels, 1996). Balancing these terms gives

$$h \asymp n^{-1/(2s+d)}, \qquad h^s + (nh^d)^{-1/2} \asymp n^{-s/(2s+d)},$$

the usual nonparametric regression rate (Stone, 1982). Consequently, $2s > d$ is a sufficient smoothness–dimension condition for the common $o_p(n^{-1/4})$ nuisance rate used in (R4), provided that this rate also holds in the required weighted norm. The condition concerns each nuisance regression separately: $g_\gamma$ is conditioned on $X_1$, whereas $m$ and the outcome regressions are conditioned on $X = (X_1, X_2)$. Under (R1) the training-sample sizes are of order $N$. These bandwidths do not generally make the plain estimators' first-order biases $o(N^{-1/2})$; those biases are retained in Theorems S1–S2.

The numerical studies use the second-order choice $s = 2$: Gaussian local-constant smoothers with bandwidths of order $\hat{\sigma} n^{-1/(d+4)}$, where $\hat{\sigma}$ is the coordinate-wise scale estimate. When $X_1$ is categorical, $g_\gamma$ is estimated by exact cell matching; continuous coordinates are smoothed with a product kernel and discrete coordinates are matched exactly. Here $s = 2$ describes the smoothness exploited by the numerical smoothing rule, not an upper bound on the smoothness of the data-generating functions. This specialization does not certify (R4) in every reported setting: boundary contributions, low-propensity tails, and the dimension of each fitted regression still require verification. Boundary-adaptive smoothing or a direct integrated boundary bound is needed on a compact nonperiodic support. The general smoothness conditions and their scope are detailed in Section S5.

### S2.2 Constants used in the numerical studies

All kernel estimators use a Gaussian kernel. The treatment of $X_1$ differs according to whether it is categorical or continuous, and this distinction governs which competitors are

available in each study. The two studies were implemented separately, and the settings below differ accordingly.

For the continuous-$X_1$ synthetic designs of Section 5.1, a product kernel is used throughout, with $h_1 = 1.06\,\hat{\sigma}(X_1)\,n^{-1/5}$ for the sums that depend on $X_1$ alone, and $h_1$ together with $h_2 = 1.06\,\hat{\sigma}(X_2)\,n^{-1/6}$ for those that depend on $X = (X_1, X_2)$. These are the mean-squared-error-optimal rates for one and two smoothing dimensions respectively, giving smoothing bias $O(h^2) = O(n^{-2/5})$ and $O(n^{-1/3})$. These bias orders are $o(n^{-1/4})$, but verification of (R4) additionally requires the stochastic and weighted bounds for every fitted component; neither bias order is $o(N^{-1/2})$, which is the stronger requirement that (14) but not (18) imposes.

When $X_1$ is categorical, as in the sensitivity designs of Section 5.3, the profile component $g_\gamma(\cdot)$ is estimated by exact cell matching: the sums defining $\hat{g}_\gamma$ run over the units sharing the value of $X_1$, with no smoothing in that coordinate. Smoothing is applied only to $X_2$, with bandwidth $h = n_k^{-1/5}\hat{\sigma}(X_2 \mid X_1)$, where $n_k$ is the number of units in the cell and $\hat{\sigma}$ is the design-weighted within-cell standard deviation computed from sample $A$. When $X_2$ has several coordinates a product kernel is used and the exponent becomes $-1/(d_{2c} + 4)$, where $d_{2c}$ counts the continuous coordinates; discrete coordinates of $X_2$ are matched exactly rather than smoothed.

Exact cell matching has one consequence for the choice of competitors. When $X_1$ is categorical with $C$ levels, a parametric selection model with $C - 1$ dummy variables represents $g$ exactly, so nothing is gained by estimating it nonparametrically. That is why $\hat{\theta}_{\mathrm{PMLd}}$ is correctly specified in the sensitivity designs of Section 5.3 and has no counterpart in the designs of Section 5.1.

Two settings differ between the two studies. In the designs of Section 5.1 the reference sample has $n_A = 2{,}000$ from $N = 20{,}000$ and the bandwidths above are used at their reference values. In Section 6.1 the reference sample has $n_A = 1{,}000$ from $N = 50{,}000$, and bandwidths are indexed by the achieved sample size $n_A + n_B$ rather than by $N$. The multiplier used there is given below with the other settings. Section S7.5 reports the sensitivity to that multiplier and the diagnostics behind the choice. In both studies the reference sample is drawn by simple random sampling without replacement, which is the design for which Section S1.3 gives an explicit variance formula.

Cross-fitting uses $K = 5$ folds, assigned by simple random sampling stratified on $\delta_i$, with the two records of a unit appearing in both samples forced into a common fold, as Condition (R6) requires. The number of such units is recorded in every replicate; in the data application it averages about 50, against $N = 50{,}000$ and $n_A = 1{,}000$. The tilting parameter $\hat{\gamma}$ solves a single pooled orthogonal score equation rather than fold-wise equations, since the linearization variance in Section S1.3 is derived for the pooled solution; solving fold-wise and averaging leaves a non-zero residual in the pooled score, so the two are not interchangeable for inference.

The score is evaluated on a grid over $\Gamma$, and sign-change brackets are refined by Brent's method. Two features of this procedure matter in practice. First, bracketing on the endpoints of $\Gamma$ alone is not sufficient: the orthogonal score need not be monotone over a wide interval, and a global minimization of the squared score in that situation returns a boundary point rather than a root. Second, when more than one root is found a rule fixed in advance selects one; the two studies use different rules, stated below with their other settings. Boundary solutions are counted and reported in Sections 5.2 and 5.3 for the simulation designs, and in Sections 6.2 and 6.3 for the study based on screening data.

The interval $\Gamma$ is set from the scale of the outcome rather than chosen arbitrarily. Under a Gaussian respondent law, the exponential tilt displaces the non-respondent conditional mean by $\gamma\sigma^2$. We use this relation more generally as a scale-based rule, taking $|\gamma| \leq 1/\hat{\sigma}_1$ to represent a one-standard-deviation shift. In the designs of Section 5 the respondent conditional standard deviation is $\hat{\sigma}_1 = 1$, giving $\Gamma = [-1, 1]$, with grid spacing 0.2 in the categorical designs and 0.4 in the continuous ones. When several interior roots occur, the root with the largest absolute score derivative is taken, ties broken by the smallest root. In the study of Section 6 the outcome is $\log(\mathrm{TG})$ and the data-generating value $\gamma = 1.047$ was set so that $\gamma\,\mathrm{sd}(y) = 0.6$, matching the signal strength of the Section 5 designs; the same rule then gives an upper limit near 1.7, and we use $\Gamma = [-0.5, 3.0]$ with spacing 0.1, which contains that limit with margin. The interval is not symmetric because the selection mechanism in (24) is specified with a positive tilt, so the informative region lies on one side of zero. Brackets are refined by Brent's method to a tolerance of $10^{-7}$ on the scale of $\gamma$, six to seven orders of magnitude below the Monte Carlo dispersion of $\hat{\gamma}$ in every design. When several interior roots occur, the bracket whose grid endpoint carries the score closest

to zero is refined and its root taken.

For the NHISS study, fitted conditional propensities are floored at $5 \times 10^{-3}$ and the grid spacing is 0.1. The reported bandwidth rule for this study is $3\hat{\sigma}_X(n_A + n_B)^{-1/(d+4)}$, with $d$ the smoothing dimension. SEX is matched exactly and the rule is applied to the remaining coordinates, so $d = 3$ under $\mathrm{M}_1$ and $d = 4$ under $\mathrm{M}_2$; with $s = 2$ the sufficient condition $2s > d$ therefore holds under $\mathrm{M}_1$ and fails under $\mathrm{M}_2$ by one dimension. The floor is applied to the conditional propensity $\hat{\pi}^B_m(x)$, to the marginal propensity $\hat{\varpi}^B(x_1)$, to the profiled propensity inside the orthogonal score, and to $\hat{\pi}^B(x_1, y)$ itself, which carries the inverse-probability weight in (13) and the residual term in (18); the reported count of floored records is the number of $B$-records at which the last of these binds. The denominators of the nuisance ratios are separately guarded at machine epsilon, which is a numerical safeguard rather than a floor. That guard is distinct from the truncation of the count in (S2), whose occurrences are reported in Section S7.5. A fixed active floor is not guaranteed to have negligible effect under (R3). The bandwidth and boundary diagnostics therefore describe finite-sample behavior, without verifying all the theoretical conditions.

In the synthetic designs the finite population is generated as a mixture. For each unit, $m_1(X)$ is computed, $p(X)$ is formed from $g(X_1)$ and $\gamma m_1(X)$, and $\delta \sim \mathrm{Bernoulli}\{p(X)\}$ is drawn; $Y$ is drawn afterwards, from $\mathcal{N}\{m_1(X), \sigma^2\}$ if $\delta = 1$ and from $\mathcal{N}\{m_1(X) + \gamma\sigma^2, \sigma^2\}$ if $\delta = 0$, so that the respondent law is exactly $\mathcal{N}\{m_1(X), \sigma^2\}$. This covers MNAR-I to MNAR-III and the MAR, GMAR and INV-IV designs. NONLIN-Y is generated in the opposite order, with $Y$ drawn for every unit before selection, so that the respondent law carries the quadratic tilt.

## S2.3 Likelihood constructions

The parametric benchmark of Liu et al. (2026) specifies

$$\pi(x, y; \phi) = \mathrm{expit}(x^\top \phi_1 + y\phi_2). \tag{S5}$$

The exclusion restriction is imposed by omitting the instrument from the selection covariates. The marginal propensity is obtained from the respondent law $f_1(y \mid x)$:

$$\pi(x; \phi) = \left[\int \{\pi(x, y; \phi)\}^{-1} f_1(y \mid x)\, d\mu(y)\right]^{-1}.$$

Write $\hat{\pi}(x;\phi)$ for this expression with an estimated respondent law. Replacing the census covariate total in the selection likelihood by its design-weighted estimate from $A$ gives

$$\ell_p(\phi) = \frac{1}{N}\sum_{i\in B}\log\frac{\hat{\pi}(x_i;\phi)}{1-\hat{\pi}(x_i;\phi)} + \frac{1}{N}\sum_{i\in A} d_i^A \log\{1-\hat{\pi}(x_i;\phi)\}. \tag{S6}$$

This construction, introduced under MAR by Chen et al. (2020), requires no record linkage. Its consistency requires the relevant selection and respondent-law assumptions and regularity conditions.

The semiparametric profile quantities in the main paper also define

$$\ell_p(\gamma) = \frac{1}{N}\sum_{i\in B}\log\frac{\hat{\pi}_m^B(x_i;\gamma)}{1-\hat{\pi}_m^B(x_i;\gamma)} + \frac{1}{N}\sum_{i\in A} d_i^A \log\{1-\hat{\pi}_m^B(x_i;\gamma)\}, \tag{S7}$$

and the pseudo-profile likelihood estimator

$$\widetilde{\gamma}_{\mathrm{PL}} = \arg\max_{\gamma\in\Gamma} \ell_p(\gamma). \tag{S8}$$

For each candidate, estimate $g_\gamma$ and $m(\cdot;\gamma)$ and form the fitted marginal propensity. This likelihood construction motivates profiling, but the final estimator analyzed and denoted OP in the main paper solves the orthogonal score equation (9). The two estimators are distinct.

## S2.4 Additional competing estimators

In the categorical-$X_1$ study, PMLd uses $C-1$ dummy variables and therefore represents the covariate component exactly. This isolates the effect of saturating $g$ within the parametric working model.

For the oracle comparator based on Shetty et al. (2025), we adapt the missing-data estimating equation to the two-sample setting as

$$S_{\text{eff}}(\gamma) = \sum_{i\in A} d_i^A c^*(x_i) - \sum_{i=1}^{N} \delta_i\{1+\exp(-g^*(x_{1i})+\gamma y_i)\}c^*(x_i), \tag{S9}$$

where $c^*(x)$ is as in their Theorem 1. We supply $g^*(x_1) = g(x_1)$ and denote the resulting estimator by SMZ$^*$. The efficiency result in that paper concerns its own missing-data experiment; we do not establish attainment of the efficiency bound for this two-sample adaptation. We evaluate this implementation only in the categorical-$X_1$ study. The underlying method is not restricted to categorical covariates.

Our implementation uses Gaussian working moments for the required conditional expectations. Under NONLIN-Y, the respondent density has the form

$$f_1(y \mid x) \propto \phi(y; m_x, \sigma^2)\, \operatorname{expit}(\phi_0 + \phi_1 x_1 - \gamma y^2),$$

which is not Gaussian. The implemented PML and SMZ* calculations therefore incur Gaussian-moment misspecification in addition to the incorrect linear tilt. The kernel variant of Shetty et al. (2025) avoids the Gaussian-moment assumption, but its performance under this quadratic tilt is not evaluated here.

Under MAR, SMZ* has RMSE 0.1068, compared with 0.0816 for OP.mi⊥. Supplying a true nuisance function does not ensure the smallest finite-sample variance; fitted weights can induce calibration effects (Henmi and Eguchi, 2004). This comparison is not a claim of semiparametric efficiency of the proposed method. Under NONLIN-Y, the implemented SMZ* has bias 0.1724 and mean $\hat{\gamma} = -0.05$. These results describe the implemented working model, not every variant of that method.

# S3 Regularity conditions

The following conditions distinguish the nuisance approximation needed for the orthogonal estimators from the integrated linearizations additionally needed for the plain estimators. All rates are under the joint superpopulation and sampling law. Let $Pf = E_0(f)$ and $\mathbb{P}_N f = N^{-1} \sum_{i=1}^{N} f(W_i)$. Expectations applied to a fitted function integrate a new argument against $P_0$, holding the fitted function fixed. For notational simplicity, take $N$ divisible by the fixed number $K$ of folds. Write

$$\mathbb{P}_k f = \frac{K}{N} \sum_{i \in U^{(k)}} f(W_i), \qquad \mathbb{P}_{-k} f = \frac{K}{N(K-1)} \sum_{i \notin U^{(k)}} f(W_i).$$

Only sampled records contribute to the estimating equations; this population indexing specifies their normalization. Equal fold sizes up to a bounded rounding error give the same representations. Other allocations require the corresponding evaluation and training weights, not a substitution of the pooled average for each training average.

(R1) *Sampling and asymptotic regime.* The population values $Z_i = (X_i, Y_i)$ are i.i.d. from $P_0$ and $N \to \infty$. Conditional on these values, $\delta_i$ are independent Bernoulli variables

with probabilities $\pi_i^B$, and the reference design is independent of the selection of $B$. Its known inclusion probabilities satisfy $0 < c_A \le \pi_i^A \le 1$, and $n_A/N \to_p \rho_A \in (0,1)$ and $E(n_B)/N \to \rho_B \in (0,1)$. Thus $I_i^A d_i^A$ is uniformly bounded and $E_A(I_i^A d_i^A \mid Z_1, \ldots, Z_N, \delta_1, \ldots, \delta_N) = 1$. The fold-conditional requirements in (R6), the oracle CLT in (R9), and variance consistency in (R10) are separate requirements on a dependent reference design. None follows from unbiasedness of the design weights alone.

(R2) *Identification and selection of the root.* The selection model (1), exclusion restriction, conditional completeness, and associated nondegeneracy and square-integrability conditions in Section 2.2 hold, and $\gamma_0$ is interior to a compact $\Gamma$. These conditions identify the population model by Lemma S5. For the selected scalar estimating equation, additionally assume that the continuous moment

$$M(\gamma) = P\Phi_\perp\{\gamma, \eta_0(\gamma)\}$$

has a separated zero: $M(\gamma_0) = 0$ and $\inf_{\gamma\in\Gamma:\,|\gamma-\gamma_0|\ge\epsilon} |M(\gamma)| > 0$ for every $\epsilon > 0$ for which the set is nonempty. The cross-fitted moment $\hat{M}_N(\gamma) = K^{-1}\sum_k \mathbb{P}_k\Phi_\perp\{\gamma, \hat{\eta}_\gamma^{[-k]}\}$ satisfies $\sup_{\gamma\in\Gamma} |\hat{M}_N(\gamma) - M(\gamma)| = o_p(1)$. The data-based rule in Section 3.4 returns an interior approximate root with probability tending to one and satisfies $\hat{M}_N(\hat{\gamma}) = o_p(N^{-1/2})$. Together with the separated zero, the global uniform convergence gives consistency of this selected root. These estimating-equation requirements are additional to completeness; completeness identifies the population law but need not exclude other stationary points of a scalar criterion.

(R3) *Overlap, tails, and derivative envelopes.* The probabilities $p(X) = P(\delta = 1 \mid X)$ and $q(X_1) = P(\delta = 0 \mid X_1)$ and the conditional exponential moments used as denominators are strictly positive and finite almost surely. No common positive lower bound on $p$, $q$, or $\pi^B(X_1, Y)$ is imposed.

The following specifies precisely the weighted norms used below. For pointwise differentiation, regard the already defined nuisance functions as the evaluation coordinates

$$\eta_\gamma(X) = \big(g_\gamma(X_1), m_\gamma(X), c_\gamma(X), \xi_\gamma(X_1), q(X_1), p_0(X),$$
$$\mu_\gamma(X), a_{g,\gamma}(X_1), \mu_1(X), \mu_{0,\gamma}(X)\big),$$

where $p_0 = \pi^B_{m,0}$. Coordinates not used by a functional are omitted. The functions of primitive moments in Section S1.1 induce these coordinates; they do not require additional fits. Consider the sample contributions $\Phi_\perp$, $\theta_0 - \Psi_{\mathrm{MI},\perp}$, $\delta Y/\pi^B$, and $I^A d^A\{p_\gamma \mu_1 + (1-p_\gamma)\mu_{0,\gamma}\}$, as needed for the result in question. Each has the form $I^A d^A F_A(X;\gamma,\eta) + \delta F_B(X,Y;\gamma,\eta)$. On a deterministic admissible class containing the population profile and, with probability tending to one, every line segment from that profile to a fitted nuisance, suppose the first and second ordinary derivatives of $F_A, F_B$ in the evaluation coordinates have envelopes $L_1(X,Y)$ and $L_2(X,Y)$. The envelopes are uniform on a fixed neighborhood $\Gamma_0$ of $\gamma_0$. For some $\epsilon > 0$,

$$E\{L_1^{2+\epsilon} + L_2^{1+\epsilon} + |Y|^{2+\epsilon}\} < \infty,$$

and the true sample-specific contributions in (R9) have finite $(2+\epsilon)$th moments. Define

$$w(X) = 1 + E\{L_1(X,Y)^2 + L_2(X,Y) \mid X\},$$
$$\|\eta - \eta_0\|^2_{\mathcal{H}} = E\left[w(X)\sum_\nu |\eta_\nu(X) - \eta_{0,\nu}(X)|^2\right].$$

The envelopes include all inverse-denominator and exponential factors in the displayed score derivatives. These weighted moments allow small propensities without imposing a common lower bound. Estimated denominators must be positive on the evaluation domain. A growing-domain or stabilized implementation is allowed only if the omitted or altered aggregate contribution is $o_p(N^{-1/2})$, and the same negligibility holds for the variance at scale $N^{-1}$. A fixed active propensity floor has no such guarantee.

(R4) *Nuisance rates at the true tilting parameter.* For the coordinates used by the two orthogonal scores, including $c$, $\xi$, $\mu$, and $a_g$, assume

$$\max_{1\le k\le K} \|\hat\eta^{[-k]}_{\gamma_0} - \eta_0\|_{\mathcal{H}} = O_p(r_N), \qquad r_N \downarrow 0, \quad \sqrt{N}\, r_N^2 \to 0,$$

where $r_N$ is deterministic. In particular, each derived ratio and derivative coefficient must satisfy its stated weighted rate; consistency of unweighted primitive moments alone is not enough when denominators can be small. For Theorems S1–S2, the same bound is required for the coordinates entering the relevant plain functional. The rates in this condition are evaluated at $\gamma_0$. Substitution of the estimated $\hat\gamma$ is handled by (R8), not by assuming orthogonality throughout a neighborhood of $\gamma_0$.

(R5) *Population remainder bounds.* Differentiation under $P$ is justified by (R3). For either orthogonal score $F$ at its true finite-dimensional parameter, its evaluation-coordinate representation and the induced primitive-moment paths are compatible. The envelopes in (R3) then give

$$\|F(\gamma_0, \eta) - F(\gamma_0, \eta_0)\|_2 \le C\|\eta - \eta_0\|_{\mathcal{H}},$$
$$|P\{F(\gamma_0, \eta) - F(\gamma_0, \eta_0)\} - D_\eta PF(\gamma_0, \eta_0)[\eta - \eta_0]| \le C\|\eta - \eta_0\|^2_{\mathcal{H}}.$$

The linear term is zero by Proposition 1. The same Taylor remainder bound, without deleting its linear term, is used for the plain functionals. These are population bounds only; the empirical term is controlled separately by (R6). No uniform $O_p(r_N^2)$ population-drift assertion is made for $\gamma \neq \gamma_0$. If such a neighborhood bound is used, the first derivative generally contributes an additional $O_p(|\gamma - \gamma_0| r_N)$ term.

(R6) *Outer folds and conditional evaluation bounds.* For each fixed candidate $\gamma$, all fits used on $U^{(k)}$, including pseudo-outcomes entering $\hat{a}_g^{[-k]}$, are measurable with respect to the outer training information $\mathcal{T}_{-k}$. Inner fits, bandwidth selection, and tuning use only that information. For linked overlapping records, both appearances of a population unit receive the same fold. Independent unit-level fold labels give the sufficient cases proved in Lemma S2. Stratification on observed sample membership, PSU splitting, and unlinked overlap require verification of the following bounds for their actual allocation.

For each functional $F$ used in the result, put $\Delta F_k = F(\gamma_0, \hat{\eta}_{\gamma_0}^{[-k]}) - F(\gamma_0, \eta_0)$ and $L_{k,N} = (\mathbb{P}_k - P)\Delta F_k$. Require

$$|E(L_{k,N} \mid \mathcal{T}_{-k})| = O_p(r_N/\sqrt{N}), \qquad \mathrm{Var}(L_{k,N} \mid \mathcal{T}_{-k}) = O_p(r_N^2/N).$$

For an unlinked implementation, it is sufficient that its pooled moments, plain-estimator linearizations, and relevant derivatives differ from a linked construction satisfying these bounds by $o_p(N^{-1/2})$, $o_p(N^{-1/2})$, and $o_p(1)$, respectively. Merely assigning different records to different folds does not establish this condition. A joint CLT for oracle variables in (R9) is not a substitute for these conditional bounds for fitted random functions.

(R7) *Bias centers and foldwise integrated linearizations.* This condition is additional and is used only for the plain estimators. The coefficients and outcome-law contribution used below are defined in Section S4.1. For each fold and $\nu \in \{g, m, \mu_1, \mu_0\}$, define the population smoother $\nu_N^{[-k]}$ by replacing all training sums in the corresponding nonparametric fit by their population expectations at its deterministic reference bandwidth. Set

$$b_{\nu,N}^{[-k]} = \nu_N^{[-k]} - \nu, \qquad b_{\nu,N} = \frac{1}{K}\sum_{k=1}^{K} b_{\nu,N}^{[-k]}.$$

These are exact smoothing centers, not an unspecified first term of a bandwidth expansion. Lemma S3 gives their formulas for (S1)–(S2) and the two conditional means. Data-dependent bandwidths are allowed when their replacement by the reference bandwidths changes the relevant integrated centers and stochastic linearizations by $o_p(N^{-1/2})$.

For $a_1 = -q\kappa_g$ and $a_2 = E(k \mid X_1)$, require, uniformly over the fixed collection of folds,

$$\begin{aligned} &P[a_\ell(X_1)\{\hat{g}^{[-k]} - g - b_{g,N}^{[-k]}\}] \\ &\quad = \mathbb{P}_{-k}\left[\frac{a_\ell(X_1)}{q(X_1)}\{\delta e^{\gamma_0 Y - g(X_1)} - (I^A d^A - \delta)\}\right] + o_p(N^{-1/2}), \qquad \ell = 1, 2, \end{aligned}$$

and

$$\begin{aligned} P\big[p(\hat{\mu}_1^{[-k]} - \mu_1 - b_{1,N}^{[-k]}) + (1-p)(\hat{\mu}_0^{[-k]} - \mu_0 - b_{0,N}^{[-k]}) \\ - k(\hat{m}^{[-k]} - m - b_{m,N}^{[-k]})\big] = \mathbb{P}_{-k}\Lambda_f + o_p(N^{-1/2}). \end{aligned}$$

All functions are evaluated at $\gamma_0$. The exact identity $K^{-1}\sum_k \mathbb{P}_{-k} f = \mathbb{P}_N f$ converts these training-fold expansions to the pooled representations. Their left-hand sides are not equal to pooled empirical averages separately for each fold.

The relevant coordinates, including their bias centers, satisfy the weighted small-error and quadratic-remainder bounds in (R3)–(R5). Lemma S3 supplies a sufficient smooth compact-support kernel case. Bandwidth orders alone, without smoothness, density, boundary, tail, and design conditions, are not a sufficient verification. For a correctly specified finite-dimensional $f_1(y \mid x; \xi)$, require

$$\hat{\xi}^{[-k]} - \xi_0 = \mathbb{P}_{-k}\varphi_\xi + o_p(N^{-1/2}),$$

with $\varphi_\xi$ as defined in Section S4.1; the second display is then replaced by $\Gamma_\xi^\top \mathbb{P}_{-k}\varphi_\xi + o_p(N^{-1/2})$ and its outcome-model smoothing centers can be set to zero. This specialization retains the empirical ratio fit of $g$ in (S2). Substituting model-based expectations into that fit as well would require a different joint linearization.

(R8) *Local differentiability of the fitted profiles.* On a fixed neighborhood $\Gamma_0$ of $\gamma_0$, the population profile and its moments are continuously differentiable, and $\mathcal{J} = M'(\gamma_0) \neq 0$ is finite. The fitted pooled moment is continuously differentiable there with probability tending to one and

$$\sup_{\gamma \in \Gamma_0} |\hat{M}'_N(\gamma) - M'(\gamma)| = o_p(1).$$

For each required mean estimator, viewed as a function $\hat{\theta}_j(\gamma)$ before substituting $\hat{\gamma}$, require similarly

$$\sup_{\gamma \in \Gamma_0} |\hat{\theta}'_j(\gamma) - T'_j(\gamma)| = o_p(1),$$

where $T_j$ is the population functional along the population profile. At $\gamma_0$ the derivatives are respectively $H$, $H_\gamma$, and $G_{\mathrm{MI},\gamma}$. This condition is only needed for the estimators covered by the given theorem. It avoids a circular assumption that $\hat{\gamma}$ already belongs to a shrinking root-$N$ neighborhood before its rate has been established.

(R9) *Oracle joint CLT and moment convergence.* The vector of oracle linearized variables required in the given theorem, together with $Y_i - \theta_0$ when passing between population targets, satisfies a joint CLT and has a convergent covariance matrix under the combined law. In the unrestricted outcome-law case, the finite-population contributions can be written as

$$(I_i^A d_i^A - 1)u_A(X_i) + (\delta_i - \pi_i^B)u_B(X_i, Y_i).$$

The conditional design covariance of their normalized sums converges in probability to a finite deterministic matrix, and the design Lindeberg condition holds. Its scalar version is

$$\frac{1}{N}\left\{\sum_{i,j=1}^{N} \frac{\pi_{ij}^A - \pi_i^A \pi_j^A}{\pi_i^A \pi_j^A} u_{A,i} u_{A,j} + \sum_{i=1}^{N} \pi_i^B (1 - \pi_i^B) u_{B,i}^2\right\} \longrightarrow_p V_j.$$

The variance relevant for an asserted nondegenerate normal limit is positive. For independent Poisson sampling with inclusion probabilities depending only on the

unit's covariates, (R3) and covariance convergence give the usual independent-array CLT. SRSWOR is covered by its finite-population CLT under the same moment and covariance conditions. An unequal-probability or systematic design must meet this requirement separately. For the parametric-$f_1$ specialization, include also the model-centering residual $r_{\xi,i}$ in Section S1.3; its superpopulation variation is part of the joint-law variance.

(R10) *Variance estimation and substitution.* The reference-sample variance estimator is valid for the actual design. The pairwise formula in Section S1.3 requires known positive $\pi_{ij}^A$ for the pairs it uses. Otherwise a design-specific consistent replication or alternative estimator must be supplied. In particular, a randomized systematic start does not alone verify variance consistency. For each required oracle coefficient pair, its variance estimate differs from its conditional variance by $o_p(N^{-1})$. The Bernoulli variance formula and, where applicable, the model-centering variance satisfy the corresponding law of large numbers. Substitution of cross-fitted coefficients changes these variance estimates by $o_p(N^{-1})$; this is a weighted quadratic condition and is not implied by unweighted prediction consistency. Consequently $N\widehat{V}_j \to_p V_j$ for the mean estimators. If a consistent bias estimate is used without changing this variance, additionally require $\widehat{\mathcal{B}}_{j,N} - \mathcal{B}_{j,N} = o_p(N^{-1/2})$; otherwise its first-order stochastic contribution and covariances must also be included.

The absence of a uniform bound on $p(X)$ in (R3) is deliberate. For the Gaussian respondent model of Section 5.2,

$$p(X) = \operatorname{expit}\{g(X_1) - \gamma_0\mu_1(X) - \gamma_0^2/2\}.$$

It can approach zero as the Gaussian instrument increases. Nevertheless, polynomial moments times exponentials of linear Gaussian predictors are finite; inverse-probability moments of fixed order therefore need not fail. Small-denominator and tail effects are controlled by the explicit envelopes and weighted rates above, rather than a common lower bound. These moment observations do not by themselves verify the fitted-smoother rates, membership-stratified fold bounds, or systematic-design variance conditions. The compact-support verification below is a sufficient case, not a claim that every Gaussian-kernel implementation satisfies all the conditions. Coverage reported for an implementation outside

these verified conditions is a finite-sample diagnostic rather than a consequence of the theorems.

# S4 Proofs of the main results

Throughout, expectations are taken under the superpopulation distribution of Condition (R1), and all nuisance functions are evaluated at the truth $\eta_0$ unless otherwise noted. Conditions (R1)–(R10) refer to Section S3.

## Proof of Proposition 1: tilt score

Write the decomposition

$$\Phi_\perp(W;\gamma,\eta) = \Phi_0(W;\gamma,\eta) + \Phi_g(W;\gamma,\eta) + \Phi_m(W;\gamma,\eta),$$

where

$$\begin{aligned}
\Phi_0(W;\gamma,\eta) &= \{\delta - I^A d^A \pi_m^B(X;\gamma)\}\, c(X;\gamma),\\
\Phi_g(W;\gamma,\eta) &= \frac{\xi(X_1;\gamma)}{1-\varpi_{g,0}^B(X_1)}\Big\{(I^A d^A - \delta) - \delta \exp\{\gamma Y - g(X_1;\gamma)\}\Big\},\\
\Phi_m(W;\gamma,\eta) &= \delta\, \frac{c(X;\gamma)\{\pi_m^B(X;\gamma)\}^2}{\exp\{g(X_1;\gamma)\}\, \pi_{m,0}^B(X)}\Big\{\exp(\gamma Y) - \exp\{m(X;\gamma)\}\Big\}.
\end{aligned}$$

At $(\gamma_0,\eta_0)$, we have $E(\delta \mid X) = \pi_{m,0}^B(X) = \pi_m^B(X;\gamma_0)$ and $E(I^A d^A \mid X) = 1$ by Condition (R1), so

$$E\{\Phi_0(W;\gamma_0,\eta_0)\} = E\big[\{E(\delta \mid X) - \pi_m^B(X;\gamma_0)\}c(X;\gamma_0)\big] = 0.$$

For $\Phi_g$, conditional on $X_1$,

$$\begin{aligned}
E\big\{(I^A d^A - \delta) \mid X_1\big\} &= 1 - \varpi_{g,0}^B(X_1),\\
E\big[\delta \exp\{\gamma_0 Y - g(X_1;\gamma_0)\} \mid X_1\big] &= \exp\{-g(X_1;\gamma_0)\}\, r_g^{(0)}(X_1;\gamma_0) = 1 - \varpi_{g,0}^B(X_1),
\end{aligned}$$

where the last equality uses $\exp\{g(X_1;\gamma_0)\} = r_g^{(0)}(X_1;\gamma_0)/\{1-\varpi_{g,0}^B(X_1)\}$ from Section 3.2. Hence $E\{\Phi_g(W;\gamma_0,\eta_0)\} = 0$. Similarly, $E\{\exp(\gamma_0 Y) - \exp\{m(X;\gamma_0)\} \mid X, \delta = 1\} = 0$ by the definition of $m$, so $E\{\Phi_m(W;\gamma_0,\eta_0)\} = 0$.

Let $t \mapsto \eta_t$ be a regular path through $\eta_0$, and denote pathwise derivatives by dots. The derivative of $E\{\Phi_0(W;\gamma_0,\eta_t)\}$ involves $\dot\pi^B_m(X) = \pi^B_m(X)\{1-\pi^B_m(X)\}\{\dot g(X_1) - \dot m(X)\}$, yielding

$$\frac{d}{dt}E\{\Phi_0\}\Big|_{t=0} = -E\big[c(X)\pi^B_m(X)\{1-\pi^B_m(X)\}\dot g(X_1)\big] + E\big[c(X)\pi^B_m(X)\{1-\pi^B_m(X)\}\dot m(X)\big]. \tag{S10}$$

For $\Phi_g$, by the definition of $\xi(X_1)$ in Section 3.2,

$$\frac{d}{dt}E\{\Phi_g\}\Big|_{t=0} = E\big[\xi(X_1)\dot g(X_1)\big] = E\big[c(X)\pi^B_m(X)\{1-\pi^B_m(X)\}\dot g(X_1)\big], \tag{S11}$$

which cancels the first term of (S10). For $\Phi_m$,

$$\frac{d}{dt}E\{\Phi_m\}\Big|_{t=0} = -E\big[c(X)\pi^B_m(X)\{1-\pi^B_m(X)\}\dot m(X)\big], \tag{S12}$$

which cancels the second term of (S10). More explicitly, differentiating the multiplier functions produces terms multiplied by $(I^A d^A - \delta) - \delta\exp\{\gamma_0 Y - g(X_1;\gamma_0)\}$ or by $\delta[\exp(\gamma_0 Y) - \exp\{m(X;\gamma_0)\}]$. The first residual has conditional mean zero given $X_1$, and the second has conditional mean zero given $(X, \delta = 1)$ after multiplication by $\delta$; hence all such terms vanish. Summing (S10)–(S12) gives $(d/dt)E\{\Phi_\perp\}|_{t=0} = 0$. □

## Proof of Proposition 1: mean score

At $(\theta_0, \eta_0)$,

$$E\left\{\frac{\delta}{\pi^B(X_1, Y;\gamma_0)}Y\right\} = E\big[E(Y \mid X)\big] = E(Y) = \theta_0$$

and

$$E\left[\left(I^A d^A - \frac{\delta}{\pi^B(X_1, Y;\gamma_0)}\right)\{\mu(X;\gamma_0) + a_g(X_1)\}\right] = 0$$

because $E\{\delta/\pi^B(X_1, Y;\gamma_0) \mid X\} = 1$ and $E(I^A d^A \mid X) = 1$. Hence $E\{\Psi_{\mathrm{MI},\perp}(W;\theta_0,\eta_0)\} = 0$.

Along a regular path $t \mapsto \eta_t$ with $\gamma$ held fixed at $\gamma_0$, the derivative of $E\{\Psi_{\mathrm{MI},\perp}(W;\theta_0,\eta_t)\}$ equals

$$E\left[\left\{\frac{\delta}{\pi^B(X_1, Y;\gamma_0)} - I^A d^A\right\}\{\dot\mu(X) + \dot a_g(X_1)\}\right] + E\big[\dot g(X_1)\, E\big\{\delta O^B(X_1, Y;\gamma_0)\{Y - \mu(X;\gamma_0) - a_g(X_1)\} \mid X_1\big\}\big]. \tag{S13}$$

The first expectation in (S13) is zero because $E\{\delta/\pi^B(X_1,Y;\gamma_0) - I^A d^A \mid X\} = 0$. The inner conditional expectation in the second term is zero by the definition of $a_g(X_1)$ in (16):

$$E\big\{\delta O^B(X_1,Y;\gamma_0)(Y-\mu(X;\gamma_0)) \;\big|\; X_1\big\} = a_g(X_1)\, E\big\{\delta O^B(X_1,Y;\gamma_0) \;\big|\; X_1\big\},$$

so the bracket cancels. Hence (S13) is zero, and the score is Neyman-orthogonal at $(\theta_0,\eta_0)$. This establishes orthogonality with respect to the infinite-dimensional nuisance component $\eta$; the finite-dimensional parameter $\gamma$ is handled through the joint expansion in the proof of Theorem 1 below. □

## Proof of Theorem 1: tilting parameter

First, (R2) and the data-based root-selection rule give

$$|M(\hat\gamma)| \le \sup_{\gamma\in\Gamma} |\hat M_N(\gamma) - M(\gamma)| + |\hat M_N(\hat\gamma)| = o_p(1).$$

The separated zero therefore implies $\hat\gamma \to_p \gamma_0$; proximity to the unknown true value is not part of the numerical definition.

For each outer fold, add and subtract the population expectation holding its fitted nuisance fixed:

$$\hat M_N(\gamma_0) - \mathbb{P}_N\Phi_\perp(\gamma_0) = \frac{1}{K}\sum_{k=1}^{K}\big[(\mathbb{P}_k - P)\{\Phi_\perp(\gamma_0,\hat\eta^{[-k]}_{\gamma_0}) - \Phi_\perp(\gamma_0,\eta_0)\} \\ + P\{\Phi_\perp(\gamma_0,\hat\eta^{[-k]}_{\gamma_0}) - \Phi_\perp(\gamma_0,\eta_0)\}\big].$$

Lemma S1 bounds this by $O_p(r_N/\sqrt{N} + r_N^2) = o_p(N^{-1/2})$. The first term uses the conditional evaluation bounds of (R6); an oracle CLT alone would not control it. The second term uses Proposition 1 only at $\gamma_0$.

On the consistency event, the entire interval between $\hat\gamma$ and $\gamma_0$ lies in $\Gamma_0$. The fundamental theorem of calculus gives

$$\hat M_N(\hat\gamma) = \hat M_N(\gamma_0) + \left[\int_0^1 \hat M_N'\{\gamma_0 + t(\hat\gamma-\gamma_0)\}\,dt\right](\hat\gamma-\gamma_0).$$

The bracket is $\mathcal{J} + o_p(1)$ by consistency and (R8), and $\mathcal{J} \ne 0$. Since $\hat M_N(\hat\gamma) = o_p(N^{-1/2})$ and $\mathbb{P}_N\Phi_\perp(\gamma_0) = O_p(N^{-1/2})$ by (R9), solving this equation first establishes the root-$N$ rate and then gives (20). No root-$N$ rate was assumed in the Taylor remainder.

Finally, the joint oracle CLT in (R9) gives $N^{-1/2}\sum_i \Phi_{\perp,i}(\gamma_0) \Rightarrow \mathcal{N}(0,\Omega_\gamma)$. Slutsky's theorem gives the asserted distribution. In the i.i.d.-Poisson specialization, $\Omega_\gamma = E\{\Phi_\perp(W;\gamma_0,\eta_0)^2\}$. □

## Closed form of the Jacobian $\hat{\mathcal{J}}_\gamma$

Differentiating the population moment along the profile and using Condition (R1) gives

$$\mathcal{J} = E\big[I^A d^A \pi_m^B(X;\gamma_0)\{1-\pi_m^B(X;\gamma_0)\}c(X;\gamma_0)\{\nabla_\gamma m(X;\gamma_0) - \nabla_\gamma g(X_1;\gamma_0)\}\big]\,.$$

Each augmentation term has expectation zero for every $\gamma$ along the population profile, so the derivative of its expectation is zero. This is separate from Neyman orthogonality, which concerns nuisance perturbations at a fixed $\gamma$. The plug-in estimator $\hat{\mathcal{J}}_\gamma$ in (S4) replaces the expectation by its design-weighted sample analogue and the truth by cross-fitted estimates.

## Proof of Theorem 1: population mean

Define

$$h_{\mathrm{MI}}(W;\gamma,\eta) = \frac{\delta}{\pi^B(X_1,Y;\gamma)}Y + \left\{I^A d^A - \frac{\delta}{\pi^B(X_1,Y;\gamma)}\right\}\{\mu(X;\gamma)+a_g(X_1)\},$$

so that $\Psi_{\mathrm{MI},\perp}(W;\theta,\gamma,\eta) = \theta - h_{\mathrm{MI}}(W;\gamma,\eta)$, and let

$$G_{\mathrm{MI},\gamma} = \frac{d}{d\gamma}E\{h_{\mathrm{MI}}(W;\gamma,\eta_0(\gamma))\}\bigg|_{\gamma=\gamma_0}.$$

The joint estimating equations are

$$\mathbb{P}_N\Phi_\perp(W;\hat\gamma,\hat\eta_{\hat\gamma}) = 0, \qquad \mathbb{P}_N\Psi_{\mathrm{MI},\perp}(W;\hat\theta_{\mathrm{MI},\perp},\hat\gamma,\hat\eta_{\hat\gamma}) = 0.$$

Their population Jacobian with respect to $(\gamma,\theta)$ is

$$\mathcal{A} = \begin{pmatrix} \mathcal{J} & 0 \\ -G_{\mathrm{MI},\gamma} & 1 \end{pmatrix}.$$

Lemma S1 gives the oracle replacement at $(\gamma_0,\theta_0)$. Theorem 1 first establishes $\hat\gamma - \gamma_0 = O_p(N^{-1/2})$; (R8) then gives $\hat\theta_{\mathrm{MI},\perp}(\hat\gamma) - \hat\theta_{\mathrm{MI},\perp}(\gamma_0) = G_{\mathrm{MI},\gamma}(\hat\gamma-\gamma_0) + o_p(N^{-1/2})$. Together these yield the joint superpopulation expansion

$$\sqrt{N}\begin{pmatrix} \hat\gamma - \gamma_0 \\ \hat\theta_{\mathrm{MI},\perp} - \theta_0 \end{pmatrix} = -\mathcal{A}^{-1}\frac{1}{\sqrt{N}}\sum_{i=1}^N \begin{pmatrix} \Phi_{\perp,i}(\gamma_0) \\ \Psi_{\mathrm{MI},\perp}(W_i;\theta_0,\gamma_0,\eta_0) \end{pmatrix} + o_p(1).$$

Subtracting $\theta_N - \theta_0 = N^{-1}\sum_i (Y_i - \theta_0)$ yields

$$\hat{\theta}_{\mathrm{MI},\perp} - \theta_N = \frac{1}{N}\sum_{i=1}^{N} \varphi^F_{\mathrm{MI},\perp,i} + o_p(N^{-1/2}), \tag{S14}$$

where

$$\begin{aligned}\varphi^F_{\mathrm{MI},\perp,i} = {} & (I_i^A d_i^A - 1)\{\mu(X_i;\gamma_0) + a_g(X_{1i})\} \\ & + \left\{\frac{\delta_i}{\pi_i^B(\gamma_0)} - 1\right\}\{Y_i - \mu(X_i;\gamma_0) - a_g(X_{1i})\} \\ & - G_{\mathrm{MI},\gamma}\mathcal{J}^{-1}\Phi_{\perp,i}(\gamma_0).\end{aligned} \tag{S15}$$

To identify $G_{\mathrm{MI},\gamma}$, differentiate along the population profile. Terms containing the derivatives of $\mu(X;\gamma)$ and $a_g(X_1;\gamma)$ vanish because $E\{I^A d^A - \delta/\pi^B \mid X\} = 0$. The derivative of $1/\pi^B$ gives

$$E\left[\delta O^B\{Y - \nabla_\gamma g_{\gamma_0}(X_1)\}\{Y - \mu(X;\gamma_0) - a_g(X_1)\}\right].$$

The term involving $\nabla_\gamma g_{\gamma_0}(X_1)$ is zero because $E[\delta O^B\{Y - \mu(X;\gamma_0) - a_g(X_1)\} \mid X_1] = 0$ by (16). Hence

$$G_{\mathrm{MI},\gamma} = E\left[\delta O^B(X_1, Y;\gamma_0) Y\{Y - \mu(X;\gamma_0) - a_g(X_1)\}\right].$$

Condition (R9) applied to (S15) gives

$$\sqrt{N}\{\hat{\theta}_{\mathrm{MI},\perp} - \theta_N\} \Rightarrow \mathcal{N}(0, V^F_{\mathrm{MI},\perp}),$$

where $V^F_{\mathrm{MI},\perp}$ is the limiting variance of $N^{-1/2}\sum_i \varphi^F_{\mathrm{MI},\perp,i}$. The term involving $\Phi_\perp$ propagates the root-$N$ variation of $\hat{\gamma}$ into the variance and does not alter the point estimator in (18). □

## S4.1 Bias and influence representations

This subsection states the bias-centered limits for the plain estimators and derives their influence representations.

**Theorem S1** (Plain IPW with first-order smoothing bias). *Assume Conditions (R1)–(R10). The plain IPW estimator in* (13) *satisfies*

$$\sqrt{N}\{\hat{\theta}_{\mathrm{IPW}} - \theta_N - \mathcal{B}_{\mathrm{IPW},N}\} \;\Rightarrow\; \mathcal{N}(0, V_{\mathrm{IPW}}), \tag{S16}$$

*where $\mathcal{B}_{\mathrm{IPW},N}$ is the first-order smoothing bias from estimating $g_\gamma$. The bias and variance are specified in Supplementary Section S4.1.*

**Theorem S2** (Plain MI with first-order smoothing bias)**.** *Assume Conditions (R1)–(R10). The plain MI estimator in* (14) *satisfies*

$$\sqrt{N}\{\hat{\theta}_{\mathrm{MI}} - \theta_N - \mathcal{B}_{\mathrm{MI},N}\} \;\Rightarrow\; \mathcal{N}(0, V_{\mathrm{MI}}), \tag{S17}$$

*where $\mathcal{B}_{\mathrm{MI},N}$ is the first-order smoothing bias from estimating the profiled selection component and the respondent outcome law. The bias and variance are specified in Supplementary Section S4.1.*

All population functions below are evaluated at $\gamma_0$. Write

$$p(X) = \pi_m^B(X; \gamma_0), \qquad q(X_1) = 1 - \varpi_{g,0}^B(X_1).$$

Let $b_{g,N}$, $b_{m,N}$, $b_{1,N}$, and $b_{0,N}$ be the averages across outer folds of the exact population-smoother biases of $\hat{g}$, $\hat{m}$, $\hat{\mu}_1$, and $\hat{\mu}_0$, respectively, as defined in Condition (R7).

For the plain IPW estimator, define

$$\kappa_g(x_1) = E(Y \mid X_1 = x_1, \delta = 0), \qquad H = E\big[\{1 - \pi^B(X_1, Y; \gamma_0)\}\{Y - \kappa_g(X_1)\}^2\big]\,.$$

The first-order bias is

$$\mathcal{B}_{\mathrm{IPW},N} = -E\{q(X_1)\kappa_g(X_1)b_{g,N}(X_1)\}. \tag{S18}$$

The finite-population influence contribution is

$$\begin{aligned} \varphi^F_{\mathrm{IPW},i} &= (I_i^A d_i^A - 1)\kappa_g(X_{1i}) \\ &\quad + \left\{\frac{\delta_i}{\pi_i^B(\gamma_0)} - 1\right\}\{Y_i - \kappa_g(X_{1i})\} - H\mathcal{J}^{-1}\Phi_{\perp,i}(\gamma_0). \end{aligned} \tag{S19}$$

Under Conditions (R1)–(R10),

$$\hat{\theta}_{\mathrm{IPW}} - \theta_N - \mathcal{B}_{\mathrm{IPW},N} = \frac{1}{N}\sum_{i=1}^{N} \varphi^F_{\mathrm{IPW},i} + o_p(N^{-1/2}). \tag{S20}$$

For the plain MI estimator, let

$$\mu_1(X) = E(Y \mid X, \delta = 1), \qquad \mu_0(X) = E(Y \mid X, \delta = 0), \qquad \Delta\mu(X) = \mu_1(X) - \mu_0(X),$$

$$k(X) = \Delta\mu(X)p(X)\{1 - p(X)\}, \qquad c_g(x_1) = -\frac{E\{k(X) \mid X_1 = x_1\}}{q(x_1)}.$$

The contribution from estimating the respondent outcome law is

$$\begin{aligned}\Lambda_f(W) = \delta\{Y - \mu_1(X)\} + \delta O^B(X_1, Y; \gamma_0)\{Y - \mu_0(X)\} \\ - \delta \frac{k(X)}{p(X)e^{m(X;\gamma_0)}} \left[e^{\gamma_0 Y} - e^{m(X;\gamma_0)}\right].\end{aligned} \tag{S21}$$

The first-order bias is

$$\begin{aligned}\mathcal{B}_{\mathrm{MI},N} = E\big[k(X)\{b_{g,N}(X_1) - b_{m,N}(X)\} \\ + p(X)b_{1,N}(X) + \{1 - p(X)\}b_{0,N}(X)\big].\end{aligned} \tag{S22}$$

With

$$H_\gamma = E[k(X)\{\nabla_\gamma g(X_1; \gamma_0) - \nabla_\gamma m(X; \gamma_0)\} + \{1 - p(X)\} \operatorname{Var}(Y \mid X, \delta = 0)],$$

the finite-population influence contribution is

$$\begin{aligned}\varphi^F_{\mathrm{MI},i} = (I_i^A d_i^A - 1)\mu(X_i; \gamma_0) + \{\mu(X_i; \gamma_0) - Y_i\} + \Lambda_f(W_i) \\ + c_g(X_{1i}) \left\{ I_i^A d_i^A - \frac{\delta_i}{\pi_i^B(\gamma_0)} \right\} - H_\gamma \mathcal{J}^{-1} \Phi_{\perp,i}(\gamma_0).\end{aligned} \tag{S23}$$

Under Conditions (R1)–(R10),

$$\hat{\theta}_{\mathrm{MI}} - \theta_N - \mathcal{B}_{\mathrm{MI},N} = \frac{1}{N} \sum_{i=1}^{N} \varphi^F_{\mathrm{MI},i} + o_p(N^{-1/2}). \tag{S24}$$

For both estimators, the asymptotic variance is

$$V_j = \lim_{N\to\infty} \operatorname{Var}\left\{ N^{-1/2} \sum_{i=1}^{N} \varphi^F_{j,i} \right\}, \qquad j \in \{\mathrm{IPW}, \mathrm{MI}\}, \tag{S25}$$

with the joint limit specified in (R9). The corresponding design-aware variance estimators are given in Section S1.3.

When the respondent density is correctly specified by a finite-dimensional model $f_1(y \mid x; \xi)$ and its outer-training-fold estimates are regular and asymptotically linear, while $g$ is still estimated by the empirical ratio in (S2), the outcome-law contribution $\Lambda_f$ in (S23) is replaced by $\Gamma_\xi^\top \varphi_\xi$, where

$$\Gamma_\xi = E\{\partial_\xi \mu(X; g_0, \gamma_0, \xi_0)\}, \qquad \varphi_\xi = -A_\xi^{-1} \delta s_\xi(Y \mid X; \xi_0),$$

$A_\xi = E\{\partial_\xi[\delta s_\xi(Y \mid X; \xi_0)]\}$, and the outcome-law bias terms $b_{m,N}, b_{1,N}, b_{0,N}$ are $o(N^{-1/2})$. In that important special case,

$$\mathcal{B}_{\mathrm{MI},N} = E\{k(X)b_{g,N}(X_1)\} + o(N^{-1/2}).$$

If $X_1$ has a fixed finite number of cells and $g$ is estimated cellwise, $b_{g,N} = o(N^{-1/2})$ under the usual regular finite-dimensional expansion, so the plain MI estimator is centered at the root-$N$ scale in this parametric $f_1$ specialization. For the finite-population target, the corresponding joint-law variance contains the additional model-centering component described in Section S1.3; it is absent for the unrestricted outcome-law linearization because the latter satisfies a pointwise cancellation. Moreover, $E\{\delta O^B(Y-\mu) \mid X\} = -k(X)$, so $c_g(X_1) = a_g(X_1)$. This identity provides a direct consistency check between Theorem S2 and the orthogonal score in Section 3.3.

## Proof of Theorems S1–S2

To prove Theorem S1, let $T_{\mathrm{IPW}}(g,\gamma) = E\{\delta Y/\pi^B(X_1, Y; g, \gamma)\}$. At $\gamma_0$, its Gateaux derivative in direction $h$ is

$$D_g T_{\mathrm{IPW}}[h] = -E\{\delta O^B(X_1, Y; \gamma_0) Y h(X_1)\} = -E\{q(X_1)\kappa_g(X_1)h(X_1)\}.$$

At fixed $\gamma_0$, Lemma S1 for the plain functional permits replacing the evaluation-fold nuisance difference by its population derivative, retaining a remainder $o_p(N^{-1/2})$. Apply (R7) separately to each $\hat{g}^{[-k]}$ with $a_1 = -q\kappa_g$. The deterministic terms average to $\mathcal{B}_{\mathrm{IPW},N}$. The stochastic terms average to

$$\begin{aligned} \frac{1}{K}\sum_{k=1}^{K} \mathbb{P}_{-k}\left[\kappa_g(X_1)\left\{I^A d^A - \frac{\delta}{\pi^B(X_1, Y; \gamma_0)}\right\}\right] \\ = \mathbb{P}_N\left[\kappa_g(X_1)\left\{I^A d^A - \frac{\delta}{\pi^B(X_1, Y; \gamma_0)}\right\}\right], \end{aligned}$$

since each population unit appears in exactly $K-1$ training folds. The training average is replaced by the pooled average only at this step.

Along the population profile, $\nabla_\gamma g(X_1; \gamma_0) = E(Y \mid X_1, \delta = 0) = \kappa_g(X_1)$, and

$$\frac{d}{d\gamma} E\left\{\frac{\delta Y}{\pi^B(X_1, Y; g_\gamma, \gamma)}\right\}\bigg|_{\gamma_0} = E\left[\{1-\pi^B\}\{Y - \kappa_g(X_1)\}^2\right] = H.$$

Condition (R8) and the now established root-$N$ rate give $\hat{\theta}_{\mathrm{IPW}}(\hat{\gamma}) - \hat{\theta}_{\mathrm{IPW}}(\gamma_0) = H(\hat{\gamma} - \gamma_0) + o_p(N^{-1/2})$. Substitution of $\hat{\gamma} - \gamma_0 = -\mathcal{J}^{-1}\mathbb{P}_N\Phi_\perp + o_p(N^{-1/2})$ therefore contributes $-H\mathcal{J}^{-1}\mathbb{P}_N\Phi_\perp$. Combining this with the empirical IPW term and subtracting $\theta_N =$

$N^{-1}\sum_i Y_i$ gives (S20). Conditions (R4), (R7), and (R8) make the quadratic remainder $o_p(N^{-1/2})$.

For Theorem S2, at fixed $X$ and $\gamma_0$,

$$\mu(X) = p(X)\mu_1(X) + \{1 - p(X)\}\mu_0(X),$$

so its first differential is

$$d\mu(X) = k(X)\{dg(X_1) - dm(X)\} + p(X)d\mu_1(X) + \{1 - p(X)\}d\mu_0(X).$$

Apply this differential separately to each outer-training fit. Lemma S1 controls the evaluation error and the quadratic Taylor remainder. The four population-smoother centers then average to $\mathcal{B}_{\mathrm{MI},N}$. For the centered part of $\hat{g}^{[-k]}$, the first linearization in Condition (R7), with $a_2(X_1) = E\{k(X) \mid X_1\}$, yields

$$c_g(X_1)\left\{I^A d^A - \frac{\delta}{\pi^B(X_1, Y; \gamma_0)}\right\}, \qquad c_g(X_1) = -\frac{E\{k(X) \mid X_1\}}{q(X_1)}.$$

The second linearization in (R7) contributes $\mathbb{P}_{-k}\Lambda_f$. Averaging the two contributions over folds gives

$$\mathbb{P}_N\left[c_g(X_1)\left\{I^A d^A - \frac{\delta}{\pi^B(X_1, Y; \gamma_0)}\right\} + \Lambda_f(W)\right],$$

using $K^{-1}\sum_k \mathbb{P}_{-k} = \mathbb{P}_N$. Thus the outcome-law contribution is retained along with the profiling contribution.

The total derivative along the population $\gamma$ profile is

$$\frac{d}{d\gamma}E\{\mu(X; g_\gamma, \gamma)\}\bigg|_{\gamma_0} = H_\gamma,$$

because $\partial_\gamma p = p(1-p)(\nabla_\gamma g - \nabla_\gamma m)$ and $\partial_\gamma \mu_0 = \mathrm{Var}(Y \mid X, \delta = 0)$. By (R8), $\hat{\theta}_{\mathrm{MI}}(\hat{\gamma}) - \hat{\theta}_{\mathrm{MI}}(\gamma_0) = H_\gamma(\hat{\gamma} - \gamma_0) + o_p(N^{-1/2})$. Substitution of the expansion of $\hat{\gamma}$ gives $-H_\gamma \mathcal{J}^{-1}\mathbb{P}_N\Phi_\perp$. Adding the Horvitz–Thompson evaluation term and subtracting $\theta_N$ proves (S24). Applying (R9) after the stated deterministic bias centering gives the normal limits in both theorems.

Finally,

$$E\{\delta O^B(Y - \mu) \mid X\} = -p(X)\{1 - p(X)\}\Delta\mu(X) = -k(X),$$

while $E(\delta O^B \mid X_1) = q(X_1)$. Hence $a_g(X_1) = c_g(X_1)$. Under a correctly specified finite-dimensional $f_1(y \mid x; \xi)$, standard parametric $M$-estimation replaces $\Lambda_f$ by $\Gamma_\xi^\top \varphi_\xi$ and makes the corresponding bias functions $o(N^{-1/2})$. □

# S5 Verification lemmas

## S5.1 Profile discrepancy criterion

Consider the profile criterion

$$\mathcal{Q}_N(\gamma) = \frac{1}{N}\left[\sum_{i\in B}\{\hat{\pi}_m^B(\boldsymbol{x}_i;\gamma)\}^{-1} + \sum_{i\in A} d_i^A \log \hat{\pi}_m^B(\boldsymbol{x}_i;\gamma)\right].$$

This choice has a population discrepancy interpretation. Let $\pi_{m,0}^B(X) = P_0(\delta = 1 \mid X)$. The population criterion is $\mathcal{Q}(\gamma) = E\{\pi_{m,0}^B(X)/\pi_m^B(X;\gamma) + \log \pi_m^B(X;\gamma)\}$. Whenever these expectations are finite,

$$\mathcal{Q}(\gamma) - \mathcal{Q}(\gamma_0) = E\left[\frac{\pi_{m,0}^B(X)}{\pi_m^B(X;\gamma)} - \log\frac{\pi_{m,0}^B(X)}{\pi_m^B(X;\gamma)} - 1\right] \geq 0.$$

The integrand is the Itakura–Saito divergence between the true and candidate marginal propensities (Févotte et al., 2009; Févotte and Idier, 2011); it measures relative rather than absolute discrepancy. Equality holds precisely when the two propensities agree almost surely. Moreover, the profile identity (4) gives $E\{\pi_{m,0}^B(X)/\pi_m^B(X;\gamma) \mid X_1\} = 1$. Thus inverse-marginal-propensity weighting of the respondent covariate measure yields a probability distribution, and the displayed contrast is also the Kullback–Leibler divergence from the population covariate distribution to this reweighted distribution. The derivation follows below. This interpretation motivates the criterion without identifying it with the Bernoulli pseudo log-likelihood in (S7).

To establish the population interpretation of $\mathcal{Q}_N$, consider any candidate $\gamma$ for which the tilted moments exist. The profile identity (4) implies

$$\begin{aligned} E\left\{\frac{\pi_{m,0}^B(X)}{\pi_m^B(X;\gamma)} \mid X_1\right\} &= E(\delta \mid X_1) + e^{-g_\gamma(X_1)}E(\delta e^{\gamma Y} \mid X_1) \\ &= 1. \end{aligned}$$

Hence the measure obtained by weighting the respondent covariates by $1/\pi_m^B(X;\gamma)$ is normalized and has density $\pi_{m,0}^B(X)/\pi_m^B(X;\gamma)$ relative to the population covariate law. Equivalently, the conditional respondent distribution is weighted by $P_0(\delta = 1)/\pi_m^B(X;\gamma)$. The Kullback–Leibler divergence from the population law to this reweighted law is therefore

$$E\log\frac{\pi_m^B(X;\gamma)}{\pi_{m,0}^B(X)} = \mathcal{Q}(\gamma) - \mathcal{Q}(\gamma_0).$$

Without using normalization, the same difference is the expected scalar Itakura–Saito divergence displayed in Section 3.2; see also Févotte et al. (2009); Févotte and Idier (2011) for this divergence. These are population identities. The fitted empirical criterion need not itself be nonnegative, and the augmented estimating equation is not its sample derivative.

Differentiation of $\mathcal{Q}_N$ gives (8), using $\partial_\gamma \log \hat{\pi}_m^B = -\hat{\pi}_m^B \hat{c}$. The final estimating equation adds nuisance corrections and is not the sample derivative of this criterion.

## S5.2 Remainder bounds and linearization lemmas

**Lemma S1** (Population and evaluation remainders)**.** *Under (R3)–(R6), for either orthogonal score $F$ at the true finite-dimensional parameter,*

$$\frac{1}{K}\sum_{k=1}^{K} \mathbb{P}_k\{F(\hat{\eta}_{\gamma_0}^{[-k]}) - F(\eta_0)\} = O_p(r_N^2 + r_N/\sqrt{N}) = o_p(N^{-1/2}).$$

*For a plain functional the same statement holds after retaining $K^{-1}\sum_k D_\eta PF(\eta_0)[\hat{\eta}_{\gamma_0}^{[-k]} - \eta_0]$ on the right-hand side.*

*Proof.* Let $e_k(X) = \hat{\eta}_{\gamma_0}^{[-k]}(X) - \eta_0(X)$. Taylor's formula with integral remainder, the bound on $I^A d^A$, and (R3) give

$$\begin{aligned}|P\{F(\eta_0 + e_k) - F(\eta_0) - \partial_\eta F(\eta_0) e_k\}| &\le CE\{L_2(X,Y)|e_k(X)|^2\} \\ &\le C\|e_k\|_{\mathcal{H}}^2.\end{aligned}$$

Likewise the first-derivative envelope gives $\|F(\eta_0 + e_k) - F(\eta_0)\|_2 \le C\|e_k\|_{\mathcal{H}}$. At $\gamma_0$ the population linear term is zero for the orthogonal scores. Conditional Chebyshev's inequality applied to (R6) gives $(\mathbb{P}_k - P)\Delta F_k = O_p(r_N/\sqrt{N})$. Averaging a fixed number of folds proves the assertions. This argument uses orthogonality only at $\gamma_0$; it does not discard the nonzero nuisance derivative at nearby values of $\gamma$. □

For the mean score one can also see the product structure without Taylor expansion. Fix $\gamma_0$ and suppress the common arguments $W$ and $\gamma_0$ in $h_{\mathrm{MI}}$. Since $E[\delta O^B\{Y - \mu(X) - a_g(X_1)\} \mid X_1] = 0$, direct subtraction gives

$$\begin{aligned}&P\{h_{\mathrm{MI}}(\tilde{g}, \tilde{\mu}, \tilde{a}_g) - h_{\mathrm{MI}}(g, \mu, a_g)\} \\ &\quad = E\big[\delta O^B\{1 - e^{-(\tilde{g}-g)(X_1)}\}\{\tilde{\mu}(X) - \mu(X) + \tilde{a}_g(X_1) - a_g(X_1)\}\big]\,.\end{aligned}$$

Thus an error in $g$ enters this drift multiplied by an error in the augmented regression. The derivative-envelope norm bounds the absolute value by a constant times

$$\|\tilde{g} - g\|_{\mathcal{H}}\{\|\tilde{\mu} - \mu\|_{\mathcal{H}} + \|\tilde{a}_g - a_g\|_{\mathcal{H}}\}.$$

This identity concerns smoothing error within the correctly specified model; it does not remove selection-model misspecification or the effect of an active fixed truncation.

**Lemma S2** (Fold-conditional bounds for two reference designs)**.** *Suppose the unit-level folds are independent of the population values and sampling indicators, linked records share their fold, and (R1), (R3), and (R4) hold. The evaluation bounds in (R6) hold for either (i) independent Poisson reference sampling with inclusion probability $\pi_A(X_i)$, or (ii) SRSWOR of a fixed $n_A$ independent of the population values.*

*Proof.* For a fitted functional difference write $\Delta F_k = I^A d^A \Delta F_{A,k}(X) + \delta \Delta F_{B,k}(X, Y)$. The envelopes in (R3) imply

$$E\{\Delta F_{A,k}^2 + \Delta F_{B,k}^2\} = O_p(r_N^2)$$

when the training functions are held fixed. Under Poisson sampling, evaluation units are independent of $\mathcal{T}_{-k}$ and of one another. The conditional mean of $(\mathbb{P}_k - P)\Delta F_k$ is zero and its conditional variance is at most $Cr_N^2/N$.

For SRSWOR let $\rho_{A,N} = n_A/N$ and let $n_{A,k}$ be the number of reference-sample units in $U^{(k)}$. Conditional on the training selection, $n_{A,k} = n_A - n_{A,-k}$ is fixed, and the evaluation selection is SRSWOR of that size from $N/K$ units. Put $\rho_{A,k} = Kn_{A,k}/N$ for this proof. The evaluation population values and their $B$ indicators remain independent of the training data. Hence

$$E\{(\mathbb{P}_k - P)\Delta F_k \mid \mathcal{T}_{-k}\} = \left(\frac{\rho_{A,k}}{\rho_{A,N}} - 1\right) P\Delta F_{A,k}.$$

Hypergeometric count variation gives $\rho_{A,k} - \rho_{A,N} = O_p(N^{-1/2})$, while $|P\Delta F_{A,k}| = O_p(r_N)$. The displayed mean is therefore $O_p(r_N/\sqrt{N})$. The SRS variance bound for the reference contribution and the independent-unit bound for the $B$ contribution give conditional variance $O_p(r_N^2/N)$; their within-unit covariance is controlled by Cauchy–Schwarz. This proves (R6) without treating the SRS indicators as i.i.d. The proof uses independent population-unit folds, and does not assert the same conditional calculation for membership-stratified or unlinked-record folds. □

**Lemma S3** (A sufficient kernel route for the plain linearizations)**.** *Consider either reference design in Lemma S2, independent unit folds, and deterministic bandwidths. Suppose $Y$ is bounded and, in each of finitely many discrete covariate cells with positive mass, the continuous covariates range over a compact set. For each component $\nu$, its covariate density and the kernel numerator and denominator densities belong to a Hölder class of smoothness $s_\nu > 0$. Specifically, derivatives through order $\lceil s_\nu \rceil - 1$ are bounded, and the highest-order derivatives are Hölder continuous of order $s_\nu - \lceil s_\nu \rceil + 1$. Covariate densities and population kernel denominators are bounded below on this set. Use bounded kernel families that reproduce polynomials of total degree less than $s_\nu$, are uniformly Lipschitz after rescaling, and have uniformly bounded absolute integrals, bounded rescaled absolute $s_\nu$th moments, and squared integrals $O(h_{\nu,N}^{-d_\nu})$. Require the approximate-identity property also for the adjoint integration operator. At a boundary use a kernel with the same polynomial reproduction and moment bounds; an uncorrected local-constant boundary fit is not covered by the stated bias order (Fan and Gijbels, 1996). These assumptions include ordinary convolution kernels on a compact periodic domain. For ordinary compact supports the stated boundary properties must be verified for the chosen correction.*

*Assume the population ratio maps are twice continuously differentiable on a common neighborhood, and*

$$h_{\nu,N} \to 0, \qquad \sqrt{N}\, h_{\nu,N}^{2s_\nu} \to 0, \qquad \frac{\log N}{\sqrt{N}\, h_{\nu,N}^{d_\nu}} \to 0.$$

*Then the population-center definitions and the foldwise integrated linearizations in (R7) hold. Their quadratic functional remainders are $o_p(N^{-1/2})$.*

*Proof.* It suffices to work in one continuous cell. For a generic training-fold kernel sum define

$$T_h(x) = E\{K_h(x - X)U(W)\}, \qquad \hat{T}_h^{[-k]}(x) = \mathbb{P}_{-k}\{K_h(x - X)U(W)\}.$$

The components of $U$ are selected from $I^A d^A$, $\delta$, $\delta e^{\gamma_0 Y}$, $\delta Y$, and $\delta Y e^{\gamma_0 Y}$; their dimension is fixed. Each smoother in (R7) is a ratio or log-ratio map $F(\hat{T}_h^{[-k]})$. In particular its exact

population center is obtained from

$$g_h(x_1) = \log \frac{E\{\delta e^{\gamma_0 Y} K_h(x_1 - X_1)\}}{E\{(1-\delta) K_h(x_1 - X_1)\}},$$
$$m_h(x) = \log \frac{E\{\delta e^{\gamma_0 Y} K_h(x - X)\}}{E\{\delta K_h(x - X)\}},$$
$$\mu_{1,h}(x) = \frac{E\{\delta Y K_h(x - X)\}}{E\{\delta K_h(x - X)\}}, \qquad \mu_{0,h}(x) = \frac{E\{\delta Y e^{\gamma_0 Y} K_h(x - X)\}}{E\{\delta e^{\gamma_0 Y} K_h(x - X)\}}.$$

The bandwidth and conditioning dimension in each line are those used by that component; each fold uses its own deterministic reference bandwidth. Set $b_{\nu,N}^{[-k]} = \nu_{h_{\nu,N}^{[-k]}} - \nu$.

The bounded design weights and kernel assumptions imply

$$E\|\hat{T}_h^{[-k]} - T_h\|_2^2 = O\{(Nh^d)^{-1}\}, \qquad \|\hat{T}_h^{[-k]} - T_h\|_\infty = O_p\left(\sqrt{\frac{\log N}{Nh^d}}\right).$$

For SRSWOR the selected and nonselected sets of i.i.d. population values can be conditioned on the sampling indicators; the same kernel bounds hold within those sets, whose sizes are of order $N$, with an additional $O_p(N^{-1/2})$ count fluctuation for a training fold. The latter is already included in the displayed orders. Thus all empirical denominators remain in the ratio-map neighborhood with probability tending to one. For a smooth bounded weight $a(x)$, Taylor expansion around $T_h$ gives

$$P\left[a\{F(\hat{T}_h^{[-k]}) - F(T_h)\}\right]$$
$$= (\mathbb{P}_{-k} - P)\ell_{a,h} + O_p\{(Nh^d)^{-1}\},$$
$$\ell_{a,h}(W) = \int a(x) \nabla F\{T_h(x)\}^\top \{K_h(x - X) U(W) - T_h(x)\}\, dP_X(x).$$

Here $P\ell_{a,h} = 0$. The approximate-identity property, smoothness, and bounded denominators yield $\|\ell_{a,h} - \ell_a\|_2 \to 0$. The independent-Poisson or SRS variance bound therefore gives $(\mathbb{P}_{-k} - P)(\ell_{a,h} - \ell_a) = o_p(N^{-1/2})$. The quadratic term is $o_p(N^{-1/2})$ by the bandwidth condition.

For $g$ the limiting variable is

$$\ell_a(W) = \frac{a(X_1)}{q(X_1)} \{\delta e^{\gamma_0 Y - g(X_1)} - (I^A d^A - \delta)\}.$$

For $(\mu_1, \mu_0, m)$ the separate limiting regression residuals are

$$\frac{\delta}{p}(Y - \mu_1), \qquad \frac{\delta e^{\gamma_0 Y}}{p e^m}(Y - \mu_0), \qquad \frac{\delta}{p e^m}(e^{\gamma_0 Y} - e^m).$$

Multiplication by $(p, 1-p, -k)$ and addition give $\Lambda_f$. This proves both displays in (R7), with $\mathbb{P}_{-k}$ on their right-hand sides. Their population smoothing centers have order $O(h^s)$ and the pointwise functional Taylor remainder has integrated order $O_p\{h^{2s} + \log N/(Nh^d)\}$, which is $o_p(N^{-1/2})$ under the stated conditions. Finite discrete cells are treated by ordinary ratio linearization and have population smoothing center zero. □

Lemma S3 is a concrete sufficient route for (R7), not a verification of a generic learner or of an uncorrected boundary smoother. A local-polynomial implementation can be treated by including its moment matrix in $T_h$ and requiring uniform nonsingularity of the normalized population moment matrix. For unbounded supports, the same argument requires weighted integrated bounds and a tail remainder $o_p(N^{-1/2})$; replacing a density lower bound by pointwise positivity alone does not prove it. Data-dependent bandwidths also require the substitution condition in (R7). The displayed bandwidth range is feasible when $2s_\nu > d_\nu$. In particular, $h_{\nu,N} \asymp N^{-1/(2s_\nu+d_\nu)}$ then satisfies both quadratic-remainder conditions, including the logarithmic factor. For general smoothness $s$ and dimension $d$, the usual unweighted regression error balance is

$$O_p\{h^s + (nh^d)^{-1/2}\} = O_p\{n^{-s/(2s+d)}\} \quad \text{when } h \asymp n^{-1/(2s+d)};$$

see Stone (1982); Fan and Gijbels (1996). A local polynomial of degree at least $\lceil s \rceil - 1$, with a uniformly nonsingular local moment matrix, can exploit this smoothness. For higher-order kernels, the corresponding moment cancellations and boundary correction are required. Larger $s$ relaxes the sufficient condition $2s > d$ only when the estimator actually exploits that extra smoothness; using a Gaussian kernel in a local-constant fit does not by itself give a bias order above two.

The numerical rule is the $s = 2$ specialization. For this specialization, the generic componentwise $o_p(n^{-1/4})$ argument requires $d < 4$; at $d = 4$ it gives only $O_p(n^{-1/4})$. This does not imply that every four-dimensional problem fails, but it does mean that the displayed rate alone does not prove the required negligible remainder. Boundary and weighted tail bounds, the sampling-design conditions, and rates for all derived nuisance quantities remain separate requirements. The common componentwise rate is sufficient rather than necessary; sharper, model-specific product bounds can sometimes be weaker (Chernozhukov et al., 2018).

For an uncorrected centered plain estimator, a sufficient additional undersmoothing condition is $\sqrt{N}h_{\nu,N}^{s_\nu} \to 0$ for every bias-contributing component. Combined with the stochastic-remainder condition in the lemma, this simple bandwidth route requires $s_\nu > d_\nu$ (up to the stated logarithmic factor). This stronger condition is not imposed on the bias-explicit expansions or on the bias-removal comparison (23).

For variance consistency in the compact-support case of Lemma S3, suppose the auxiliary conditional means defining $c, \xi, a_g$ are smooth to their corresponding kernel orders and their second-stage kernel weights have bounded absolute sums with probability tending to one. Their errors are bounded by the first-stage uniform error plus their own regression error. They are therefore uniformly consistent at the displayed rates. The same argument with the additional bounded responses $Y^j e^{\gamma Y}$ verifies profile-derivative consistency locally in $\gamma$. With separated roots, the analogous uniform bounds on $\Gamma$ verify the uniform convergence part of (R2).

For Poisson sampling the oracle versions of $N\widehat{\mathcal{V}}_A$ and $N\widehat{\mathcal{V}}_B$ are averages of bounded independent variables, so the law of large numbers gives their variance limits. For SRSWOR, $N\widehat{\mathcal{V}}_A = (1-n_A/N)(N/n_A)s_{u,A}^2$ and $s_{u,A}^2 \to_p \mathrm{Var}\{u_A(X)\}$; the independent Bernoulli $B$ part has the same law of large numbers as before. Uniform consistency of the bounded fitted coefficients makes the change in either unscaled variance $o_p(N^{-1})$ by Cauchy–Schwarz. These verify the substitution part of (R10) in this sufficient case. For unbounded outcomes or other reference designs, the corresponding weighted laws of large numbers and the actual design variance remain necessary; these conclusions cannot be inferred from the bandwidth orders alone.

**Lemma S4** (Common first-order fluctuation of the two MI estimators)**.** *At the true distribution, for the unrestricted outcome-law contribution in* (S21)*,*

$$\Lambda_f(W) = \frac{\delta}{\pi^B}(Y-\mu), \qquad c_g = a_g, \qquad H_\gamma = G_{\mathrm{MI},\gamma}.$$

*Consequently* $\varphi^F_{\mathrm{MI},i} = \varphi^F_{\mathrm{MI},\perp,i}$*. If both estimators use the same profiled fits and the conditions of Theorem S2 hold, then* (23) *follows.*

*Proof.* Suppress arguments and write $\Delta\mu = \mu_1 - \mu_0$. Since $\mu = \mu_0 + p\Delta\mu$ and $e^{\gamma_0 Y - m} =$

$pO^B/(1-p)$, direct substitution gives

$$\Lambda_f = \delta\big[(Y-\mu_1) + O^B(Y-\mu_0) - (1-p)\Delta\mu\{e^{\gamma_0 Y - m} - 1\}\big]$$
$$= \delta(1+O^B)(Y-\mu) = \frac{\delta}{\pi^B}(Y-\mu).$$

Moreover $E\{\delta O^B(Y-\mu) \mid X\} = -k(X)$ and $E(\delta O^B \mid X_1) = q(X_1)$ give $c_g = a_g$.

Conditioning the expression for $G_{\mathrm{MI},\gamma}$ on $X$ yields

$$G_{\mathrm{MI},\gamma} = E\big[(1-p)\operatorname{Var}(Y \mid X, \delta = 0) - k\mu_0 - (1-p)\mu_0 a_g\big].$$

Use $\nabla_\gamma g_{\gamma_0} = \kappa_g$ and $\nabla_\gamma m_{\gamma_0} = \mu_0$. Then

$$H_\gamma - G_{\mathrm{MI},\gamma} = E\{k\kappa_g + (1-p)\mu_0 a_g\} = 0,$$

because $E\{(1-p)\mu_0 \mid X_1\} = q\kappa_g$ and $E(k \mid X_1) = -qa_g$. Substitution of these three identities into (S23) gives exactly (S15). Subtracting (S14) from (S24) proves (23). In the parametric-$f_1$ specialization, $\Gamma_\xi^\top \varphi_\xi$ is not generally equal to $\delta(Y-\mu)/\pi^B$, so this comparison does not follow there. □

At the truth, the coefficient decomposition in Section S1.3 satisfies

$$B_\gamma(X,Y) = -A_\gamma(X)/\pi^B(X_1,Y), \qquad \Phi_\perp = A_\gamma(X)\left(I^A d^A - \frac{\delta}{\pi^B}\right).$$

This follows by combining $\Phi_0$ with $\Phi_m$ using $p_0 = p$ and $e^{m-g} = (1-p)/p$. The finite-population influence contributions therefore have the centered sample-specific representation in (R9). Conditional on the finite population, reference and nonprobability selection errors are independent. Their variances add to the scalar expression in (R9), and (R10) justifies nuisance substitution and the sample-separated estimators. In the parametric-$f_1$ case the additional conditional mean $N^{-1}\sum_i r_{\xi,i}$ is random under the superpopulation law and has zero covariance with the two conditionally centered sampling errors. This gives the model-centering variance specified in Section S1.3.

# S6 Identification under conditional completeness

The following argument adapts the complete-case completeness approach of Miao et al. (2024) to the present two-sample setting. The reference sample with known positive design

probabilities identifies the population law of $X$ through its design-weighted law. Together with the observed respondent measure and its population selection fraction, this identifies $\pi^B_{m,0}(X) = P_0(\delta = 1 \mid X)$ and $f_1(y \mid x) = f(y \mid x, \delta = 1)$. No record linkage is needed for this population identification argument.

**Lemma S5** (Identification under conditional completeness)**.** *Suppose the selection model, exclusion restriction, conditional completeness, nondegeneracy, and exponential-moment conditions in Section 2.2 hold, with $0 < \pi^B_{m,0}(X) < 1$ almost surely. Then the observed two-sample laws identify $(g_0, \gamma_0)$ and the population law of $(X, Y)$. In particular, $\pi^B_m(X;\gamma) = \pi^B_{m,0}(X)$ almost surely for a candidate $\gamma \in \Gamma$ implies $\gamma = \gamma_0$ and $g_\gamma = g_0$ almost surely. Whenever the population criterion in Section 3.2 is finite, $\gamma_0$ is its unique minimizer.*

*Proof.* Any candidate $(g, \gamma)$ compatible with the observed laws must satisfy

$$E_0[\exp\{-g(X_1) + \gamma Y\} \mid X, \delta = 1] = \frac{1 - \pi^B_{m,0}(X)}{\pi^B_{m,0}(X)}.$$

Indeed, this is the inverse-propensity identity (3) after subtracting one. If two candidates $(g_1, \gamma_1)$ and $(g_2, \gamma_2)$ induce the same observed laws, their difference obeys

$$E_0\left[e^{-g_1(X_1)+\gamma_1 Y} - e^{-g_2(X_1)+\gamma_2 Y} \mid X_1, X_2, \delta = 1\right] = 0.$$

For almost every fixed $x_1$, the integrand is square-integrable under $Y \mid X_1 = x_1, \delta = 1$ by the exponential-moment assumption. Conditional completeness therefore gives

$$e^{-g_1(X_1)+\gamma_1 Y} = e^{-g_2(X_1)+\gamma_2 Y} \quad \text{almost surely among respondents.}$$

Taking logarithms yields $(\gamma_1 - \gamma_2)Y = g_1(X_1) - g_2(X_1)$. The conditional nondegeneracy of $Y$ implies $\gamma_1 = \gamma_2$, and then $g_1 = g_2$ almost surely. Positivity transfers the equality in $X_1$ from the respondent law to the population law. The remaining distribution is recovered from

$$f(y \mid x) = \pi^B_{m,0}(x) f_1(y \mid x) \left[1 + \exp\{-g_0(x_1) + \gamma_0 y\}\right].$$

If $\pi^B_m(X;\gamma) = \pi^B_{m,0}(X)$, the profile pair $(g_\gamma, \gamma)$ satisfies the same conditional odds equation, so the preceding argument applies. The Itakura–Saito contrast in Section 3.2 is zero exactly when these marginal propensities agree. Thus its unique population minimizer is $\gamma_0$. □

Completeness here is a sufficient condition for identification of the selection model and population law. It is not a claim that each missing finite-population outcome is recovered, nor does it by itself establish regular root-$N$ estimation. The separated-moment and non-singularity requirements in (R2) and (R8) concern the particular estimating equation and are imposed separately.

# S7 Additional numerical results

## S7.1 Full simulation summaries

Table S1 records the design diagnostics for the continuous-$X_1$ study. The complete categorical-$X_1$ comparison is Table 3 of the main paper, and the instrument-set comparison, with the partial correlations of the instruments, is Table 7 of the main paper.

Table S1: The three designs of Section 5.1. Population values under $X_1 \sim \mathrm{Unif}(-1.5, 1.5)$ and $X_2 \sim N(0, 1)$. The residual is that from projecting $g(X_1)$ on $X_1$; $\mathrm{corr}(\mathrm{resid}, m_1)$ is its correlation with the respondent conditional mean $m_1(X)$.

| Design | $\phi_1$ | $\phi_2$ | $\beta_4$ | $\mathrm{sd}\{g(X_1)\}$ | sd(resid) | $\mathrm{corr}(\mathrm{resid}, m_1)$ |
|---|---|---|---|---|---|---|
| MNAR-I | $-0.45$ | 0 | 0 | 0.390 | 0.000 | 0.000 |
| MNAR-II | $-0.20$ | 0.5 | 0 | 0.410 | 0.351 | 0.000 |
| MNAR-III | $-0.20$ | 0.5 | 0.5 | 0.410 | 0.351 | $+0.394$ |

## S7.2 Comparison of the two mass-imputation estimators

Section 5.2 of the main text reports the paired comparison between the plain mass-imputation estimator (14) and its orthogonalized version (18). Two features of the continuous-$X_1$ setting account for the narrowness of the gap between them. The correction $\hat{a}_g(x_1)$ is itself estimated by kernel smoothing, so higher-order remainder terms remain at these sample sizes. Its error enters the orthogonal mean drift through products, as shown in Lemma S1; the results do not isolate a separate first-order smoothing bias of the correction. In addition, the smoothing bias of $\hat{\mu}$, which (14) does not control, is smaller at these sample sizes than the rate condition alone would suggest. The observed biases are small relative to the estimators' standard deviations. The matching estimated standard

errors in Table 2 of the main text are a consequence of using the shared linearized variables of Lemma S4 for both estimators, not an independent numerical verification of the limiting result.

## S7.3 Completeness and discrete instruments

This subsection expands on the identification status of the four instrument specifications in Section 6.3 of the main text.

Completeness is an additional requirement, not a consequence of these exclusion restrictions, and the specifications here differ in whether the sufficient condition is available at all. A discrete instrument takes finitely many values, so an instrument set consisting only of discrete variables cannot satisfy full $L^2$ completeness for a continuously distributed outcome at fixed $X_1$ (Miao et al., 2024); the same limitation applies to the discrete components SEX and AGE wherever they appear. Among the four specifications of Table 7 of the main text, $S_1$, $S_2$ and $S_3$ each retain at least one continuous instrument, while $S_4$ rests on a discrete instrument alone. When $Y$ is modeled as continuous, specifications whose instruments are all discrete are therefore sensitivity analyses outside this sufficient completeness framework rather than applications of its identification result. Failure of the sufficient condition does not by itself imply nonidentification of the scalar linear-tilt model, which is far weaker than the nonparametric target the completeness condition is designed for.

## S7.4 Sensitivity to the propensity floor

The fitted conditional propensity is floored at $5 \times 10^{-3}$ in the study of Section 6. Because a fixed active floor is not covered by the negligibility requirement in (R3), the sensitivity of the reported figures to its value is an empirical question. We repeat the $M_1$ design over the same 500 realized samples at floors $10^{-3}$, $5 \times 10^{-3}$ and $10^{-2}$, holding all other settings fixed.

The number of floored records changes by a factor of more than two hundred across the three settings. Only the inverse-probability-weighting estimator responds appreciably, and it responds monotonically: flooring moves weight away from the units with the smallest fitted propensities, which under a positive tilt are the units with the largest outcomes, so

Table S2: Sensitivity to the propensity floor in the study of Section 6, design $M_1$, over the same 500 realized samples at each floor. RB is the relative bias of the average estimated standard error against the Monte Carlo standard deviation.

| | $10^{-3}$ | $5 \times 10^{-3}$ | $10^{-2}$ |
|---|---|---|---|
| Floored records, total over 500 replicates | 9 | 222 | 2006 |
| Bias of $\hat{\theta}_{\mathrm{OP.ipw}}$ | −0.1675 | −0.1730 | −0.1896 |
| Bias of $\hat{\theta}_{\mathrm{OP.mi}}$ | 0.0068 | 0.0061 | 0.0063 |
| Bias of $\hat{\theta}_{\mathrm{OP.mi}\perp}$ | −0.0052 | −0.0067 | −0.0099 |
| Monte Carlo SD of $\hat{\theta}_{\mathrm{OP.mi}\perp}$ | 0.0636 | 0.0555 | 0.0511 |
| Mean estimated SE | 0.0546 | 0.0526 | 0.0509 |
| RB (%) | −14.1 | −5.2 | −0.4 |
| Coverage (%) | 94.4 | 94.4 | 94.4 |

a larger floor makes the negative bias larger. The two mass-imputation estimators replace the propensity-weighted outcome by $\hat{\mu}(X)$ and move by less than a tenth of their Monte Carlo standard deviations over the same range, with coverage of 94.4% at every floor. A larger floor does, however, trade bias for stability in the variance estimate: the Monte Carlo standard deviation of $\hat{\theta}_{\mathrm{OP.mi}\perp}$ falls from 0.0636 to 0.0511 and the relative bias of its average estimated standard error improves from −14.1% to −0.4%, the average being 0.0546 against 0.0636 at the lowest floor and 0.0509 against 0.0511 at the highest, while its own bias grows from −0.0052 to −0.0099. The floor used in Section 6 lies between the two.

This comparison is confined to one design and one mechanism and does not establish that the floor is negligible in general; it locates the sensitivity in the estimator that weights by the fitted propensity directly.

## S7.5 Bandwidth sensitivity

The bandwidth enters the proposed estimators in two places, and a single bandwidth cannot show what either does. It controls the smoothing bias of $\hat{g}_\gamma(\cdot)$, which propagates to the fitted propensity score, and it controls the variability of $\hat{g}_\gamma(\cdot)$, which propagates to $\hat{\gamma}$ through the orthogonal score. This section varies the bandwidth in the study of Section 6, where both effects are visible at the sample sizes involved.

We repeat the $M_1$ design with the bandwidth multiplier set to 1, 2 and 3, holding

all other settings fixed, with 500 Monte Carlo replicates and the same realized samples throughout. Table S3 reports the results.

Table S3: Bandwidth sensitivity in the study of Section 6, design $M_1$. Both $h_1$ and $h_2$ are multiplied by the factor shown. Based on 500 Monte Carlo replicates with the same realized samples throughout. "$\sum \delta/\hat{\pi}$" is $N^{-1}\sum_i \delta_i/\hat{\pi}_i^B$, which estimates one. True values: $\theta = 4.687$, $\gamma = 1.047$. Bias, Monte Carlo SD and the median absolute deviation are computed over all 500 replicates; the quantities involving the estimated standard error, including coverage, are computed over the replicates for which it is defined. The count of non-positive denominators is over all replicates and all evaluation points, of which there are about $1.8 \times 10^6$.

| | Bandwidth multiplier | | |
|---|---|---|---|
| | 1 | 2 | 3 |
| *Estimation of $\gamma$* | | | |
| Mean of $\hat{\gamma}_{\mathrm{OP}}$ | 1.110 | 1.010 | 1.020 |
| SD of $\hat{\gamma}_{\mathrm{OP}}$ | 1.053 | 0.454 | 0.248 |
| Boundary outcomes (%) | 20.2 | 3.2 | 0.0 |
| *Estimation of $\theta$ by $\hat{\theta}_{\mathrm{OP.mi}\perp}$* | | | |
| Bias | 0.0040 | −0.0086 | −0.0067 |
| Monte Carlo SD | 0.1861 | 0.0892 | 0.0555 |
| Median absolute deviation | 0.2042 | 0.0561 | 0.0491 |
| Median estimated SE | 0.0850 | 0.0453 | 0.0511 |
| Mean estimated SE | 2.1569 | 0.0689 | 0.0526 |
| Largest estimated SE | 726.0 | 1.89 | 0.18 |
| Replicates with SE above 0.3 | 101 | 12 | 0 |
| Replicates with SE undefined | 24 | 6 | 0 |
| Coverage (%) | 74.6 | 90.9 | 94.4 |
| Non-positive denominators in (S2) | 510 | 71 | 4 |
| $\sum \delta/\hat{\pi}$ | 0.9679 | 0.9640 | 0.9639 |

Three quantities move together as the multiplier falls, and their common movement identifies the source.

The dispersion of $\hat{\gamma}$ rises from 0.248 to 1.053, and the search returns a boundary outcome in 20.2% of replicates at multiplier one against none at multiplier three. A narrower bandwidth makes $\hat{g}_\gamma(\cdot)$ more variable, the orthogonal score flatter near its root, and the root correspondingly less well determined.

The variance estimator inherits that instability. At multiplier one the mean estimated standard error is 2.1569 against a median of 0.0850: a small number of replicates in which the Jacobian $\hat{\mathcal{J}}_\gamma$ is close to zero produce standard errors orders of magnitude larger than

the rest. At multiplier two the same happens in 12 of 500 replicates, though less severely, with a maximum of 1.89. At multiplier three there are no such replicates, the mean and median agree to within 3%, 0.0526 against 0.0511, and the median estimated standard error, 0.0511, matches the median absolute deviation of $\hat{\theta}_{\mathrm{OP.mi}\perp}$, 0.0491, to within 5%. Coverage follows: 74.6%, 90.9% and 94.4%.

The normalization diagnostic $N^{-1}\sum_i \delta_i/\hat{\pi}_i^B$ moves from 0.9639 at multiplier three to 0.9679 at multiplier one. This quantity depends on the fitted selection model rather than the outcome regression. Its deviation from one can reflect smoothing, propensity flooring, and estimation of $\gamma$; the comparison does not isolate the contribution of each. The improvement is small relative to the increase in boundary outcomes and dispersion of $\hat{\gamma}$. We therefore use a multiplier of three in Section 6. Exact matching on the discrete coordinate sharpens this trade-off rather than softening it. Matching restricts each local average to records of the same sex, roughly halving the effective local sample, and removing that coordinate from $d$ narrows the bandwidth itself through the exponent $-1/(d+4)$. The two act in the same direction, so the multipliers that were already sparse are the ones most affected: coverage is 74.6% at multiplier one and 90.9% at multiplier two, against 94.4% at multiplier three.

These results illustrate a finite-sample trade-off over the three multipliers considered. The bandwidth strongly affects root stability and estimated standard errors. Boundary frequency and weight normalization are useful diagnostics, but neither establishes that smoothing bias or the theoretical remainder is negligible.



\end{document}